\documentclass[nofootinbib,aps,showpacs,twocolumn,amsmath,amssymb,superscriptaddress]{revtex4}
\usepackage{enumitem} 
\usepackage{endnotes}
\usepackage{algcompatible}
\usepackage{algorithm2e}

\usepackage{hyperref} 
\usepackage{graphicx}
\usepackage{dcolumn}
\usepackage{bm}
\usepackage[super]{nth}
\usepackage{subcaption}
\usepackage{ragged2e}
\DeclareCaptionJustification{justified}{\justifying}
\usepackage{hvfloat}
\extrafloats{100}

\newcommand{\bc}{\begin{center}}
\newcommand{\ec}{\end{center}}
\newcommand{\be}{\begin{equation}}
\newcommand{\ee}{\end{equation}}
\newcommand{\bea}{\begin{eqnarray}}
\newcommand{\eea}{\end{eqnarray}}
\newcommand{\beq}{\begin{eqnarray*}}
\newcommand{\eeq}{\end{eqnarray*}}
\newcommand{\bv}{\left( \begin{array}{c} }
\newcommand{\ev}{\end{array} \right) }

\newcommand{\ve}[1]{\boldsymbol{#1}} 

\renewcommand{\thetable}{\Alph{table}}

\begin{document}
\title [mode = title]{Fast Super-Paramagnetic Clustering}

\author{Lionel Yelibi}
\email{ylblio001@myuct.ac.za}

\author{Tim Gebbie}
\email{tim.gebbie@uct.ac.za}

\affiliation{Department of Statistical Science, University of Cape Town, Rondebosch, South Africa}

\begin{abstract}
We map stock market interactions to spin models to recover their hierarchical structure using a simulated annealing based Super-Paramagnetic Clustering (SPC) algorithm. This is directly compared to a modified implementation of a  maximum likelihood approach we call fast Super-Paramagnetic Clustering (f-SPC). The methods are first applied standard toy test-case problems, and then to a data-set of 447 stocks traded on the New York Stock Exchange (NYSE) over 1249 days. The signal to noise ratio of stock market correlation matrices is briefly considered. Our result recover approximately clusters representative of standard economic sectors and mixed ones whose dynamics shine light on the adaptive nature of financial markets and raise concerns relating to the effectiveness of industry based static financial market classification in the world of real-time data analytics. A key result is that we show that f-SPC maximum likelihood solutions converge to ones found within the Super-Paramagnetic Phase where the entropy is maximum, and those solutions are qualitatively better for high dimensionality data-sets.
\end{abstract}

\pacs{05.10.Ln, 75.10.Nr, 89.65.Gh}
\keywords{Econophysics, Potts Models, Unsupervised Learning, Clustering, Phase Trasitions, Simulated Annealing}

\maketitle


\section{Introduction} \label{sec:intro}

We consider the problem of unsupervised statistical learning for feature selection and classification of financial time-series data from similarity measures that can be represented as correlation matrices. Concretely, we consider Potts model \cite{wu1982potts} based \cite{blatt1996superparamagnetic,blatt1997data} methods optimized for performance \cite{giada2001data,marsili2002dissecting,hendricks2016detecting,hendricks2016high} using a Maximum Likelihood Estimation (MLE) approach based on the ground-state Noh Ansatz \cite{noh2000model} compared to the finite-temperature approach using a cooling schedule to select configurations based on the susceptibility \cite{blatt1996superparamagnetic,blatt1997data}. 

We compare the cluster configuration from the fast clustering algorithms based on the ground state configurations \cite{hendricks2016detecting,hendricks2016high} (Sec. \ref{ssec:alg_fspc}) with those based on finite temperature Simulated Annealing (SA) based Monte-Carlo Markov-Chain (MCMC) \cite{swendsen1987nonuniversal,hoshen1976percolation} (see Sec. \ref{ssec:alg_spc}) to generate the full dendrogram of cluster configurations \cite{blatt1996superparamagnetic,blatt1997data}.
It was shown that the clustering structure of financial assets are time horizon dependent in \cite{bonanno2001high}, and given a translation of correlations into the Euclidean distance, they can be represented using Minimal Spanning Trees (MST) \cite{kruskal1956shortest}.

The Super-Paramagnetic Clustering (SPC) method originally developed in the early 1990s is a universal data clustering method \cite{blatt1996superparamagnetic} and has acquired a certain popularity for its implementation of the Maximum Entropy Principle (MEP) \cite{jaynes2003probability} where no assumptions are made about the distributions found in the data, and the number of clusters is revealed rather than predefined. SPC can be used in any environment as long as the features are embedded in an appropriate similarity metric. The method has been applied to chemical data using a sequential method and the Tanimoto similarity measure  \cite{ott2004sequential}, the detection and classification of spiking activity in neuronal tissue in \cite{quiroga2004unsupervised}, the identification of regions of the brain with shared functionality in \cite{stanberry2008functional}, yeast genes profiles in \cite{getz2000super}, histone modification data \cite{li2011potts}, and image segmentation in \cite{abramov2010real}.

Marsili and Giada \cite{giada2001data} were able to developed an efficient maximum likelihood clustering method for high dimensional data using the same spin-model inspired approach used in the SPC method. However, given the ill-posed nature of clustering they chose to evaluate the likelihood model $L_c$ for robustness relative to different optimization methods in \cite{giada2002algorithms}. The method is then applied to the detection of clusters of assets, and financial markets states in \cite{marsili2002dissecting} to uncover collective behavior of agents in a complex systems.  Hendricks {\it et al} created a GPU based Parallel Genetic Algorithm (PGA) implementation to maximize $L_c$ in near real time clustering of market states for quantitative trading applications \cite{hendricks2016detecting}. 


In this work we explore the relationship between the SPC method of Domany-Wiseman-Blatt \cite{blatt1996superparamagnetic} as compared to that of Giada-Marsili \cite{giada2001data}. A key finding here is that we are able to confirm that the likelihood method is Super-Paramagnetic and conforms well to the entropy method when dimensionality is sufficiently high (see Sec. \ref{sec:disc}). We are also able to further optimize the algorithm based on the PGA implementation \cite{hendricks2016detecting}. However, more importantly, what we are really able to demonstrate is the variety of clustering problems that the SPC can quickly and easily handle, with the advantage of leveraging a mature and well-understood foundational theoretical framework from statistical mechanics that has many, if not most, viable alternative algorithms, as special cases. We are of the view that SPC type models grounded in the maximum entropy principle and within the general framework of energy-based machine learning offer a variety of research and development opportunities for building better and faster unsupervised learning algorithms.

Towards building this argument the paper is organized as follows: Section \ref{sec:potts} discusses a brief overview of Erdos-Renyi's Random Graphs (see Sec. \ref{ssec:randomgraphs}), and their connection to the Potts models as special cases. We describe the inhomogeneous Potts Model as a data-clustering engine (see Sec. \ref{ssec:pottsmodel}). In Section \ref{ssec:spc}, the implementations with the SPC algorithm (Sec. \ref{sssec:maxent}) followed by Marsili's maximum likelihood methods (Sec. \ref{sssec:mle}). We then proceed to discussion validations procedures linking maximum likelihood methods to Super-Paramagnetic Clustering in Sec. \ref{ssec:validation}. Section \ref{sec:preprocessing} goes over our choice of similarity metric (Sec. \ref{ssec:distance}), data-preprocessing (Sec. \ref{ssec:scaling}), and time-series noise filtering (Sec. \ref{ssec:noise}). Section \ref{sec:cases} provides toy test cases, and stock market applications. In Section \ref{sec:disc} a summary analysis of the results and their implications, and finally in Section \ref{sec:conclusion} the conclusion, and potential directions for further research on similar topics are mentioned. 

\section{Potts Clustering} \label{sec:potts} 

\subsection{Random Graphs} \label{ssec:randomgraphs}

A graph is a mathematical model which formalizes pairwise relationships between objects \cite{west2001introduction}. Graphs are popular in complexity sciences because they provide a framework to model large systems of interacting components by representing the system directly through the pair-wise relationships between the components. The field has seen the rise of many models, each with their own assumptions and various nuance, almost all are of the generative form based on the premise of bottom-up causation. One feature that is useful in our context is that one can observe certain types of emergent dynamics {\it i.e.} ``phase transitions'' \cite{brush1967history}.

A general class of models called the Random Cluster model \footnote{See Grimmett \cite{grimmett2006random} and references therein.} was developed by Fortuin-Kasteleyn in 1972 \cite{fortuin1972random}. It is a ``random'' graph generated by a probability distribution 
\begin{equation} \label{eq:1} W(N) = \frac{q^{C_{N}}}{Z}  \prod_{\langle i,j \rangle }^N P_{ ij }^{n_{ij}} (1-P_{ ij })^{(1 - n_{ij})} \end{equation}
where $N$ is a given edge configuration (or adjacency matrix of $n_{ij}$ ), $C_{N}$ is the number of clusters given $N$, $P_{ij}$ is the probability of nodes $i$ and $j$ being connected, and $Z$ is the normalization constant (also called the partition function in statistical mechanics \cite{sokal2005multivariate} ). The adjacency matrix is linked to the probabilities $P_{ij}$ by picking a value $P$ such that if $P_{ij} > P$ then $n_{ij}=1$ or $0$ otherwise. $W$ is essentially the probability of the graph being connected given an adjacency matrix $N$.

If we now set $q = 1$, the random cluster model reduces to a Erdos-Renyi's random graph \cite{erds1960evolution}. Given the existence of only one class on the graph, bonds are linked independently from their respective states (i.e. they all have the same state) with equal probability $P_{ij} = \frac{1}{n}$ with $n$ the number of nodes on the graph \cite{erds1960evolution}. An important generalization of this idea is the Barabási-Albert model \cite{barabasi1999emergence}; this model works slightly differently: it starts with a low number of connected nodes $m_0$, adds new nodes one at a time, one new node is able to connect to $m < m_0$ nodes, and every time a node $i$ is connected its degree $k_i$ increases. The probability of a node connecting to another is $P_{ij} = \frac{k_i}{\sum k_j}$. This means as a node $i$ makes connections it becomes ``popular'' and succeeding nodes have a higher likelihood of connecting to that same node: this is the principle of ``preferential attachment'' which hopes to explains how some social networks are formed. These models are all based on a generative model that builds on microscopic causal relationships from the system components to the bulk.

The Fortuin-Kasteleyn random cluster model is closely related to the Potts model via its distribution \begin{equation} \label{eq:2}
P(S,N) = \prod_{\langle i,j \rangle } \big[ \left(1- P_{ ij }\right) \left(1- n_{ij}\right) + P_{ ij } n_{ij} \delta_{s_i,s_j} \big]\end{equation} which is the conditional probability of the spin configuration $S$ given the edge configuration $N$ \cite{edwards1988generalization}. The marginal probability $W$ is recovered by summing over all spin configurations. The major difference the Potts model brings is entropy maximization which assumes an exponential Boltzmann distribution of edges connections $P_{ij} = 1-e^{-J_{ij}}$ with $J_{ij}$ as the pairwise probabilities. The strength $J_{ij}$ captures the closeness between nodes, and, with the clusters membership, defines the topology of the graph. It's a central variable of the model. This is similar to the Bianconi-Barabási model \cite{bianconi2001bose} which introduces a fitness $\eta_i$ which plays a related role as an add-on to the Barabási-Albert model \cite{barabasi1999emergence}.

\subsection{The Potts Model} \label{ssec:pottsmodel}

The Ising model \cite{brush1967history} simulates the existence of phase transitions in large systems of interacting particles. The model consists in the representation of a n-Dimensional plane. Ising's PhD Thesis \cite{brush1967history} solved the 1D problem, which showed no phase transition, while Onsager provided an analytical solution using a transfer-matrix method \cite{onsager1944crystal} for the 2D case. If we consider observations in our data sets as nodes with edges which link nodes together it becomes natural to consider the data-set in the context of a Potts model \cite{wu1982potts}. An edge is active or inactive with probability dependent on the distance between two nodes. The collection of nodes and edges form the graph which is navigated for clustering. Every node can be assigned, for example, a +1 or -1 spin (for the Ising model). Interactions are permitted by randomly changing the spin values in the graph, and then accepting or rejecting new configurations is implemented using the Swendsen-Wang MCMC algorithm at every step \cite{swendsen1987nonuniversal}.

The Potts model \cite{wu1982potts} is a generalization of the Ising \cite{brush1967history} model allowing the system to accept a higher $q$ spin values instead of 2. The parameter $q$ can be compared to the $K$ value in K-means used to fix the number of clusters. $q$ is the maximum number of classes: it must be chosen to be big enough to avoid clusters forcefully merged together. The only inconvenience to a relatively high $q$ is the additional computational cost needed to perform the statistics after the system reaches thermal equilibrium.

The model is governed by a Hamiltonian Energy \footnote{ \label{ft:rbm} The more general Potts Hamiltonian can be contrasted with one of its special cases in the form of the energy of a Boltzmann machine with bias $\ve b$ and weight matrix $W$ for features $\ve x$: $E_{b,W}= - \ve b^T \ve x + \ve x^T W \ve x$ with partition function $Z_S= \sum_x e^{-E_{b,W}(x)}$ \cite{hinton2006reducing,osogami2017boltzmann,ackley1985learning,smolensky1986information}} equation \cite{wu1982potts}

\begin{equation} \label{eq:hs} H_S=\sum_{\langle i,j \rangle}J_{ij}\left(1-\delta_{s_i,s_j}\right) \end{equation}

with: $S=[s_i,...,s_N]$ the spin vector assigned to our data, spins $s_i \in [1,...,q]$, and $N$ nodes. The Kronecker delta which is 1 for equal spins and 0 otherwise. For data embedded in a metric space the Euclidean distance function $ d_{ij} = ||x_i-x_j|| $ is computed between two nodes.

$d_{ij}$ is fed to the strength function which, in turn, measures similarity. Many models for strength exist but their central feature is they must decrease with distance. This is typically achieved with a function of the type $e^{- d_{ij}}$ or a power law as seen in \cite{blatt1996superparamagnetic}:

\begin{equation} \label{eq:4} J_{ij} = \frac{1}{\hat{K}} \exp \bigg\{ -\frac{1}{2} \bigg[ \frac{d_{ij}}{a} \bigg]^2 \bigg\}  \end{equation}

where $\hat{K}$ is the average number of neighbors per node, and $a$ is a local length scale: the average $d_{ij}$ of all nearest-neighbors. 

There are alternative choices for local characteristic length scale\cite{blatt1996superparamagnetic}. We only report the results obtained with the previous definition, and note that the adjustments to $a$ are problem dependent: higher values of $a$ ensure the \nth{1} phase of the simulation is ferromagnetic while lower values start the simulation in the Super-Paramagnetic (SP) phase.

The objective is to compute averages of thermodynamic quantities after simulating the system at a given temperature for a set number of MCMC iterations $M$ until thermal equilibrium. 

The first simulation serves to uncover the existence of a critical temperature $T_c$ at which a first transition occurs. At $T < T_c $ all spins are strongly correlated, $\langle m \rangle \approx 1 $ and all have the same state: It is called spontaneous magnetization (ferromagnetic phase). At $T = T_c$ the single cluster breaks down into smaller ones (SP-phase). Furthermore, inside the temperature range where the SP-phase exists, a system can go through additional transitions: These reflect the different hierarchical structures present in the data. Finally at $T \gg T_c $ we go through a final transition into complete disorder (Paramagnetic phase): The energy $H_S$ is high, all clusters dissolve, and $\langle m \rangle \approx 0$.

The magnetization $m$ of the system is given by $$m(S)=\frac{qN_{max} (S)-N}{(q-1)N}$$ This quantity, which ranges from 0 to 1, expresses how dominated the system is by the largest cluster. The order parameter of the system is the average magnetization $\langle m \rangle $, and its variance $\chi\frac{T}{N}=\langle m^2 \rangle -\langle m \rangle^2$ is called the susceptibility density. Both can be used to detect a phase transition: $\langle m \rangle $ dives down while $\chi$ peaks at every transition.

For a quantity $A$ the thermal average will be
\begin{equation} \label{eq:5}
\langle A(S) \rangle = \sum A(S)P(S).
\end{equation}
Here each $S$ represents a single MCMC step. If $M$ is large enough, Eqn. \eqref{eq:5} is equivalent to the arithmetic mean $\langle A(S) \rangle \approx \frac{1}{M}\sum_{i=1}^{M}A(S)$. The probability of a system being in a particular state (referring to the energy of the system $H_S$) is: 

\begin{equation} \label{eq:exp} P(S) = \frac{e^{-H_S/T}}{Z}
\end{equation}
where $e^{-H_S/T}$ is the Boltzmann factor, $Z$ is the partition function $Z = \sum_{S} e^{-H_S/T}$ and the normalization constant of the Gibbs-Boltzmann distribution. 

Numerically, we use a mean-field mode of the Hamiltonian such that $H_S=\frac{1}{N}\sum_{\langle i,j \rangle}J_{ij}(1-\delta_{s_i,s_j})$. The motivation being that high levels of $H$ lead to Boltzmann factors close to 0, $Z$ also $\approx 0$ which by definition makes the computation of $P(S)$ impossible, also the value of $H_S$ impacts the temperature range explored.

\subsubsection{Maximum Entropy}

We briefly remind ourselves of the MEP \cite{jaynes2003probability}. We define a statistical mechanical system as an ensemble of objects each in their respective micro-states (spin values $s_i$) so that the resulting in microscopic state of the entire system $S=[s_i,...,s_N]$ can be used to derive parameters which characterize the distributions for macroscopic variables of interests (here the internal energy $H_S$). We assume that at equilibrium, thermodynamic systems obey conservation of energy which sets the constraints of the system such that on average $H_S$ is a constant, and then from Eqn. (\eqref{eq:5}) it follows that:

\begin{equation}\langle H_S \rangle = \sum H_S P\left(H_S\right),~\mbox{and}~\sum P\left(H_S\right) = 1. \end{equation}

We then consider that the distribution representative of the energy of the system as the one which incorporates our constraints and assumes nothing else. This maximizes the Shannon's Entropy as defined by: 

\begin{equation} \label{eq:s_entropy} S = - \sum P\left(H_S\right)\ln\left(P\left(H_S\right)\right)\end{equation}

The problem can then be reduced to a Lagrange optimization task for which the exponential family of distributions is a well known solution (see Eqn. \eqref{eq:exp} ) with the inverse temperature as its Lagrange multiplier \cite{jaynes2003probability}.

\subsection{Super-Paramagnetic Clustering (SPC)} \label{ssec:spc}

\subsubsection{A Maximum Entropy Method} \label{sssec:maxent}

We define a neighbor on a lattice to be a node located in the vicinity of another node such that a node $s_{i,j}$ will have neighbors $s_{i+1,j},s_{i-1,j},s_{i,j+1}$, and $s_{i,j-1}$. A neighborhood generated using these rules is valid for a 2D lattice with a fixed $J$ for all nodes (the interaction strength is said to be homogeneous). This is the original method used in simulating Ising/Potts models of ferromagnets. As a generalization to the problem of data clustering, we will consider a neighborhood which emerges from the inhomogeneous interaction strength $J_{ij}$. Every neighborhood is a mini-graph, and their aggregation constitutes a graph whose topology is determined by the matrix $J_{ij}$: For two nodes to be neighbors they, each, must be included in their respective K-nearest neighbors.

We first define the neighborhood of size K. This is implemented following steps in Table (\ref{tab:sizeK}) below.
\begin{table}
	
	\noindent\fbox{\parbox{0.45\textwidth}{
			\begin{enumerate}[label=\ref{tab:sizeK}\arabic*]
				\item {\bf Pick a $K$:} Build the {\tt nodenext} matrix as the array containing the locations of every node's neighbors. We use $K = 10$ (except when otherwise explicitly stated).
				\item {\bf Form the MST:} Add edges to {\tt nodenext} to  make every graph connected regardless of $K$.
	\end{enumerate}}}
	\caption{\label{tab:sizeK} Setting the neighborhood size K for SPC: The neighbor determines the scope of the algorithm. It effectively cancels the pairwise interaction strengths of the spins outside the spins respective neighbors thus producing a speed-up in the computation of the Hamiltonian. \cite{blatt1996superparamagnetic}}
\end{table}

The graph is traveled via nodes, and edges can be set active or inactive with probability

\begin{equation} \label{eq:bolt_link_dis} p_{ij} = 1 - e^{-\frac{J_{ij}}{T}\delta_{s_i,s_j}} \end{equation}

The next array is called $\texttt{links}$. It is the adjacency matrix where the activation status of edges is stored such that $\texttt{links}_{ij} = 1$ if $p_{ij} > $ rand, and 0 otherwise.

The original Hoshen-Kopelman (HK) algorithm \cite{hoshen1976percolation} is the standard for labeling clusters in many statistical mechanics problems. The 2D version is mostly restricted to two neighbors per node: $s_{i-1,j}$, and $s_{i,j-1}$. SPC (Sec. \ref{ssec:alg_spc}) deals with problems where $J_{ij}$ is not fixed, and K can be large so we apply the extension of HK to non-lattice environments found in \cite{al2003extension}.

The clusters are labeled using the extended HK algorithm (See Table \ref{tab:extendedHK} below).
\begin{table}
	\noindent\fbox{\parbox{0.45\textwidth}{
			\begin{enumerate}[label=\ref{tab:extendedHK}\arabic*]
				\item {\bf Initialize counter labels}: {\tt nodel} as 1xN array to store labels set to 0.
				\item {\bf Check for activated edges}: If none (unlinked to occupied nodes) (2a.) create new cluster label, else (2b.) find the root node and its labelled neighbour and store smallest root and {\tt nodel} and replace {\tt nodelp}
				\item {\bf Sequentialize the recorded node labels}: {\tt nodelp}
				\item Relabel nodes.{\tt nodel} $\leftarrow$ {\tt nodelp}
	\end{enumerate}}}
	\caption{\label{tab:extendedHK} Labeling: The extended Hoshen-Kopelman (HK) algorithm: reads data from a matrix of bonds indicating spins pairwise associations and creates the clusters (also see Appendix for code patterns \ref{alg:hk}). The HK algorithm is inspired by the union-find algorithm which ``finds'' the root class of nodes and ``unites'' nodes belonging to the same class. \cite{al2003extension}}
\end{table}

Once the graph is fully constructed, the next step is implement the Swendsen-Wang MCMC algorithm (See C1 in \ref{tab:sw} below.
\begin{table}
	\noindent\fbox{\parbox{0.45\textwidth}{
			\begin{enumerate}[label=\ref{tab:sw}\arabic*]
				\item {\bf Assign Spin Values:} The resulting clusters are all flipped independently from each other and each get assigned a new spin value between 0 and $q$.
	\end{enumerate}}}
	\caption{\label{tab:sw} Flipping Clusters using Swendsen-Wang MCMC \cite{swendsen1987nonuniversal} (see code-pattern in \ref{alg:sw} )}
\end{table} 

Finally, the spin-spin correlation $G_{ij}$ is the average probability of two spins being in the same cluster is computed in two steps (See Table \eqref{tab:g_ij}).

\begin{table}
	\noindent\fbox{\parbox{0.45\textwidth}{
			\begin{enumerate}[label=\ref{tab:g_ij}\arabic*]
				\item {\bf Increment two-point correlations}: At every MCMC step $c_{ij}$, the two-point connectedness, is incremented if $i$ and $j$ are clustered together.
				\item {\bf Compute spin-spin correlations:}: Once the simulation ends for the temperature explored we compute $$G_{ij} = \frac{(q-1)c_{ij}+1}{q}$$
	\end{enumerate}}}
	\caption{\label{tab:g_ij} Compute the Spin-Spin Correlation $G_{ij}$ from the incremented two-point correlations $c_{ij}$. \cite{blatt1996superparamagnetic}}
\end{table}

The spin-spin correlation $G$ is probably the most important quantity as it is used to build the final clusters. The threshold $\theta$, for which two nodes are members of the same cluster, is picked to be higher than $\frac{1}{q}$ but less than $1 - \frac{2}{q}$. The bounds on that range are explained by the distribution of $G_{ij}$ which peaks at those two values: They are respectively the peak inter and intra cluster correlations \footnote{For $q=20$ results in $\rho_{sa}=0.05$, and $\rho_{sb}=0.9$. Uncorrelated nodes have $\rho_{ij} \approx \rho_{sa}$, and correlated nodes $\rho_{ij} \approx \rho_{sb}$.}.  It is typical to use $\theta = 0.5$ as it does not significantly affect the results in previous examples \cite{blatt1996superparamagnetic}.

\subsubsection{A Maximum-Likelihood Method} \label{sssec:mle}

Based on an analysis of the spectral properties of stock market correlations matrices, Noh \cite{noh2000model} makes the following statistical {\it ansatz}: let's assume an existing market hierarchy where individual stocks dynamics are dependent on clusters of similar stocks. This can be illustrated by a simple model as follows: 
\begin{equation} 
\label{eq:7} x_i = f_i + \epsilon_i 
\end{equation} 
where $x_i$ are the stock's features, $f_i$ the cluster-related influence, and $\epsilon_i$ the node's specific effect.

In \cite{giada2001data} Giada, and Marsili formally develop a Potts model using Noh's idea, and in \cite{hendricks2016detecting} Hendricks, Gebbie, and Wilcox solved the optimization problem using a PGA for unsupervised learning for quantitative trading. Let's consider a group of $N$ observations, embedded in a space with dimensionality $D$ as the features, and as with SPC (Sec. \ref{ssec:alg_spc}), every observation is assigned a spin value. One version of the ansatz models the observation features such that 
\begin{equation} 
\label{eq:ansatz_1} x_i = g_{s_i}\eta_{s_i} + \sqrt{1- g_{s_i}^2}\epsilon_i 
\end{equation} 
where $x_i$ is one feature, $g_{s_i}$ the intra-cluster coupling parameter \footnote{ The thermal average $\langle g_s \rangle$ can be used to reconstruct data-sets sharing identical statistical features of the original time-series by using Eqn. \eqref{eq:ansatz_1} \cite{giada2001data}} , $\eta_{s_i}$ the cluster-related influence, and $\epsilon_i$ the observation's specific effect, and measurement error. A covariance analysis yields additional terms such as $n_s$ the size of cluster $s$, and $c_s$ the intra-cluster correlation \footnote{Here $n_s = \sum_{i=1}^{N} \delta_{s_i,s}$, $c_s = \sum_{i=1}^{N}\sum_{j=1}^{N}C_{ij}\delta_{s_i,s}\delta_{s_j,s}$, and $g_s = \sqrt{ \frac{c_s-n_s}{n_s^2-n_s}}$ \cite{giada2001data,hendricks2016detecting}.}.

We explicitly mention that $n_s < c_s < n_s^2$ must be enforced: the lower bound is required because $g_s$ is undefined for values of $c_s \leq n_s$, and the upper bound requires a strict inequality because Eqn. \eqref{eq:lc} is undefined when $c_s = n_s^2$. We introduce a Dirac-delta function \footnote{Let $ y_i = x_i - \big(g_{s_i}\eta_{s_i} + \sqrt{1- g_{s_i}^2}\epsilon_i\big)$, and $\delta(y)$ a Dirac delta function of $y$ which is 1 when $y=0$, and 0 otherwise.} to model the probability of observing data in a configuration $S$ close to criticality:

\begin{equation} 
\label{eq:12} P = \prod_{d=1}^{D}\prod_{i=1}^{N}\Bigg \langle \delta\left(x_i - \left( g_{s_i}\eta_{s_i} + \sqrt{1 - g_{s_i}^2}\epsilon_i\right)\right) \Bigg \rangle~.
\end{equation}

This joint likelihood is the probability of a cluster configuration matching the observed data for every observations, and for every feature. The log-likelihood derived from $P$ can be thought of the Hamiltonian of this Potts system: \begin{equation} \label{eq:lc} L_c = \frac{1}{2} \sum_{s:n_s>1} \ln \frac{n_s}{c_s} + \left(n_s-1\right) \ln \frac{n_s^2-n_s}{n_s^2-c_s}~.\end{equation}
The sum is computed for every feature, and represents the amount of structure present in the data. The value of $L_c$ is indirectly dependent on spins via the terms $n_s~,\mbox{and}~c_s$. 

A-priori advantages of this method over industry standard alternatives: First, that $L_c$ is completely dependent on $C_{ij}$, and the dimensionality of the dataset only plays a part in computing $C_{ij}$, and Second, it is unsupervised: There are no preset number of clusters. Clustering configurations are randomly generated, and that which maximizes $L_c$ provides us with the number of clusters, and their compositions.

Further modification of the model can be made to reduce the Hamiltonian to that of the standard K-means algorithm \cite{lloyd1982least}: 
\begin{equation} 
H_{KM} = \sum_{s:n_s>0} \left({n_s - \frac{n_s}{c_s}}\right).
\end{equation}

The f-SPC algorithm (Sec. \ref{alg:ga}) uses a PGA to find the global optimum of the likelihood $L_c$ \eqref{eq:lc}.

The principles of our GA are given in Table (\ref{tab:fSPC-PGA}) below. 

\begin{table}
	\noindent\fbox{\parbox{0.45\textwidth}{
			\begin{enumerate}[label=\ref{tab:fSPC-PGA}\arabic*]
				\item {\bf Generate Populations}: Generate the populations as a set of randomly generated Potts configuration with spin values ranging form $0$ to $N$, 2.) 
				\item {\bf Evaluate Fitness}: Use the computation of $L_C$
				\item {\bf Select the Best Individuals}
				\item {\bf Mutate}: A set number of individuals in the populations are mutated
				\item {\bf Recombine}: The parent and child generations are recombined and again selection of the best individuals takes place.
				\item {\bf Iterative Convergence}: Repeat 2.) to 5.) until sufficient convergence has been achieved.
	\end{enumerate}}}
	\caption{\label{tab:fSPC-PGA} f-SPC PGA Implementation: This Genetic Algorithm has no crossing step where parents would be mated. The mutations are the main genetic diversity operator. Mutations and Likelihood computations are evaluated in parallel \cite{hendricks2016detecting} (see code-pattern \ref{alg:ga}).}
\end{table}

The original PGA algorithm implemented in \cite{gebbie2010spin} contained a mating step involving a bespoke cross-over function, and a restriction: Only parents with the same number of clusters could be mated, and the resulting children should maintain the same characteristic. The enforcement of this rule restricted clusters merging and splitting through mutations only. Our f-SPC implementation removes this intermediary step. This was implemented in order to decrease the computational cost \footnote{ The algorithm is able to double the number of generations from 250 to 500 for a 5 mins simulation (Iris see Sec. \ref{ssec:iris}): $L_c = 100,~\mbox{and without the cross-over function}~L_c = 105$.}

In addition to the diversity of individuals present in the population, mutations serve as GA diversity operators: They increase the genetic diversity, and send the system onto another path toward higher local maxima. We used six equally weighted types of mutations: i) {\it New}: A complete new individual, ii) {\it Split}: a random cluster split into two, iii) {\it Merge}: two clusters merged at random, iv) {\it Swap}: two spin labels are exchanged, v) {\it Scramble}: where a sequence of spins have their labels re-assigned in reverse order, and, vi) {\it Flip}: cluster (spin) labels are randomly re-assigned using the total cluster number (See SW Table \ref{tab:sw} ).

At last, the simulation has converged once the fitness of the best individual hasn't increased for a pre-determined number of iterations called ``Stall Generations''. It should also be noted that although we did not recode a CUDA implementation for direct GPU implementation of our refined PGA algorithm this can be implemented using a modified version of the original CUDA implementation \cite{hendricks2016detecting}.

\subsection{Super-Paramagnetic Phase Validation} \label{ssec:validation}

The goal of this project was to effectively validate the existing link between the solutions obtained using $L_c$ and the original Potts Hamiltonian $H_S$. We proceeded by comparing clustering configurations obtained using the two models for the data set in Sec. \ref{ssec:financial}.

\begin{figure*}
	\centering
	\begin{subfigure}[]{0.3\textwidth}
		\includegraphics[width=\textwidth]{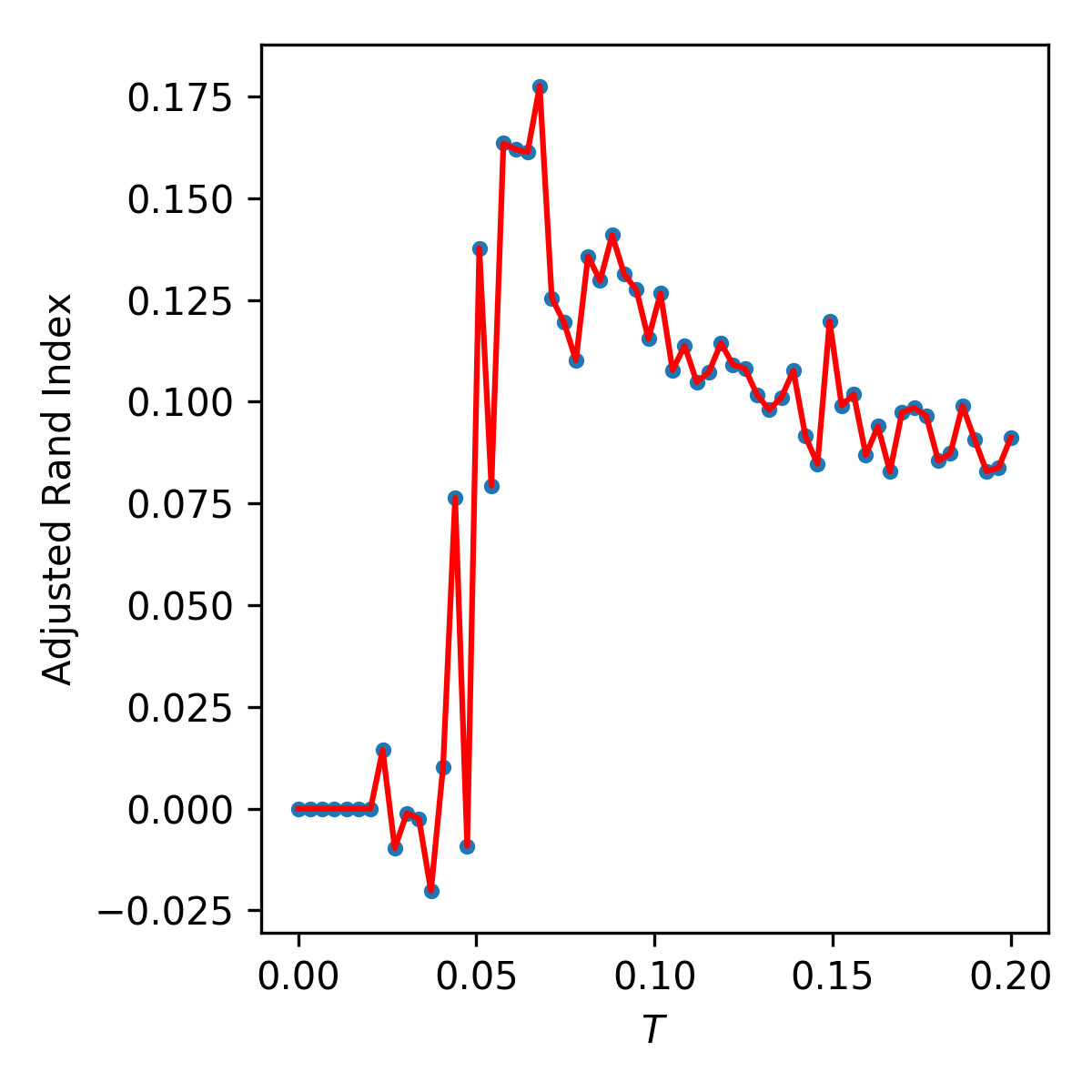}
		\caption{Market Mode}
		\label{fig:full_per}
	\end{subfigure}
	~ 
	\begin{subfigure}[]{0.3\textwidth}
		\includegraphics[width=\textwidth]{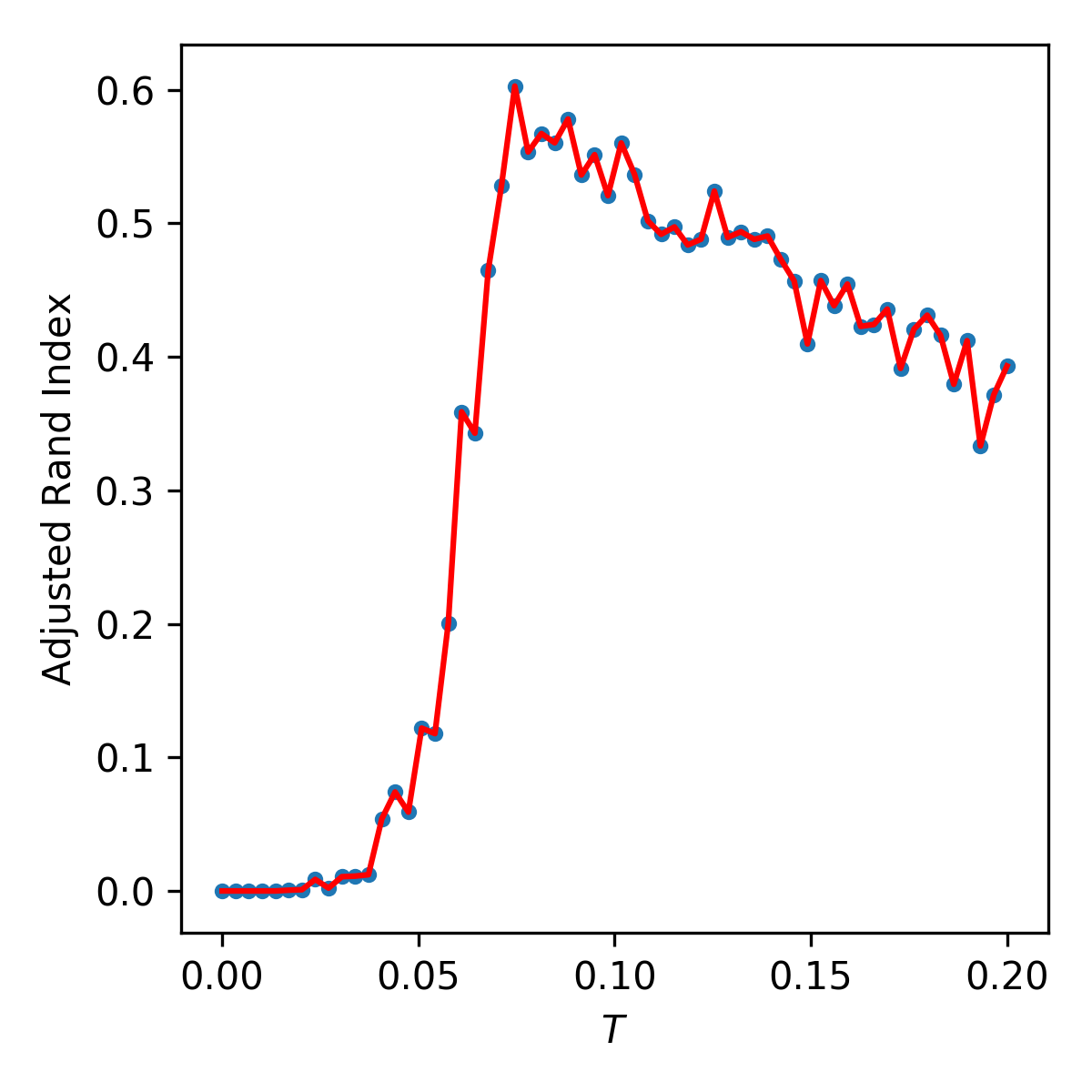}
		\caption{IMN}
		\label{fig:norm_per}
	\end{subfigure}
	~
	\begin{subfigure}[]{0.3\textwidth}
		\includegraphics[width=\textwidth]{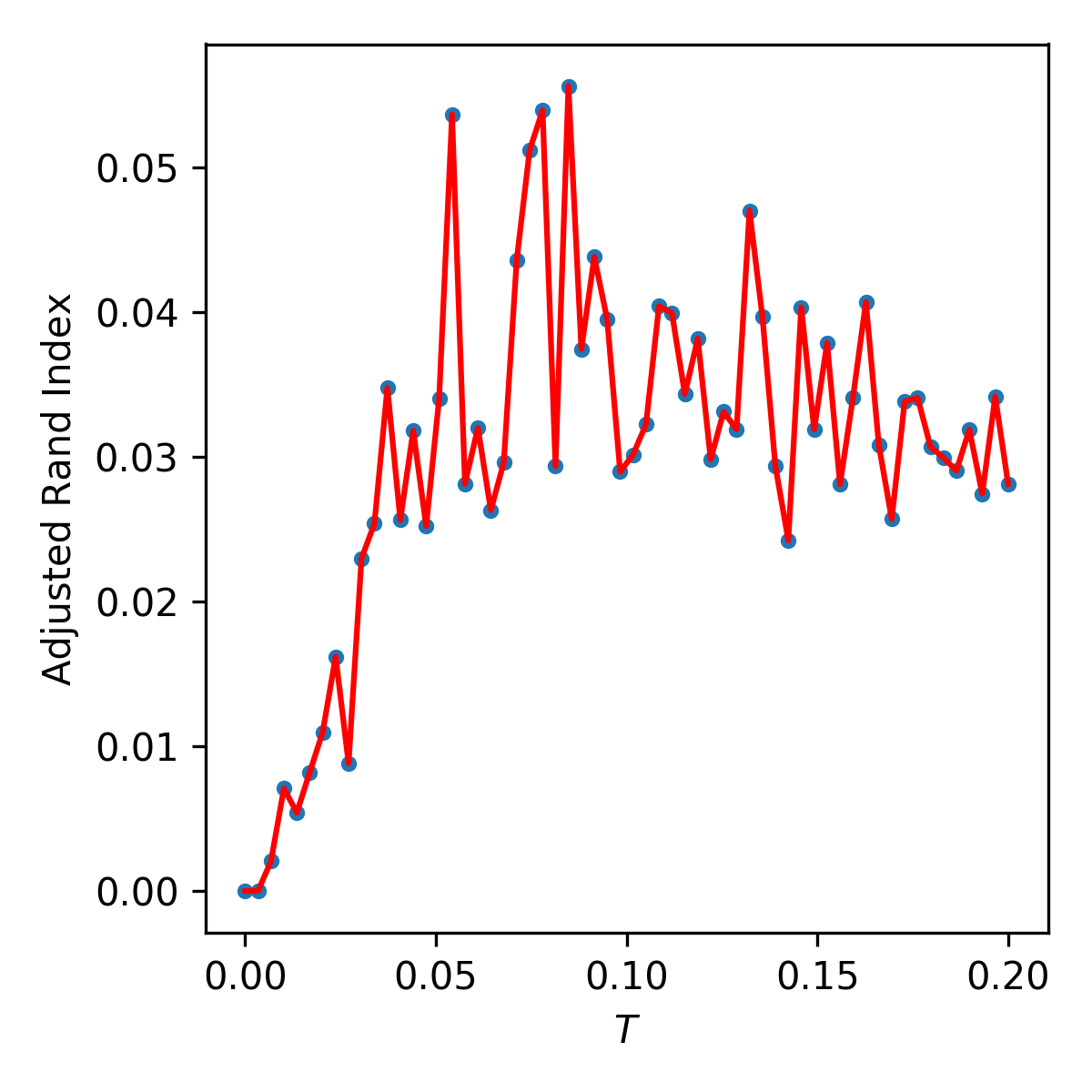}
		\caption{RMT}
		\label{fig:rmt_per}
	\end{subfigure}
	
	\caption{In figures (a), (b) and (c) find the ARI (See Sec. \ref{sec:cases}) for the following cases : (a), with a market mode (See Sec. \ref{ssec:noise}), (b) de-noising with IMN (see Sec. \ref{sssec:succnorm}) and (c) when a RMT method is used to clean the correlation matrix (See Sec. \ref{sssec:rmt}). The ARI index expresses configuration similarity on [0,1] \cite{hubert1985comparing}. Blue dots represent ARI values, and the red line the curve liking them all. We looked at NYSE S\&P500 447 Stocks Data. In all 3 cases we compare the f-SPC method (See Sec. \ref{sssec:mle}) to each of SPC candidates (See Sec. \ref{ssec:financial}). This demonstrated that in all 3 cases the maximum likelihood candidates are close to solutions recovered within the Super-Paramagnetic Phase.}\label{fig:sp500_persistence}
\end{figure*}

Shown in Fig. \eqref{fig:full_per} , Fig. \eqref{fig:norm_per}  , and Fig. \eqref{fig:rmt_per} are the Adjusted Rand Indices as functions of temperature for our three cases. We compute the ARI at every temperature taking f-SPC as the true classification, and SPC as the candidates. Maximal ARI values are respectively 0.175, 0.6, and 0.05 for temperatures $T=0.068$, $T=0.071$, and $T=0.08$. These temperature values are all located in the SP-phase where the susceptibility is non-zero confirming the claim in \cite{giada2001data} that the maximization of $L_c$ should recover a clustering configuration of the a system in the vicinity of the phase transition. We also note that despite SPC neighborhood search restrictions the Normalized f-SPC solution has the highest similarity to its SPC's counterparts. The ``Market Mode'' case of f-SPC has a large mixed cluster, and the RMT one has a its largest cluster mixed with securities from every economic sector. This would require further analysis but we think the main difference between that and SPC's equivalent is this cluster which could be of low correlation.

We push further the analysis by considering $L_c$ as a clustering quality evaluator. As was demonstrated in \cite{giada2002algorithms} $L_c$ is a consistent objective function which if maximized can discriminate between clustering algorithms. Similarly to the ``Silhouette Coefficient'', and the ``Calinski-Harabaz Index'' which are methods used to evaluate the clusters definition when the ground truth class is not known, $L_c$ plays a similar role.

\begin{figure*}
	\centering
	\begin{subfigure}[b]{0.3\textwidth}
		\includegraphics[width=\textwidth]{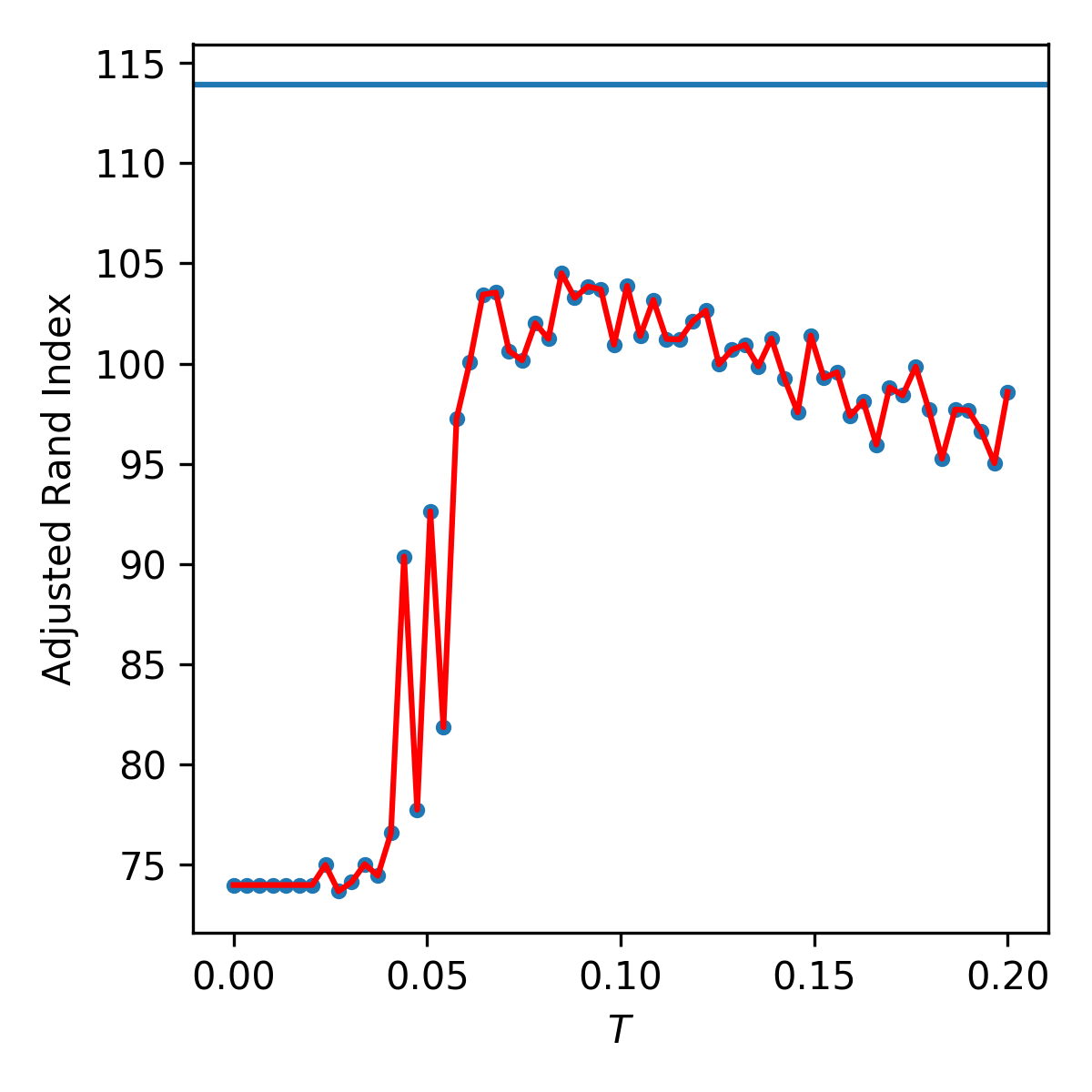}
		\caption{Market Mode}
		\label{fig:full_lc}
	\end{subfigure}
	~ 
	\begin{subfigure}[b]{0.3\textwidth}
		\includegraphics[width=\textwidth]{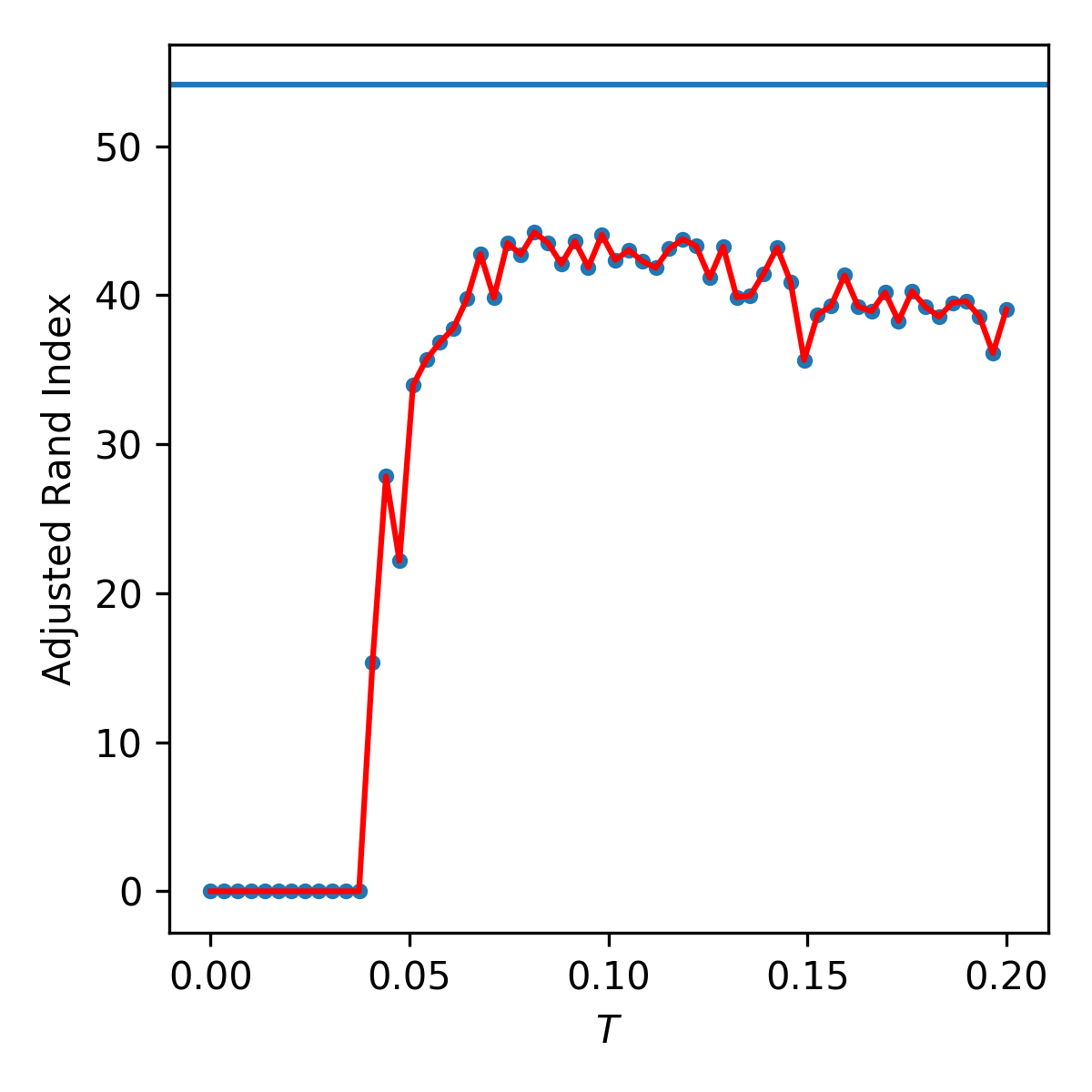}
		\caption{IMN}
		\label{fig:norm_lc}
	\end{subfigure}
	~
	\begin{subfigure}[b]{0.3\textwidth}
		\includegraphics[width=\textwidth]{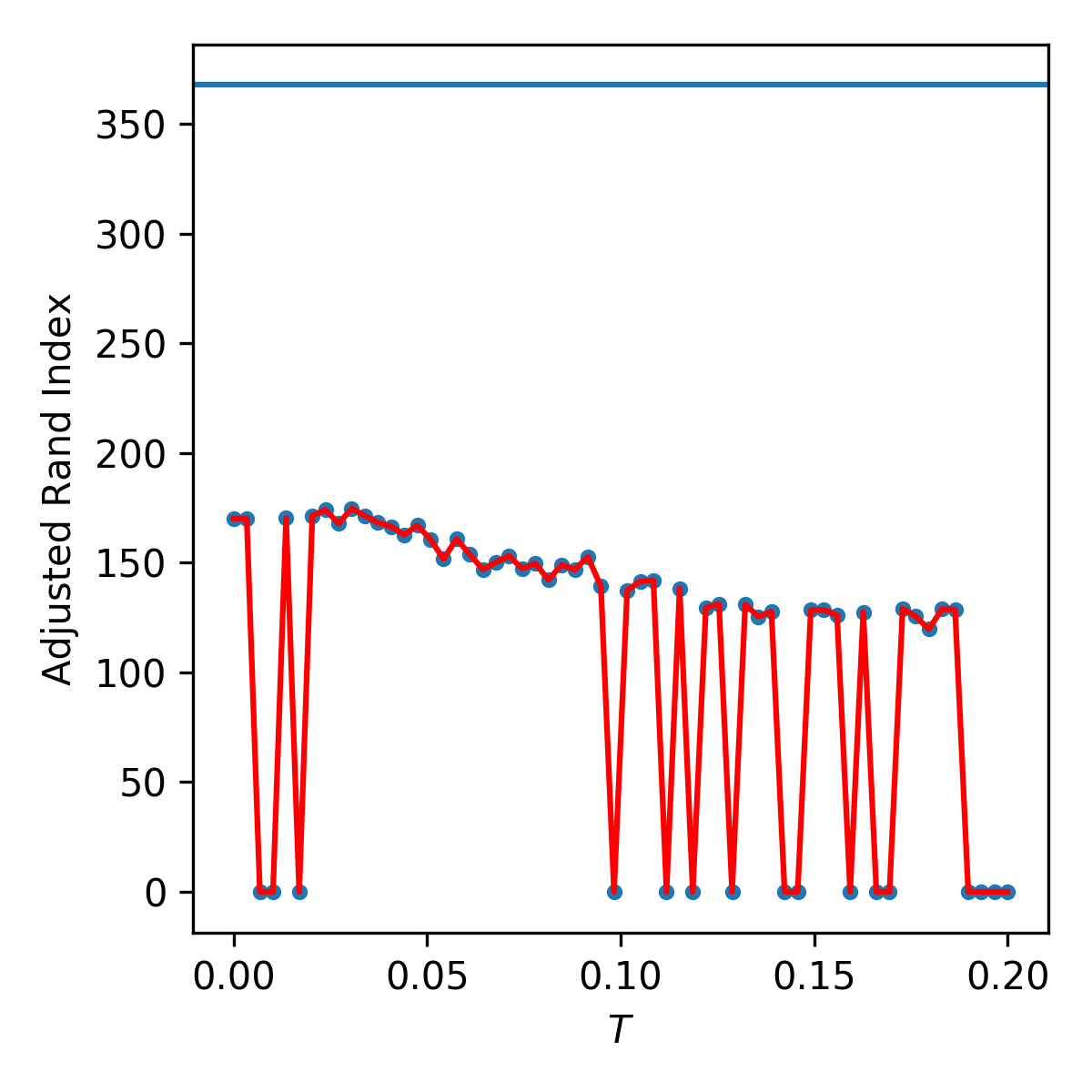}
		\caption{RMT}
		\label{fig:rmt_lc}
	\end{subfigure}
	
	\caption{S\& P500 (Sec. \ref{ssec:financial}) N = 447 Stocks, D = 1249 trading days: We computed the Likelihood \cite{giada2001data} $L_c$ \eqref{eq:lc} of every SPC solutions for all temperatures (red curve, and blue dots), and Likelihood $L_c$ of f-SPC's solution (blue horizontal line). in a) the Market Mode case, in b) the Normalized case, and in c) the RMT case. Every f-SPC solutions has higher likelihood than the SPC entire temperature range in every case. f-SPC solutions are composed of clusters with higher correlation than SPC candidates. }\label{fig:sp500_lc}
\end{figure*}		

We test for this by evaluating all SPC candidates for their $L_c$ values, and we add a horizontal line on each plot indicating the respective f-SPC's $L_c$. In every case, SPC's $L_c$ start low at low T, reaches a maximum at intermediate T and decreases slowly at T increases into Paramagnetic territory. This is yet another confirmation that higher $L_c$ values are located in a intermediate temperature regime which coincides with the SP-phase when the system is critical.  SPC's $L_c$ maxima are 105, 43, and 170 respectively in Fig. \eqref{fig:full_lc}  Fig. \eqref{fig:norm_lc}  Fig. \eqref{fig:rmt_lc}, and we observe that solutions recovered using f-SPC all have higher likelihoods than SPC's. Based on the result in this paper, and in \cite{giada2002algorithms} one could argue that f-SPC produces better clustering candidates than SPC at least in this case.

\subsubsection{Free-Energy Validation} \label{ssec:free}

We now define the Helmholtz Free Energy $F$ for a thermodynamic process of an isolated system.

\begin{equation} 
\label{eq:hem1} F = U - TS 
\end{equation}

where $U$ is the internal energy of the system (see Eqn. \eqref{eq:hs}, and \eqref{eq:lc} ), $T$ is the temperature of the heat bath or reservoir in contact the system, and $S$ is the entropy.

\begin{equation} 
\label{eq:hem2} F = - T \ln Z 
\end{equation}

The free energy can also be computed using \eqref{eq:hem2} with Z as the partition function like in Eqn. \eqref{eq:exp}. \footnote{Here we are implicitly using a course-grained approach to truncation over the monte-carlo replications; numerically computing the thermodyanimc averages over $\ln Z$ can often be more effectively computed using the replica method
$\ln Z = \lim _{n \to \infty} \left( {\frac{Z^n -1}{n}} \right)^n$ \cite{Castellani_2005} with a prudent choice of the number of replications $n$.}

We consider an isothermal process a system exchanging energy with a reservoir at constant $T$ by absorbing heat until its own temperature converges to that of the reservoir. For processes such as the one just described Eqn. \eqref{eq:hem2} tells us that some of the energy needed for the system to be realized can be spontaneously transferred from the reservoir by heating ``$TS$''. In this sense, for systems on which no work is done $\Delta F \leq 0 $ and thermal equilibrium is reached if the free energy reaches a minimum.

Using Mean Field Models in \cite{blatt1996superparamagnetic}, and \cite{giada2001data} It was shown that the free energy reaches a local minimum within the Super-Paramagnetic or Clustered Ferromagnetic Phase, and a maximum at the Paramagnetic Phase transition. We argue that the temperature at which the previously mentioned minimum happens is synonymous to the heat-bath inside of which the system is in its ``lowest level''.

\begin{figure}
	\centering
	\begin{subfigure}[b]{0.45\textwidth}
		\includegraphics[width=\textwidth]{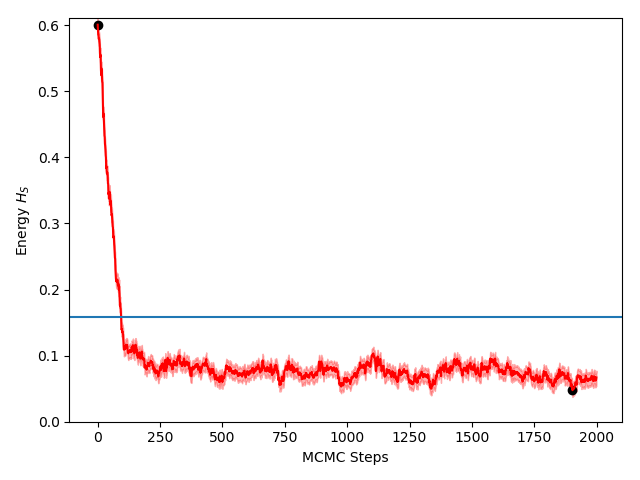}
		\caption{Evolution of the internal energy at $T=0.13$ as MCMC steps occur. No convergence per se to a unique value, but oscillations around a level. The blue line is the thermal average $ \langle H_S \rangle $}
		\label{fig:h_curve}
	\end{subfigure}

	\begin{subfigure}[b]{0.45\textwidth}
		\includegraphics[width=\textwidth]{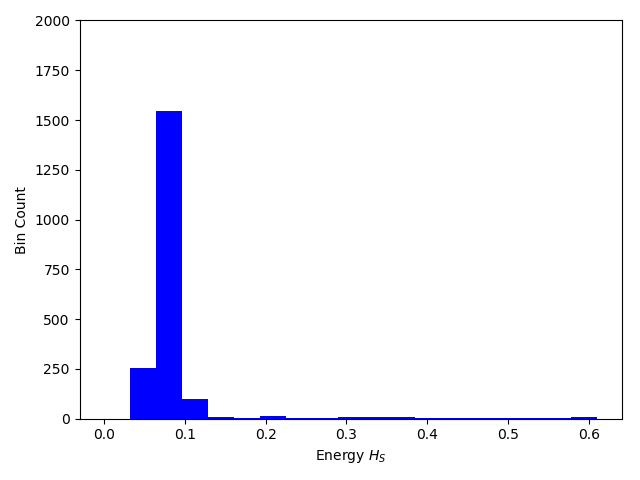}
		\caption{Histogram of the distribution of the internal energy at $T=0.13$ with the number of bins calculated using \eqref{eq:k_bins}}
		\label{fig:h_dist}
	\end{subfigure}
	
	\caption{BRICS Data: SPC's internal energy at $T=0.13$. In a) the energy as it converges, and in b) its binned distribution.}\label{fig:not_sure_}
\end{figure}	

Eqn. \eqref{eq:hem2} requires computing $Z$ whereas Eqn. \eqref{eq:hem1} needs the average energy $\langle H \rangle$, and the entropy $S$. Although we don't show it these two methods agree. At any given temperature the system doesn't converge to a specific energy level but displays a distribution (see Fig. \eqref{fig:h_curve}). The task at hand is now about picking a number of bins for our MCMC simulation which is consistent and not arbitrary. We borrow a ``low bias'' methods from \cite{HACINEGHARBI20121302} which follows:

\begin{equation} \label{eq:k_bins} k_X = \mbox{round} \bigg\{ \frac{\epsilon}{6}+\frac{2}{3\epsilon}+\frac{1}{3} \bigg\} \end{equation} where $k_X$ is the number of bins, and $$\epsilon = \sqrt[3]{8+324n+12\sqrt{36n+729n^2}}$$ with $n$ as the number of samples, here the total number of MCMC steps. Because we fix $n$ to 2000 the number of bins remains set for every temperatures, and the bin edges are set on the minimum and maximum possible energies depending on the problem. The Hamiltonian $H_S$ minimum energy is always 0 (for the ferromagnetic case), and in the case of the BRICS data the maximum was $ \approx 0.61 $  (see Fig. \eqref{fig:h_dist} ). We then determine the distribution energy levels by picking the $k_X$ bin centers which we compute by taking the mean of the distribution inside each bin. Once obtained we can now compute $Z$, the Boltzmann distribution of energies, which we use to compute the thermal average Energy $\langle H \rangle$, the entropy $S$ using Eqn. \eqref{eq:s_entropy} and the free energy $F$.

\begin{figure}
	\centering
	\begin{subfigure}[b]{0.48\textwidth}
		\includegraphics[width=\textwidth]{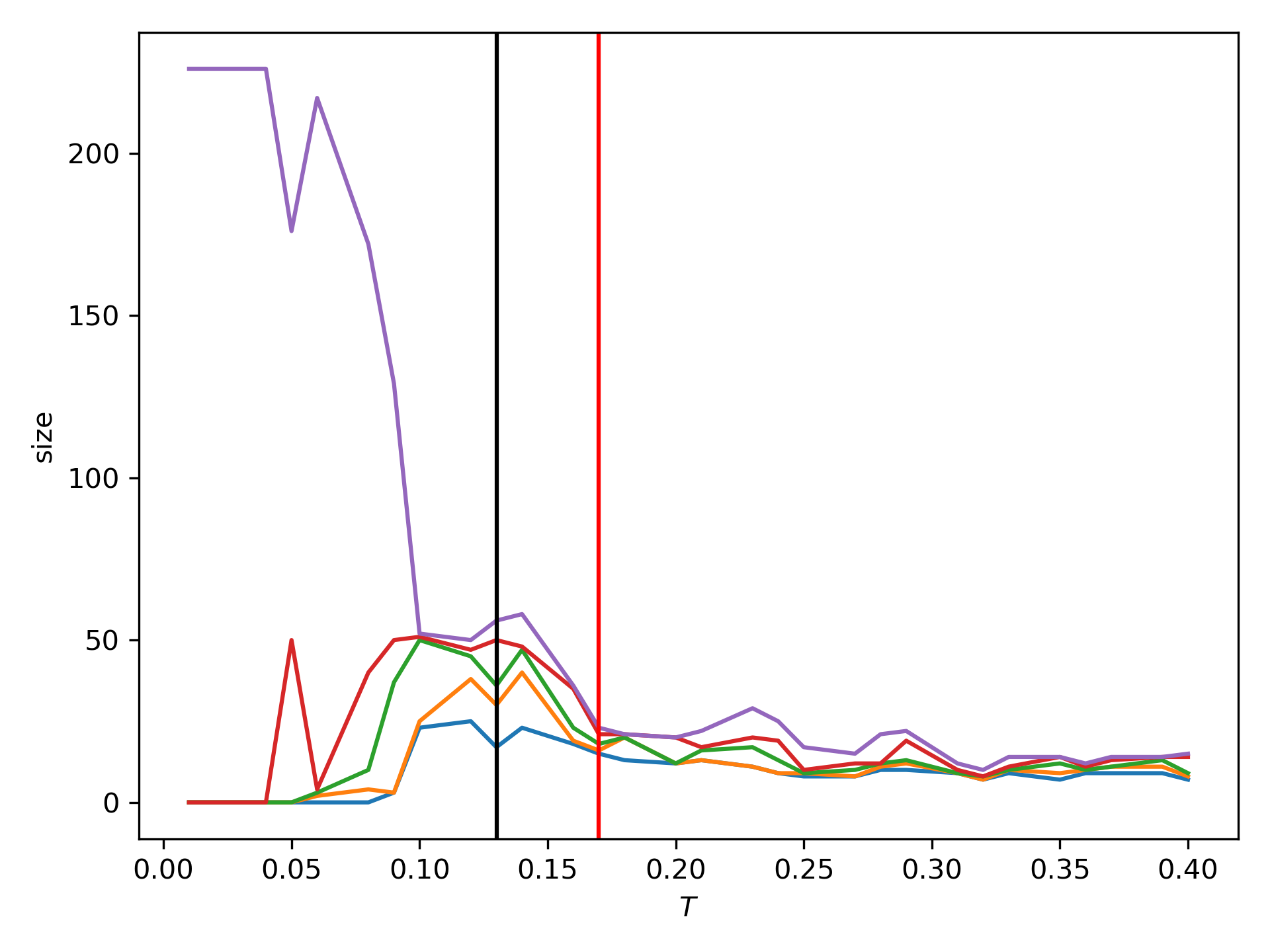}	
		\caption{BRICS DATA, Cluster sizes as function of Temperature for SPC solutions. The vertical lines respectively show the temperatures of maximum ARI, and minimum Free Energy / maximum entropy.}\label{fig:brics_clusters_verticals}
		
	\end{subfigure}
	~
	\begin{subfigure}[b]{0.48\textwidth}
		\includegraphics[width=\textwidth]{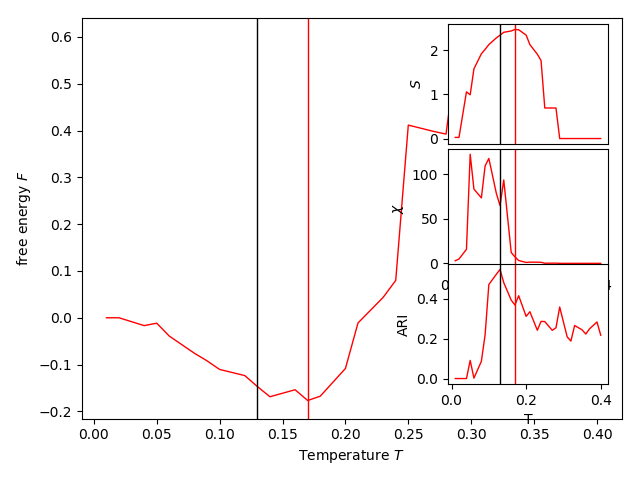}	
		\caption{BRICS DATA, Free Energy as function of Temperature for SPC solutions (in red). Insets: (top right) Entropy, (center right) Susceptibility, (bottom right) ARI.
			The vertical lines respectively show the temperatures of maximum ARI, and minimum Free Energy / maximum entropy.}\label{fig:brics_free}
	\end{subfigure}
\end{figure}	

In Figures \eqref{fig:brics_clusters_verticals}, and  \eqref{fig:brics_free} we show the results of a SPC simulation on the BRICS data (See Sec. \ref{ssec:brics} ): We show the free energy as a function of temperature for the SPC simulation, and we also plot the Adjusted Rand Index of the f-SPC solutions against the SPC ones. We quickly describe the behavior of the free energy which decreases and reaches a minimum at $T \approx 0.17$, and a maximum at $T \approx 0.25$. A quick look at Fig. \eqref{fig:brics_clusters_verticals} shows that before $T \approx 0.17$ the giant cluster is breaking down inside the Super-Paramagnetic Phase until their sizes are comparable and seemingly stable. After $T \approx 0.17$ The clusters sizes are unstable, start decreasing and it's become impossible to significantly distinguish clusters which signals a transition into the Paramagnetic Phase. More importantly the ARI curve peaks at $T \approx 0.13$, close to the minimum free energy temperature within the Super-Paramagnetic Phase thus revealing that f-SPC's algorithm and objective function minimizes the free energy (maximizes entropy) of the system as it maximizes $L_c$.

\subsubsection{ {\it Sci-Kit learn}: Varying Density Clusters } \label{ssec:blobs}

The problem consists of 3 clusters using \texttt{Sci-kit learn} samples generator \footnote{B. Thirion, G. Varoquaux, A. Gramfort, V. Michel, O. Grisel, G. Louppe, J. Nothman, 'make\_blobs', 2017. [Online]. Available: \url{http://scikit-learn.org/stable/modules/generated/sklearn.datasets.make_blobs.html}. [Accessed: 12-Jun-2018]} with $N = 500$, and $\sigma = 0.25, 0.5, 1$. We observed two cases of this problem: a 3D case, and a 500D case.

\begin{figure*}
	\centering
	\begin{subfigure}[b]{0.31\textwidth}
		\includegraphics[width=\textwidth]{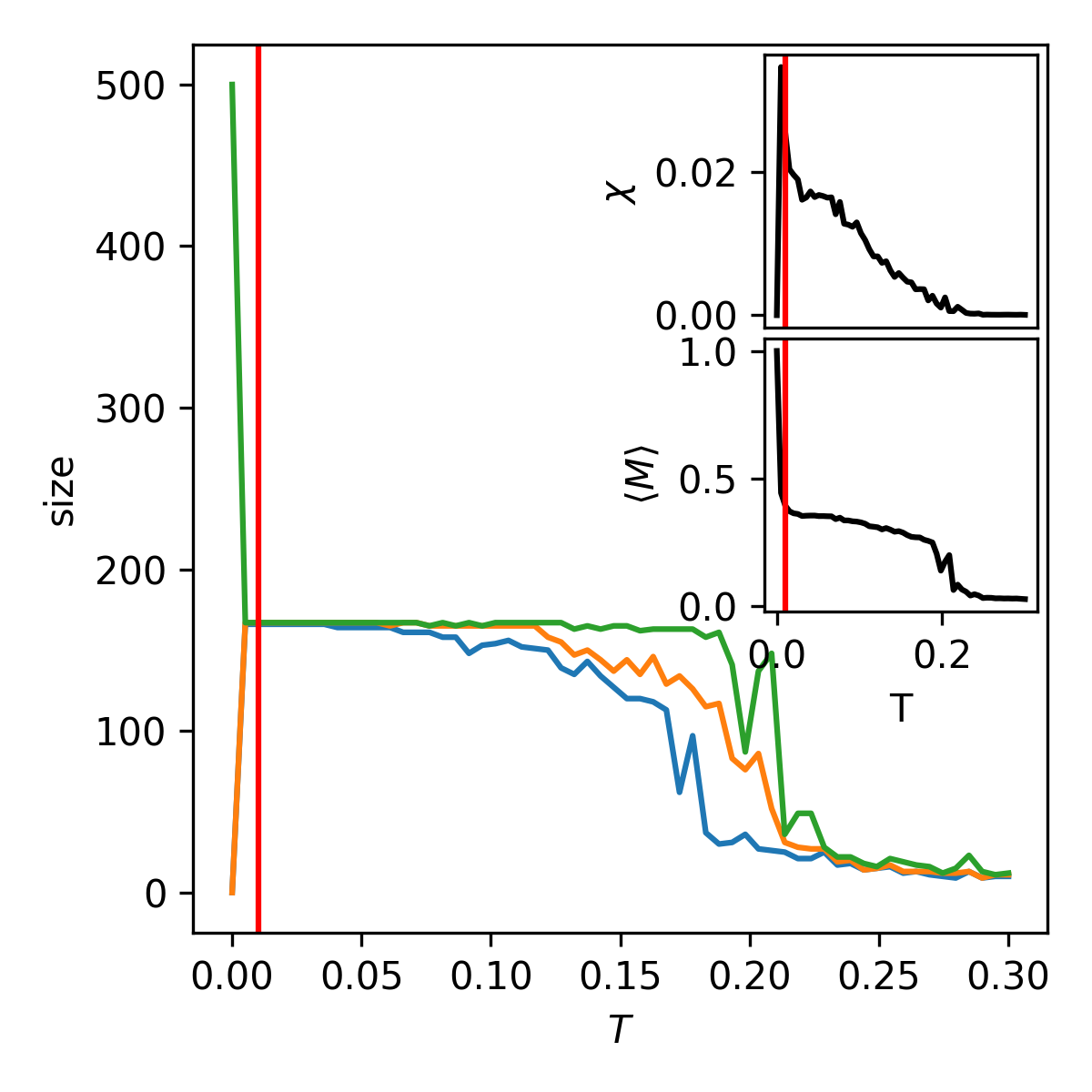}
		\caption{Size vs $T$}
		\label{fig:b_clus}
	\end{subfigure}
	~
	\begin{subfigure}[b]{0.31\textwidth}
		\includegraphics[width=\textwidth]{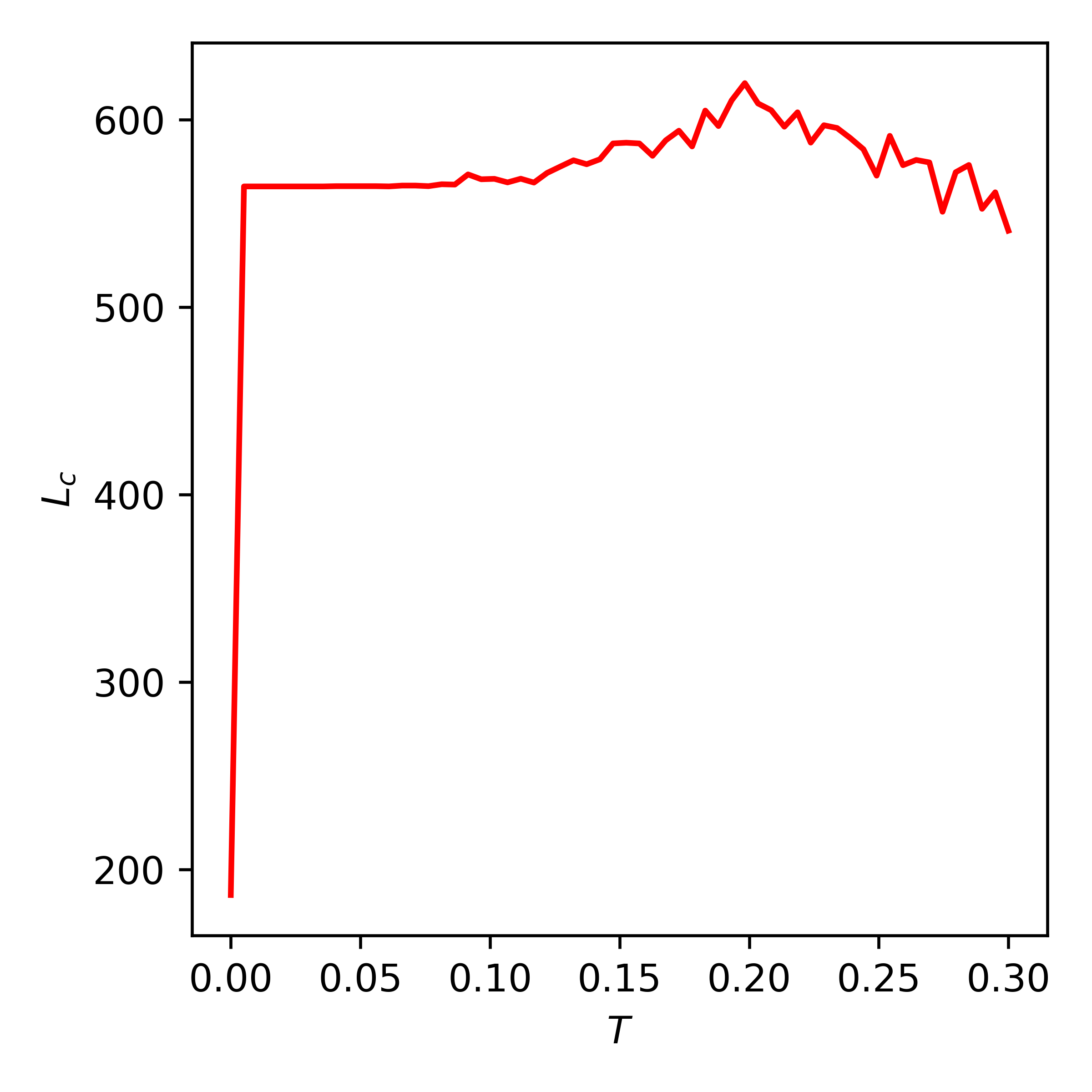}
		\caption{3D Case: $L_c$ vs $T$}
		\label{fig:b_lc3d}
	\end{subfigure}
	~ 
	\begin{subfigure}[b]{0.31\textwidth}
		\includegraphics[width=\textwidth]{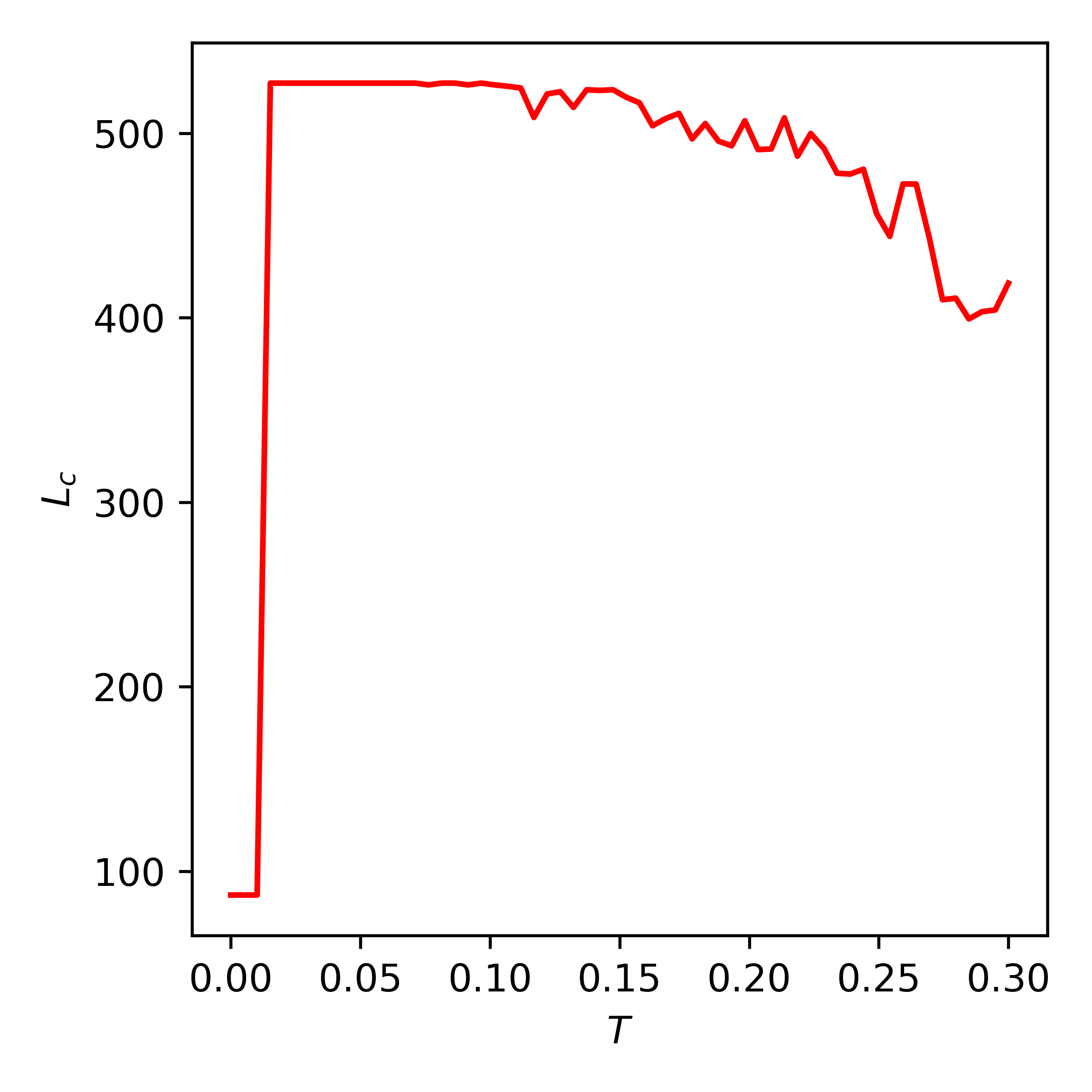}
		\caption{500D Case: $L_c$ vs $T$}
		\label{fig:b_lc500d}
	\end{subfigure}
	\caption{in a) 3D blobs (Sec. \ref{ssec:blobs}): Size vs Temperature $T$ using SPC (Sec. \ref{sssec:maxent}). 1 giant cluster at $T=0$, and 3 stable clusters from $T=0.007$ to $T \approx 0.05$. Insets: (in a) above right) Susceptibility $\chi(T)$, and (in a) below right) Average Magnetization $\langle M \rangle$ at $T=0.01$. $\chi$ peaks at $T=0.007$ into the SP-phase, remains stable until $T \approx 0.09$, and then slowly decreases to 0 into the Paramagnetic Phase.
	in b) the likelihood $L_c (T)$ of SPC for 3D clusters, and 500D in c). When D is low, $L_c$ is stable until $T \approx 0.05$ and transitions into the Paramagnetic Phase. Where we would expect a decrease in $L_c$, we see an increase as $T$ goes up until $T \approx 0.20$ which as can be seen from a) $\chi \approx 0$ signaling the Paramagnetic Phase. The 500D case in c), on the other hand, peaks early within the SP-phase then transitions into the Paramagnetic Phase at $T \approx 0.10$. Once $T > 0.10$, a net decrease in $L_c$ happens, and as $T$ goes up the slope of $L_c$ remains negative. }\label{fig:b_sol}
\end{figure*}

Figure Fig. \eqref{fig:b_clus} respectively show the susceptibility, and the clusters size as a functions of temperature. At $T = 0.01$, in the SP-phase, we observe 3 clusters in Fig. \eqref{fig:b_clus} of size 167, 166, and 166 with an ARI of 1.

We follow this with $L_c$'s solution which scores 1460.47, an ARI of 0.20, while the expected likelihood was 599.91. Yet again our solution's likelihood is higher than our expectation, and the real clusters are divided in smaller ones. In comparison K-Means and DBSCAN respectively achieve ARI of 1, and 0.8. In light of this, we decided to try again the same problem but with the dimensionality set to $D = 500$. As expected SPC's results do not change. However in this instance f-SPC's solution quickly converges to the best classification. A further investigation was done by simulating both the 3D, and 500D cases using SPC, and computing the likelihood $L_c$ for every configurations. The assumption being that the maximum likelihood should be found within the temperature range where the system is in the SP-phase.

Figures \eqref{fig:b_lc3d} and \eqref{fig:b_lc500d} respectively show the likelihood as functions of temperature. We notice that in the 500D case in Fig. \eqref{fig:b_lc500d}, the maximum likelihood is found at temperatures $T < 0.15$ within the SP-phase, and the $L_c$ monotonously decreases at higher temperatures. The opposite happens in Fig. \eqref{fig:b_lc3d} where the best classification doesn't correspond to the maximum likelihood of $L_c$ which in this case is found at high temperatures $T \approx 0.25$ which by looking at $\chi$ in Fig. \eqref{fig:b_clus} means we are effectively in the Paramagnetic phase. We provide additional comments in the discussion section of this paper. 

\section{Data Pre-Processing} \label{sec:preprocessing}

\subsection{The Distance Function} \label{ssec:distance}

The wide variety of problems our clustering methods can tackle necessitates a careful choice of pairwise distances if we are to properly identify shared behavior. We will proceed by using the Euclidean distances whenever we assume independence of the features, and the Pearson correlations otherwise especially for problems where the features consist of time-series.

This has implications for both algorithms such that we use the Euclidean distance or the Pearson correlation distance for SPC, and for f-SPC, we use the Pearson correlation matrix, and the similarity matrix, which is the Euclidean distance matrix on $[0,1]$, and subtracted from 1.

We note that from \cite{kullmann2000identification} that our Eqn. \eqref{eq:4} can be modified to incorporate negative correlations, but the authors explain this only affects the results at the ground state ({\it i.e. $T \approx 0$}).

\subsection{Scaling} \label{ssec:scaling}
Raw data sets often contain features on different order of magnitude of scales, outliers, and missing data which can have significant impact on Machine Learning algorithms. One way to deal with these issues is through normalization of the features. This was achieved using the Min-Max Scaling technique which puts all features on a 0 to 1 scale by performing the following operation:

\begin{equation} x_{scaled}=\frac{x - x_{min} }{ x_{max} - x_{min}~. } \end{equation}
Scaling has significant effects on the feature space: one example is seen in Fig. \eqref{fig:win_raw}, and Fig. \eqref{fig:win_minmx} which respectively show the unscaled and scaled plots of the 3 wines problem. The unscaled data set has two classes completely inseparable whereas scaling the data effectively dissociates all three classes with minimal overlap.

\begin{figure*}
	\centering
	\begin{subfigure}[b]{0.3\textwidth}
		\includegraphics[width=\textwidth]{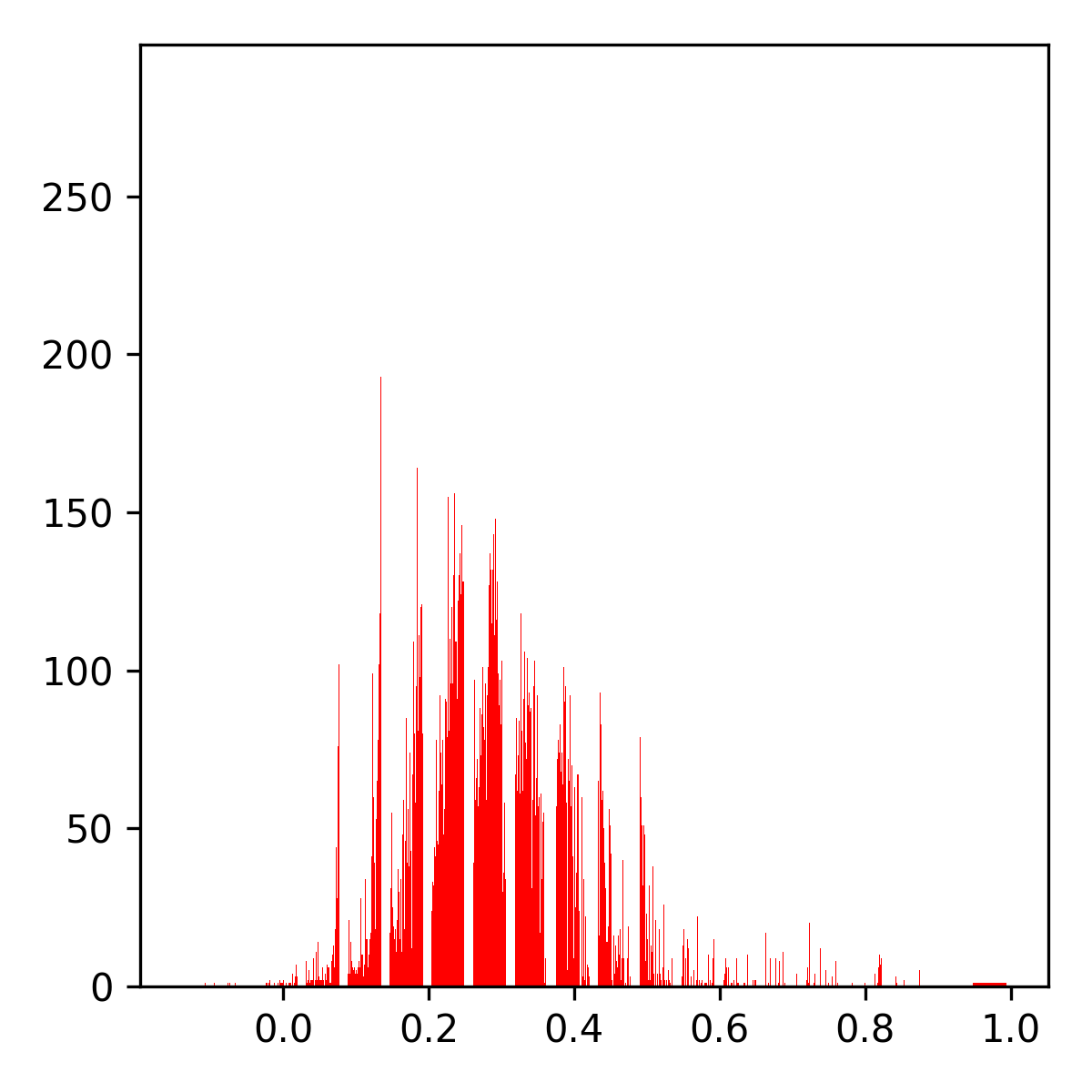}
		\caption{Market Mode}
		\label{fig:rho_full}
	\end{subfigure}
	~ 
	\begin{subfigure}[b]{0.3\textwidth}
		\includegraphics[width=\textwidth]{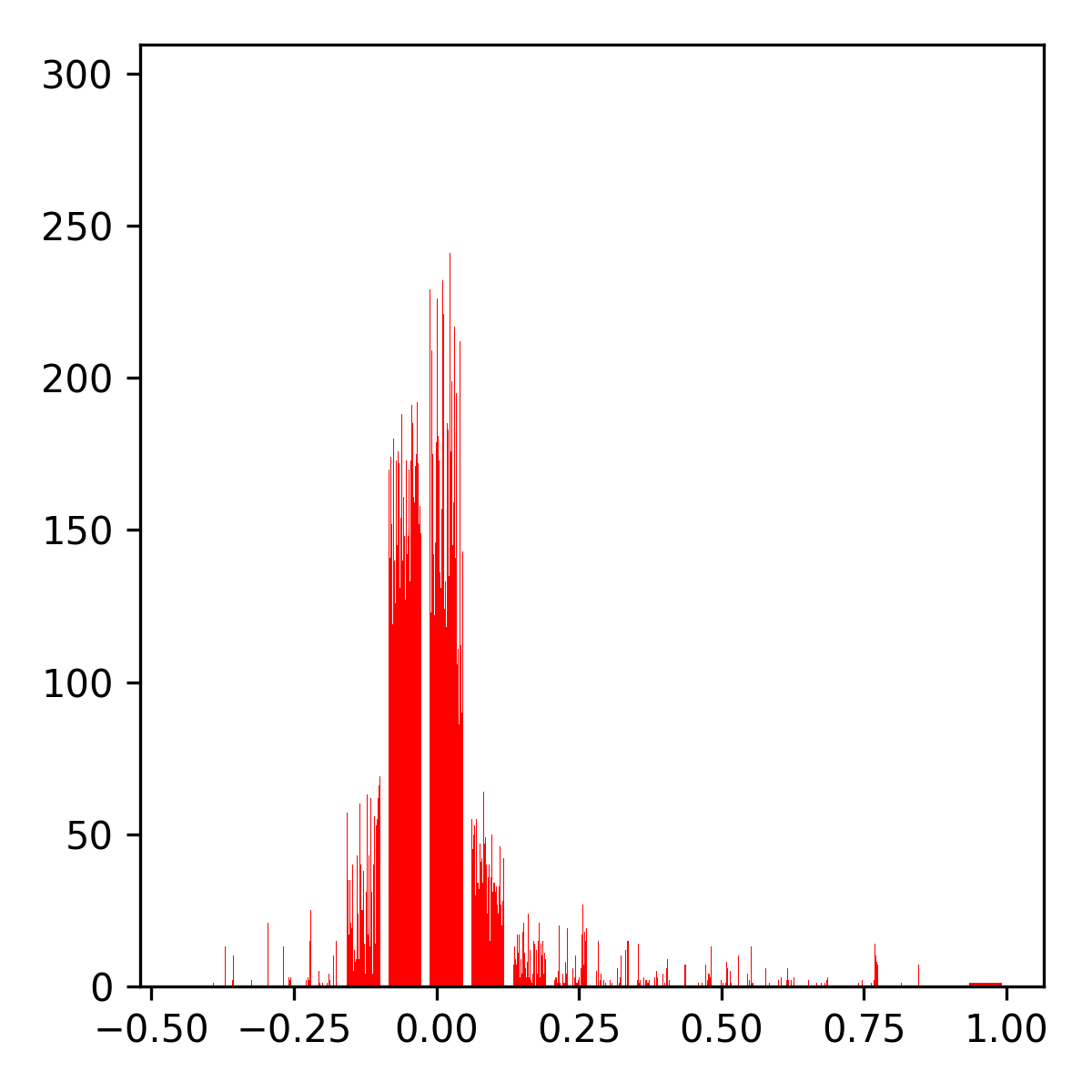}
		\caption{IMN}
		\label{fig:rho_norm}
	\end{subfigure}
	~
	\begin{subfigure}[b]{0.3\textwidth}
		\includegraphics[width=\textwidth]{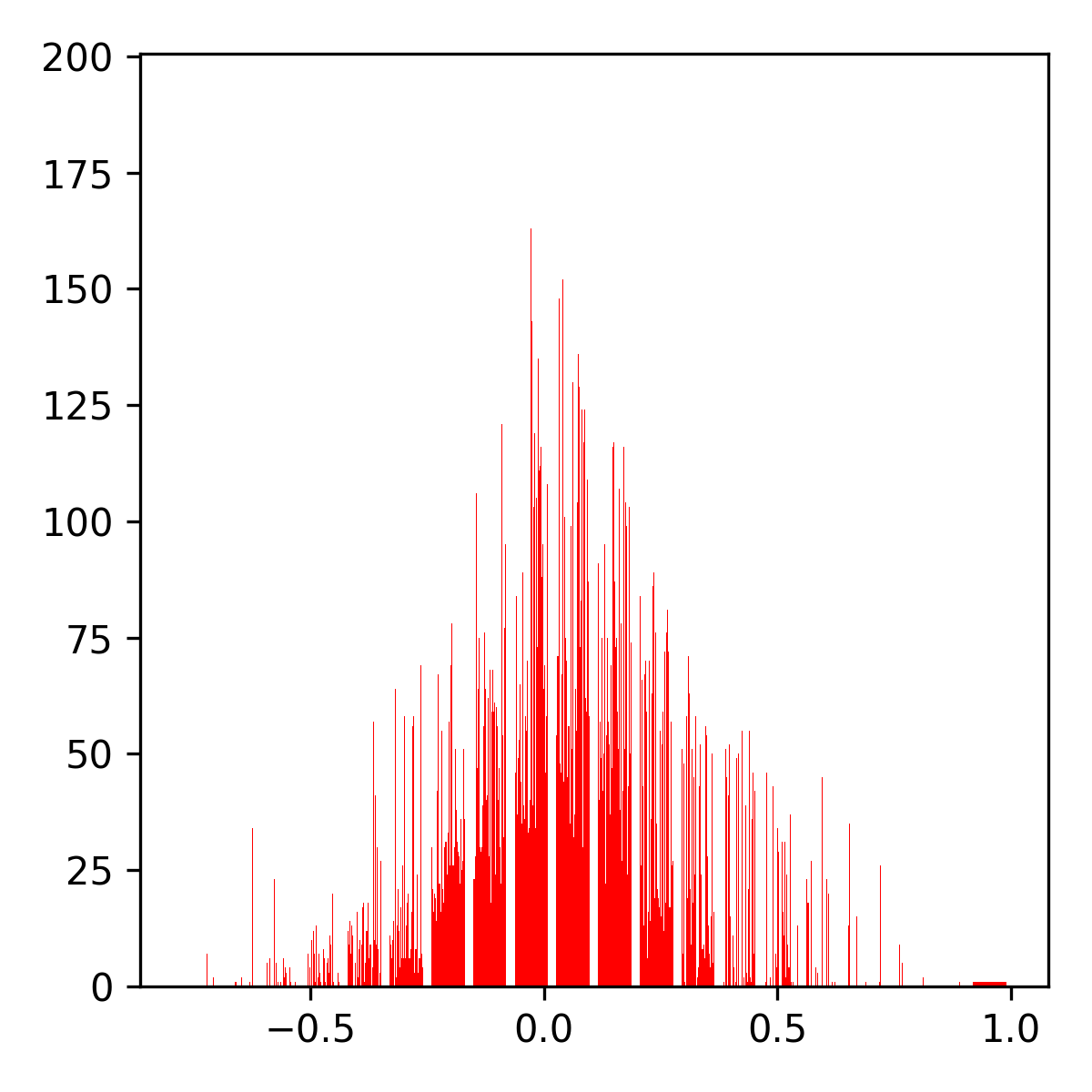}
		\caption{RMT}
		\label{fig:rho_rmt}
	\end{subfigure}
	\caption{ Distribution of Pearson correlations of daily returns for 447 publicly traded companies on the S\&P 500 stock exchange from 8/13/2012 to 8/11/2017 (Sec. \ref{ssec:financial}). The ``Market Mode'' (Sec. \ref{ssec:noise}) : Noisy markets like in a) are highly correlated with most $\rho_{ij} > 0$. The noise is cleaned by removing the ``Market Mode'' either by IMN (Sec. \ref{sssec:succnorm}) in b) or RMT methods (Sec. \ref{sssec:rmt}) in c), and produces distributions centered around 0. }\label{fig:rhos}
\end{figure*}

\subsection{Noise} \label{ssec:noise}

The next and final pre-processing task consist in removing any noise present in our data. This is especially important for financial market time-series which exhibit extreme randomness and possibly chaotic behavior. Stock market correlation matrices are noisy, and positively skewed Fig. \eqref{fig:rho_full} which translates into what is referred as the ``Market Mode''. We consider an intermediary step which consist in ``removing'' the market mode, thus ensuring we are able to recover the underlying correlation structures, if any, present in the system.

\subsubsection{Random Matrix Theory (RMT)} \label{sssec:rmt}

We follow the predictions of RMT in \cite{wilcox2007analysis} by assuming that stock market returns are IID random variables with zero mean and unit variance. These assumptions lead us to the conclusion that stock market correlations should all be zeros, and if the assumptions are indeed true, RMT predicts that the eigenvalues of the random matrices are Wishart distributed such that:

\begin{equation} \label{eq:14} P\left(\lambda\right) = \frac{Q}{2\pi} \frac{ \sqrt{ \left( \lambda_{max} - \lambda\right) \left( \lambda - \lambda_{min}\right)}      }{\lambda}\end{equation}

where $Q = \frac{D}{N}$, and $\lambda_{min/max} = 1 + \frac{1}{Q} \pm 2 \sqrt{ \frac{1}{Q} }~. $

\begin{figure}
	\centering
	\includegraphics[width=0.45\textwidth,keepaspectratio]{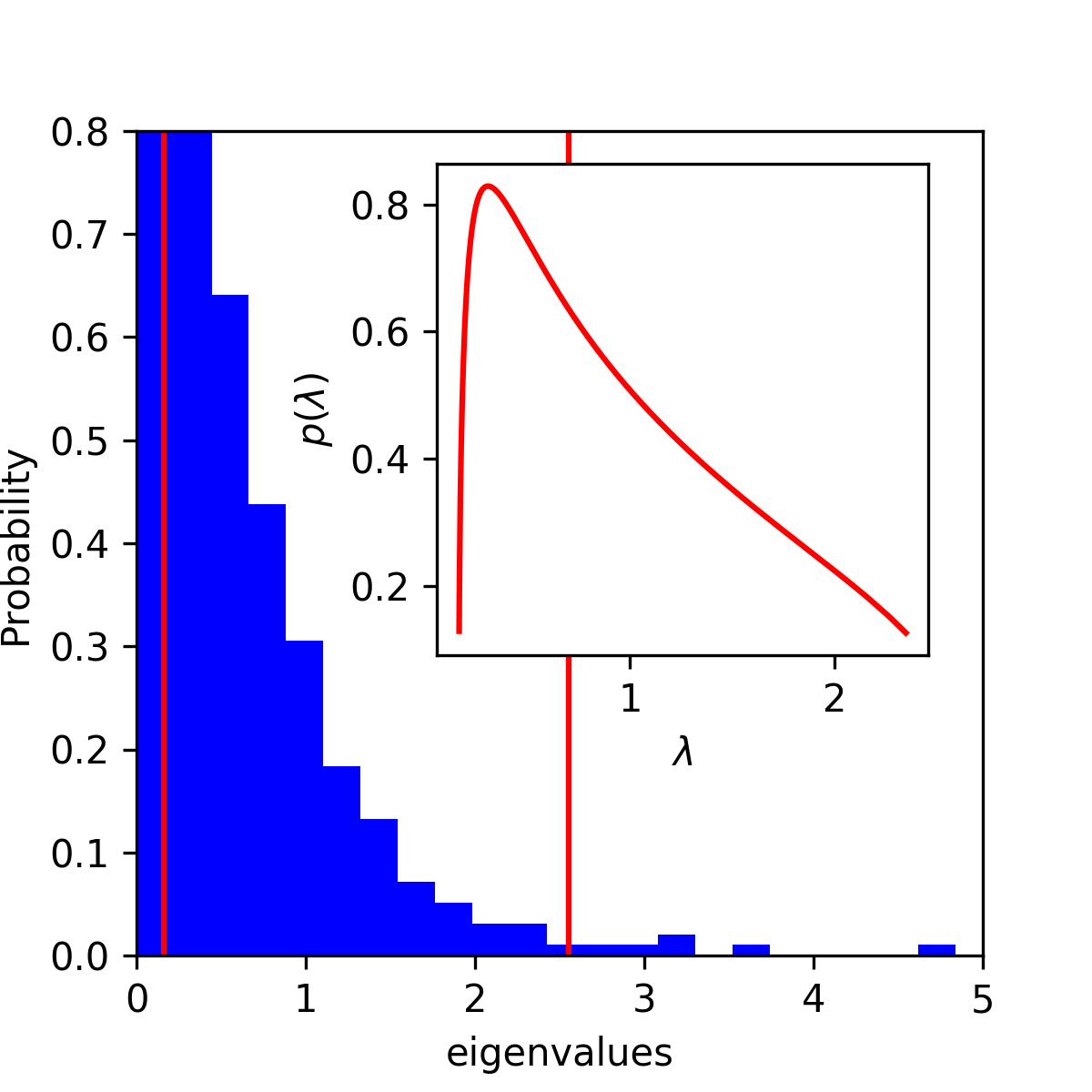}
	\captionof{figure}{The Eigenvalue distribution of the Correlation Matrix of 1249 daily returns for 447 publicly traded companies in the S\&P500 (Sec. \ref{ssec:financial}).The two vertical red lines delimit the wishart range $ \lambda_{min} = 0.16 $ and $ \lambda_{max} = 2.55 $: The eigenvalues located inside the Wishart range (see Sec. \ref{sssec:rmt} ) are noise whereas the ones outside aren't. Inset: (red curve) We show that the computed Wishart PDF of a random matrix ( using Eqn. \eqref{eq:14} ) fits well the eigenvalue distribution of our correlation matrix only inside the wishart range.  }
	\label{fig:wisdis}  
\end{figure}

Shown in Fig. \eqref{fig:wisdis} is the distribution of eigenvalues of the correlation matrix of our stock market data Sec. \ref{ssec:financial}. As can be observed the eigenvalues inside the Wishart range are responsible for the noise whereas those outside of the range are potentially correlated signal which shouldn't be discarded.

We consider the eigenvalues $\lambda > \lambda_{max}$ represent the linear, and 1st order relations between time-series while it is unclear what those on the left side ($\lambda < \lambda_{min}$) of the Wishart distribution are. The linear signals are the signals shared by assets at the sectoral level.

The RMT ``Market Mode'' removal method is implemented in the five following steps bellow in Table \ref{tab:rmt}:

\begin{table}
	\noindent\fbox{\parbox{0.45\textwidth}{
			\begin{enumerate}[label=\ref{tab:rmt}\arabic*]
				\item {\bf Compute the correlation matrix}: $c_{ij}$
				\item {\bf Extract the eigenvalues and eigenvectors}: $\Lambda$ and $U$ from $c_{ij}$
				\item {\bf Select the eigenvalues outside the Wishart Range}.
				\item {\bf Reconstruct the data}: Using the compressed signal: Let $X$ be our data, $W$ the matrix of eigenvectors found outside the Wishart Range, and $Z = W'.X$ the compressed data. The reconstructed data is then $X = W.Z$
				\item {\bf Re-compute the correlations} $\hat c_{ij}$ from the reconstructed data.
			\end{enumerate}
	}}
	\caption{ \label{tab:rmt} Implementation of RMT Noise removal methods \cite{wilcox2007analysis}}
\end{table}

We tested different time-series lengths ranging from 89 to 1249, and we note that the size of the Wishart Range increases with dimensionality, and the lower left tail decreases on the other hand. The higher the dimensionality the easier it is to rule out eigenvalues as random signals, and the more important the biggest eigenvalues are to the noise-less data reconstruction \footnote{ In \cite{giada2001data}, The authors achieve a similar result by using the model in Sec. \eqref{sssec:mle} confirming that ``noise cleaning'' mainly affects the small eigenvalues of stock market correlation matrices.}. An example of a cleaned correlation matrix resulting from this method Fig. \eqref{fig:rho_rmt}.

\subsubsection{Iterative Matrix Normalization (IMN)} \label{sssec:succnorm}

Another ``Market Mode'' removal method \cite{olshen2010successive}  IID random variables with zero mean and unit variance.

The Iterative Matrix Normalization ``Market Mode'' removal method is implemented in the three following steps bellow in \ref{tab:imn}:

\begin{table}
	\noindent\fbox{\parbox{0.45\textwidth}{
			\begin{enumerate}[label=\ref{tab:imn}\arabic*]
				\item{\bf Compute the Covariance}: $\Sigma$
				\item {\bf Standardize $\Sigma$} {\t i.e.} standardise iteratively across rows, and then columns, for a set number of iterations (i.e. 500) or until a convergence criteria is met.
				\item {\bf Extract the adjusted correlation matrix}: $\hat c_{ij}$ from $\hat \Sigma$
			\end{enumerate}
	}}
	\caption{ \label{tab:imn} Implementation of Noise Removal via Iterative Matrix Normalization \cite{giada2002algorithms} }
\end{table}

We observe in Fig. \eqref{fig:rho_norm} that the distribution of correlations is now centered around 0.

\section{The Data test-cases} \label{sec:cases}

The following examples are used as a stress test for both methods. We obtained both synthetic, and real data which enabled us to discuss the features of each models.

As a comparison tool we use the Adjusted Rand Index (ARI) \cite{hubert1985comparing} which given two classifications measures their similarity. The ARI operates on a $[-1,1]$ scale with positive values signifying increasing similarity. Where a true classification exists we will use the ARI to measure the quality of clustering of both methods but also industry standards such as ``K-Means'' \cite{lloyd1982least}, and ``DBSCAN'' \cite{ester1996density}. Using SPC (Sec. \ref{ssec:alg_spc}) we cluster a temperature range which we then compare against the $L_c$ \eqref{eq:lc} solution recovered. We then select the SPC temperature with the highest ARI for a closer comparison with the $L_c$ solution in the stock market case where a true classification is not available.

For visualization, where possible, we provide the plots or we make use of a nonlinear dimensionality reduction package called UMAP \cite{mcinnes2018umap}. The graph of the MST is also provided as it is a faithful representation of clusters on a 2D plane. The MST takes in the graph of our data, and find the unique shortest path linking every nodes.

\subsection{ {\it Sci-Kit learn}: Concentric Circles } \label{ssec:circles}
Our first problem is the identification of two concentric circles on a 2D plane using \texttt{Sci-kit learn} \cite{scikit-learn} samples generator \footnote{B. Thirion, G. Varoquaux, A. Gramfort, V. Michel, O. Grisel, G. Louppe, J. Nothman, 'make\_circles', 2017. [Online]. Available: \url{http://scikit-learn.org/stable/modules/generated/sklearn.datasets.make_circles.html}. [Accessed: 12-Jun-2018]}\label{test} with $N = 500$, 0.5 for the noise parameter, and the 2 dimensions represent the X and Y coordinates of the observations. 

\begin{figure*}
	\centering
	\begin{subfigure}[b]{0.45\textwidth}
		\includegraphics[width=\textwidth]{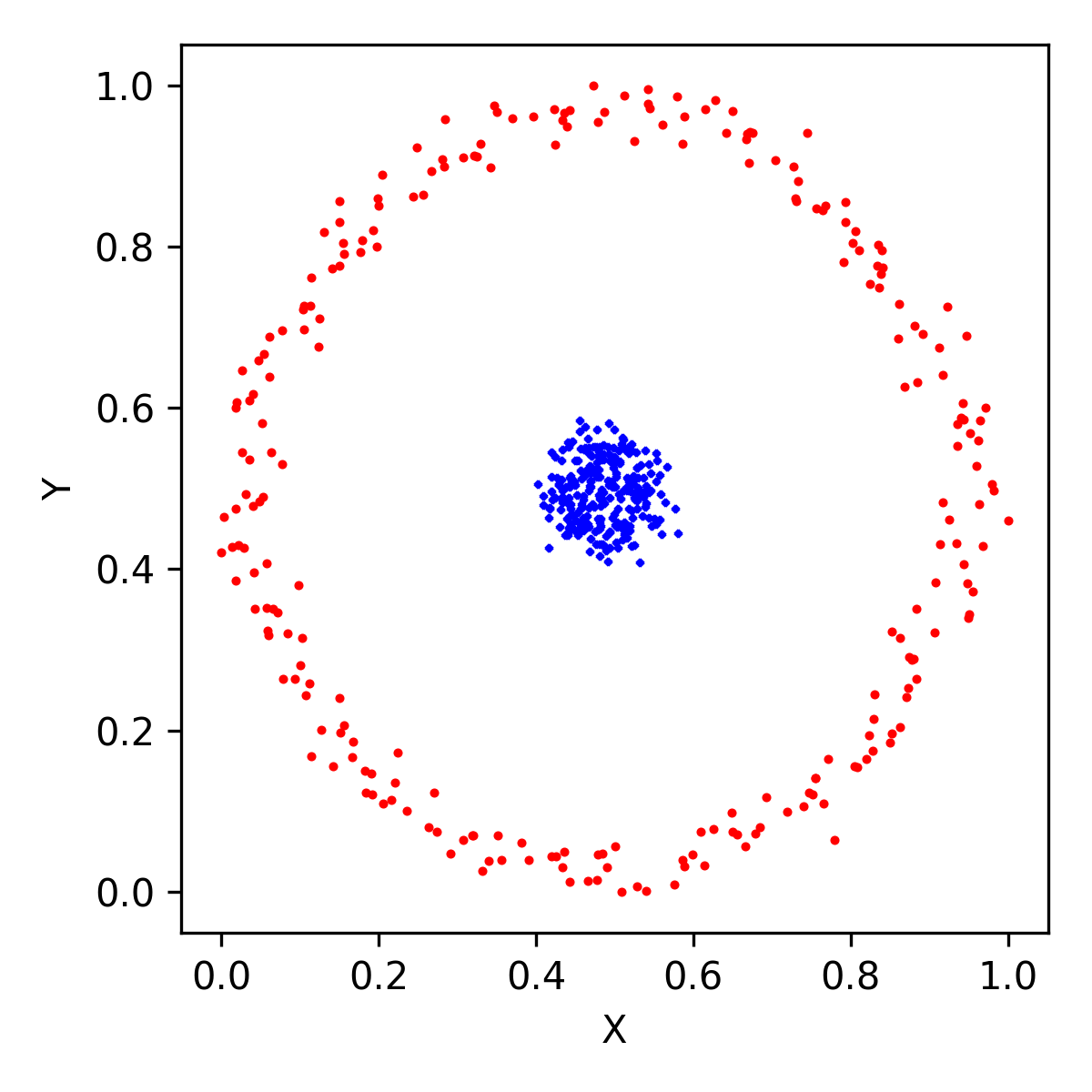}
		\caption{Scatter Plot}
		\label{fig:cir_clus}
	\end{subfigure}
	~
	\begin{subfigure}[b]{0.45\textwidth}
		\includegraphics[width=\textwidth]{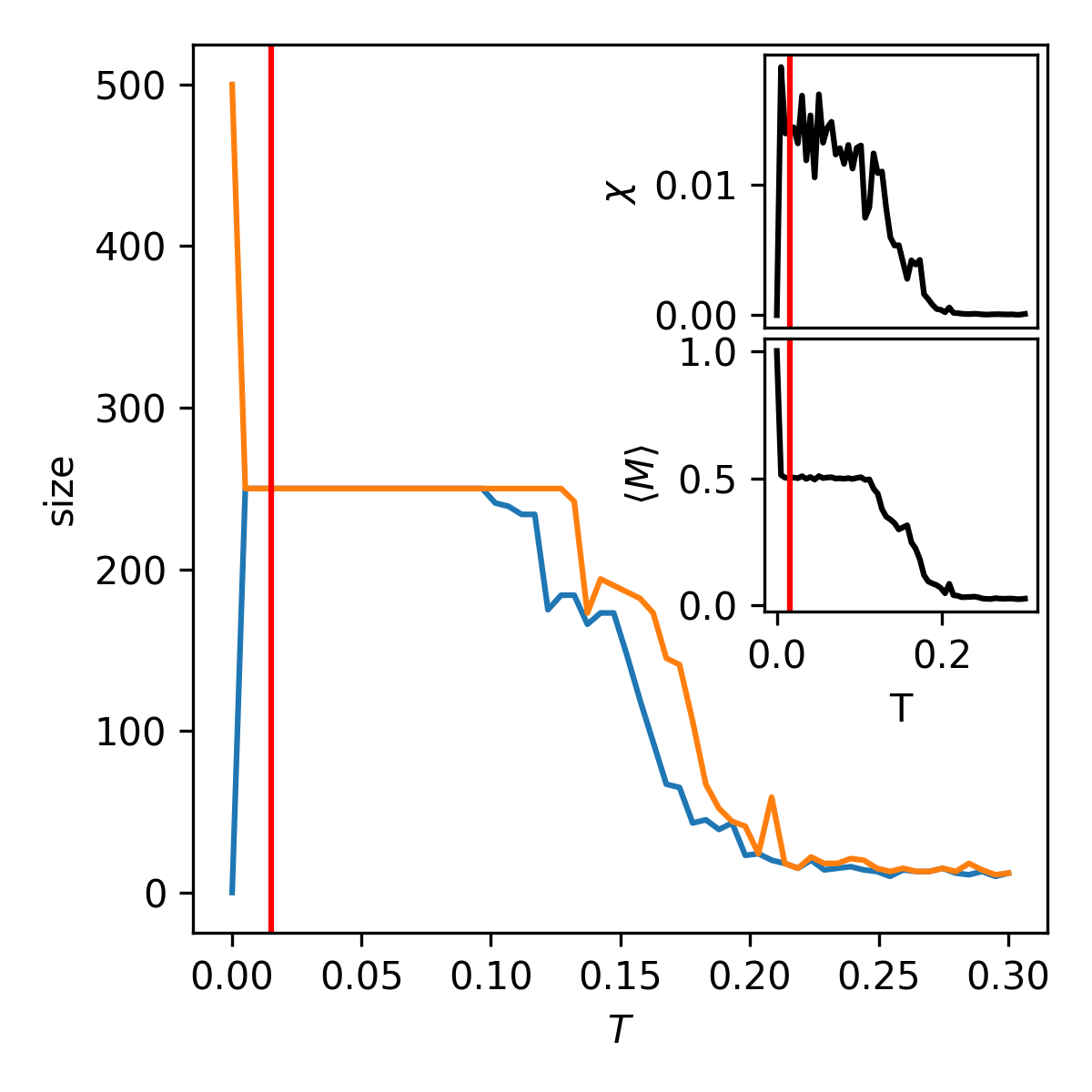}
		\caption{Cluster size vs Temperature $T$}
		\label{fig:cir_plot}
	\end{subfigure}
	\caption{in a) Two circle (Sec. \ref{ssec:circles}) shaped 2D clusters each of size $N=250$ such that the blue points have higher density than the red ones. in b) Cluster size vs Temperature $T$ using SPC (Sec. \ref{sssec:maxent}). As $T$ increases, the giant component successively breaks down: at $T = 0.007$ we can observe 2 clusters which remain stable until $T=0.10$. Insets: ( in a) above right ) Susceptibility $\chi(T)$ at $T=0.01$. $\chi$ peaks around $T=0.007$, remains stable until $T \approx 0.10$ inside the SP-phase, then dives down toward 0 for $T > 0.10$. (in a) below right) The Average Magnetization $\langle M \rangle(T)$ at $T=0.01$. $\langle M \rangle$ starts at 1 for $T=0$ then remains stable at $\langle M \rangle = 0.5$ inside the SP-phase from $T=0.01$ to $T=0.10$. This stability only occurs when clusters have uniform or identical densities, and are linearly separable. Once $T>0.10$, $\langle M \rangle$ goes down to 0 inside the Paramagnetic Phase.}\label{fig:cir1}
\end{figure*}

Judging by observing Fig. \eqref{fig:cir_plot} , and  we see no overlap between the two clusters present in the data, and we expect to recover close to perfect clusters after applying the algorithms.

\begin{figure}
	\centering
	\begin{subfigure}[b]{0.45\textwidth}
		\includegraphics[width=\textwidth]{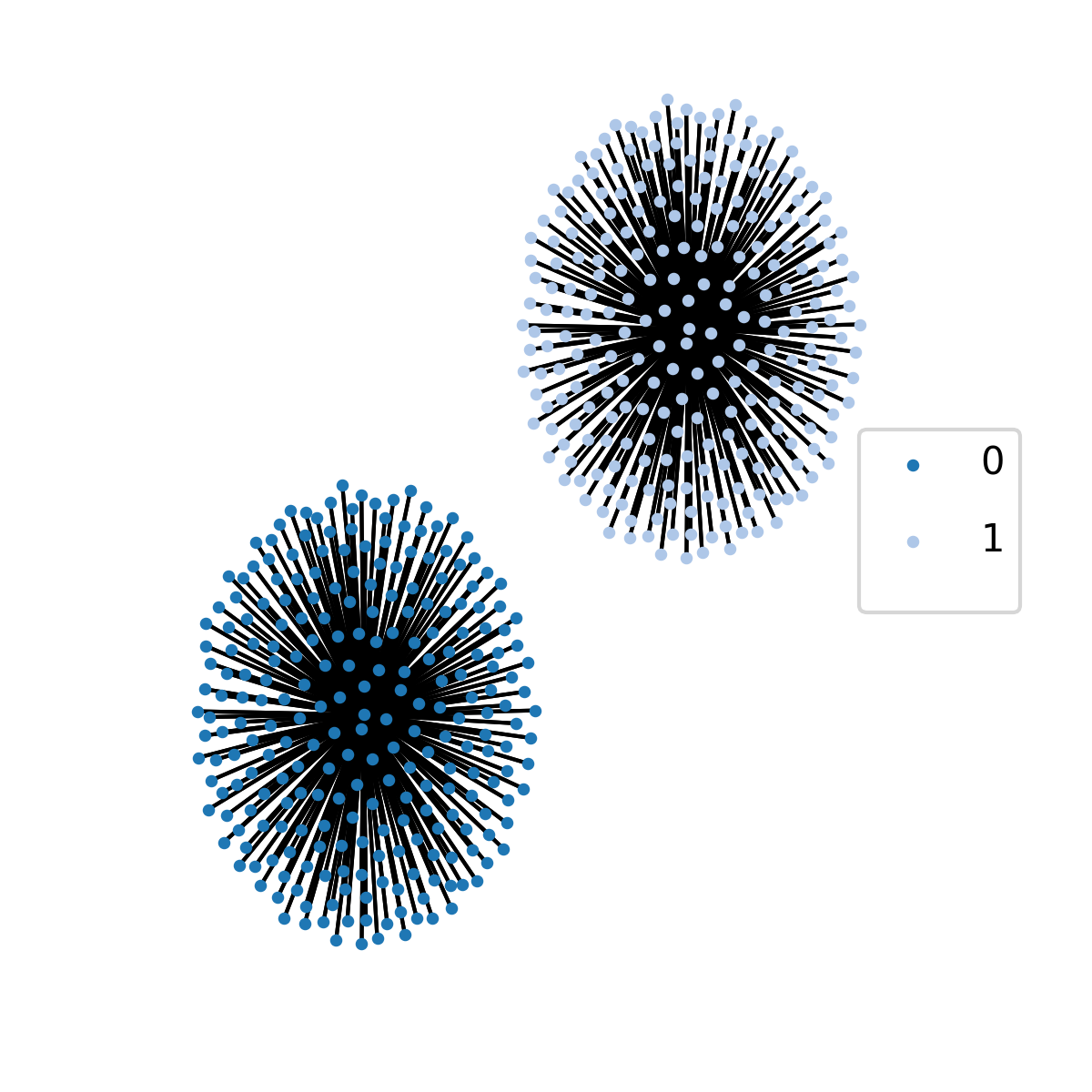}
		\caption{SPC}
		\label{fig:cir_sa}
	\end{subfigure}
	
	\begin{subfigure}[b]{0.45\textwidth}
		\includegraphics[width=\textwidth]{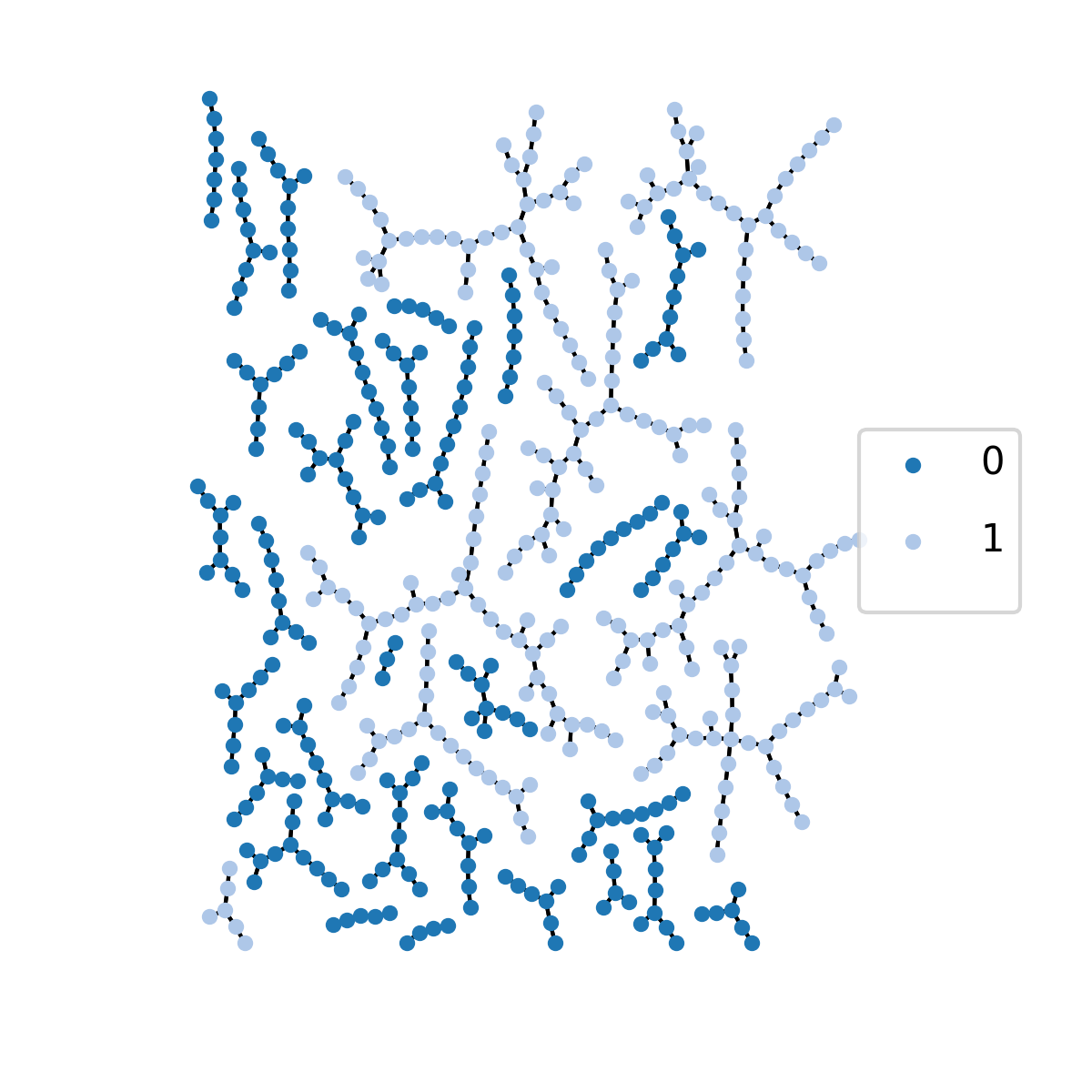}
		\caption{f-SPC}
		\label{fig:cir_mle}
	\end{subfigure}
	\caption{in a) 2 Circles (Sec. \ref{ssec:circles}): The MST of the (SPC Sec. \ref{sssec:maxent}) Solution at $T=0.01$ shows two subtrees each representing the two clusters in the data, and in b) with the f-SPC (Sec. \ref{sssec:mle}) Solution, a high number of clusters are found: There is no misclassification however the 2 original clusters are pieced apart}\label{fig:cir_sol}
\end{figure}

We obtained the susceptibility as a function of temperature in Fig. \eqref{fig:cir_clus}. Within the SP-phase at $T=0.01$ we observe two clusters in figure Fig. \eqref{fig:cir_sa} both contain 250 nodes with an ARI of 1. Once the temperature gets relatively high, near $T \approx 0.15$, the system is deemed at ``high energy'', and the clusters dissolve almost simultaneously. Unlike this particular example, this does not generally happen with real data where clusters have different densities.

On this data, f-SPC runs for 10000 generations maximizing $L_c$ to 639 while the real classification scores 317. The f-SPC configuration is presented in Fig. \eqref{fig:cir_mle} with an ARI of 0.085. In comparison K-Means and DBSCAN respectively achieve 0.16, and 1. K-Means has low clustering quality despite specifying the correct number of clusters. This is due to its inability to deal with non-spherical and non-Gaussian shaped clusters. Despite the high likelihood, Fig. \eqref{fig:cir_mle} shows a high number of clusters. The clusters are not mixed and ultimately we fail to recover the initial two clusters.

\subsection{ {\it Sci-Kit learn}: Wine data } \label{ssec:wine}

The second problem consists of a data set containing three clusters: $N = 178$, and $D=13$. It is a reputed easy problem illustrating the importance of Normalizing/Standardizing features. There are 59, 71, and 48 samples respectively for class 1, 2 and 3, and the data is generated using \texttt{Sci-kit learn} loader \footnote{D. Cournapeau, F. Pedregosa, O. Grisel 'load\_wine', 2007-2010. [Online]. Available: \url{http://scikit-learn.org/stable/modules/generated/sklearn.datasets.load_wine.html}. [Accessed: 12-Jun-2018]}.
the 13 features are quantities extracted from a chemical analysis of 3 types of italian wines: One Alcohol, Malic acid, Ash, Alcalinity of ash, Magnesium, Total phenols, Flavanoids, Nonflavanoid phenols, Proanthocyanins, Color intensity, Hue, OD280/OD315 of diluted wines, and Proline.

\begin{figure*}
	\centering
	\begin{subfigure}[b]{0.3\textwidth}
		\includegraphics[width=\textwidth]{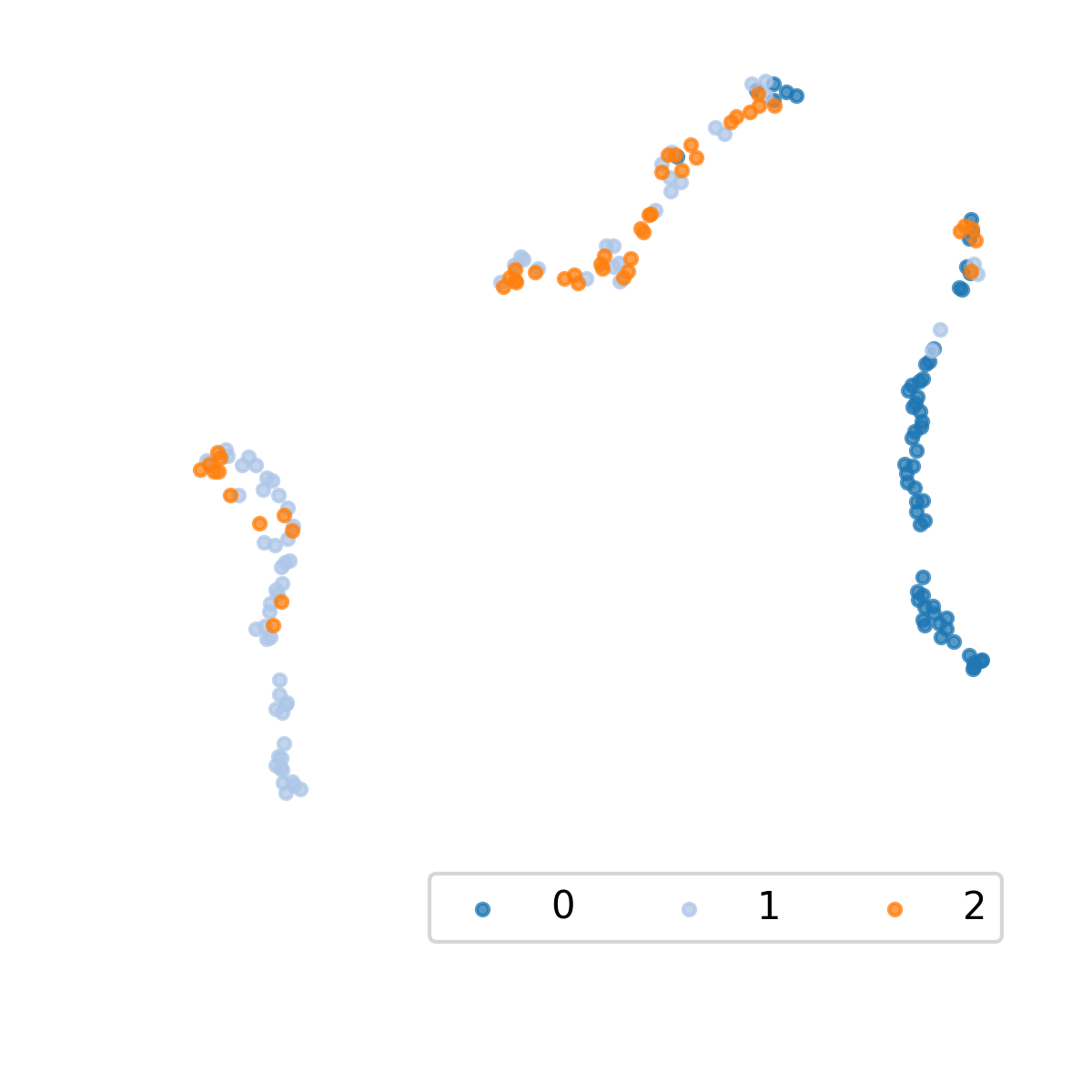}
		\caption{Raw Data}
		\label{fig:win_raw}
	\end{subfigure}
	~ 
	\begin{subfigure}[b]{0.3\textwidth}
		\includegraphics[width=\textwidth]{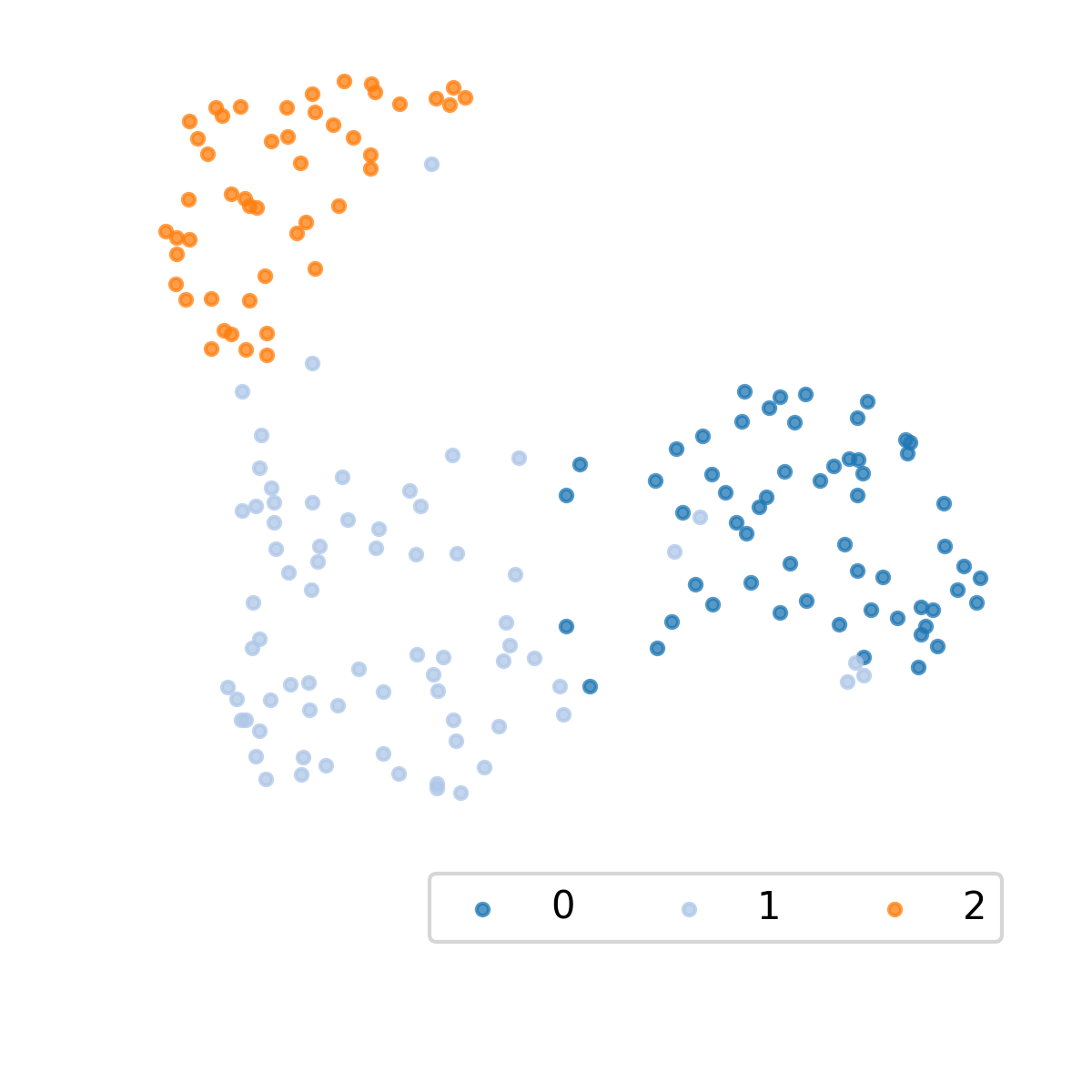}
		\caption{Normalized Data}
		\label{fig:win_minmx}
	\end{subfigure}
	~ 
	\begin{subfigure}[b]{0.3\textwidth}
		\includegraphics[width=\textwidth]{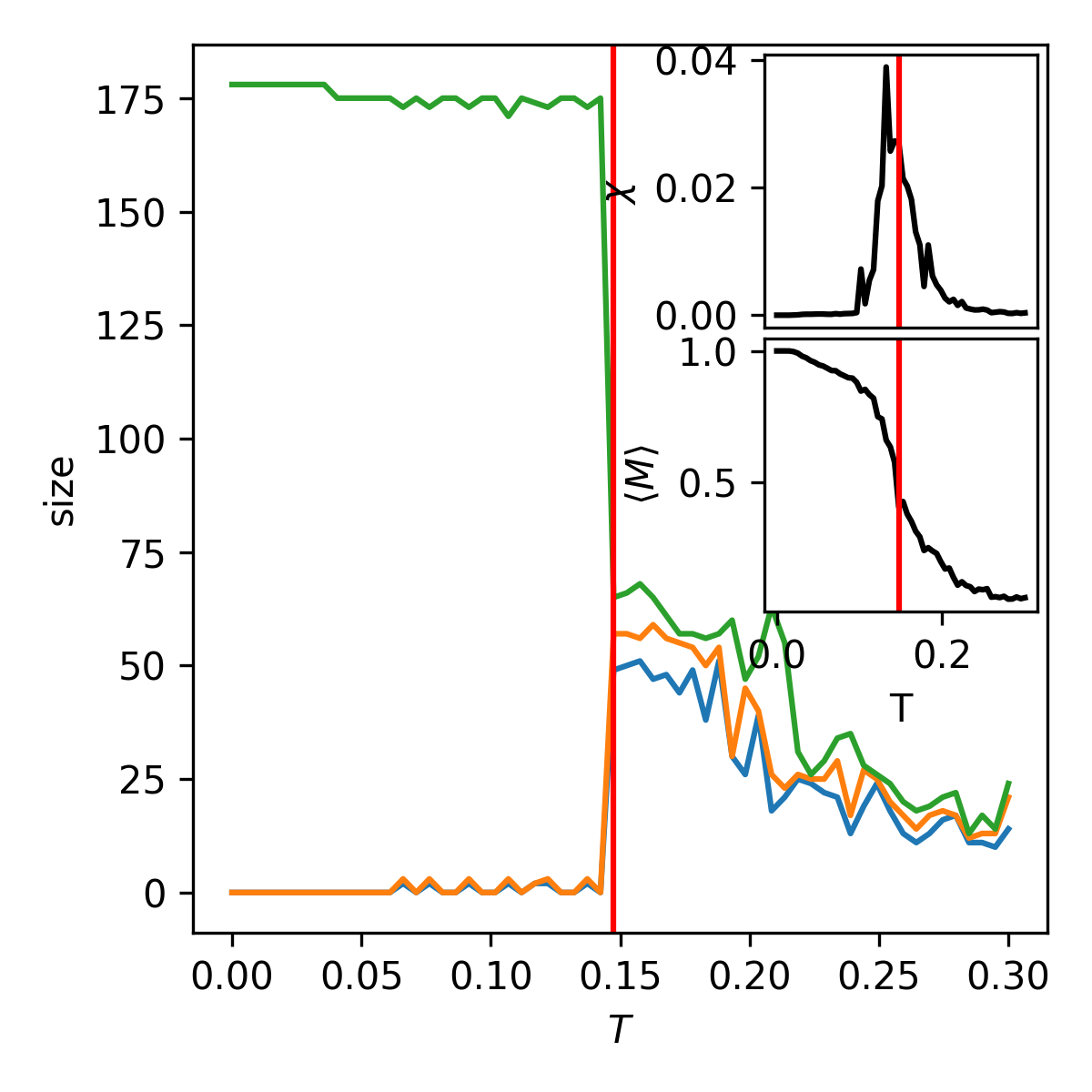}
		\caption{Clusters size vs Temperature $T$}
		\label{fig:win_clus}
	\end{subfigure}
	\caption{in a) 3 wines (Sec. \ref{ssec:wine}), 2D plot of dimensionality reduction of the 13 features using UMAP \cite{mcinnes2018umap}. No scaling and/or normalizing done to the features produces 3 clusters: Wines of type 1 and 2 are found in the same clusters while Wines in cluster 0 remain isolated. in b) MinMax features re-scaling Sec. \ref{ssec:scaling}. The original clusters spread out, and the Wine 1 and 2 clusters are now linearly separable. in c) Size vs Temperature $T$. From $T=0$ to $T \approx 0.14$, The ferromagnetic Phase with one giant cluster, then from $T \approx 0.14$ to $T \approx 0.20$, the SP-phase with 3 clusters which all start dissolving once $T > 0.20$. Insets: (in a) above right) Susceptibility $\chi(T)$, and (in a) below right) Average Magnetization $\langle M \rangle$  at $T \approx 0.15$. $\chi$ peaks at $T=0.12$ into the SP-phase, decreases, and peaks one last time at $T \approx 0.17$ to transition into the Paramagnetic Phase.}\label{fig:r}
\end{figure*}

At first sight in Fig. \eqref{fig:win_raw}, 2 clusters are merged whereas once the features have been normalized Fig. \eqref{fig:win_minmx} the 3 clusters occupy separate regions of the space. Each cluster has one extremity close to its neighboring cluster which may induce some misclassification error, and because of this we expect to recover 3 imperfect clusters.

\begin{figure}
	\centering
	\begin{subfigure}[b]{0.45\textwidth}
		\includegraphics[width=\textwidth]{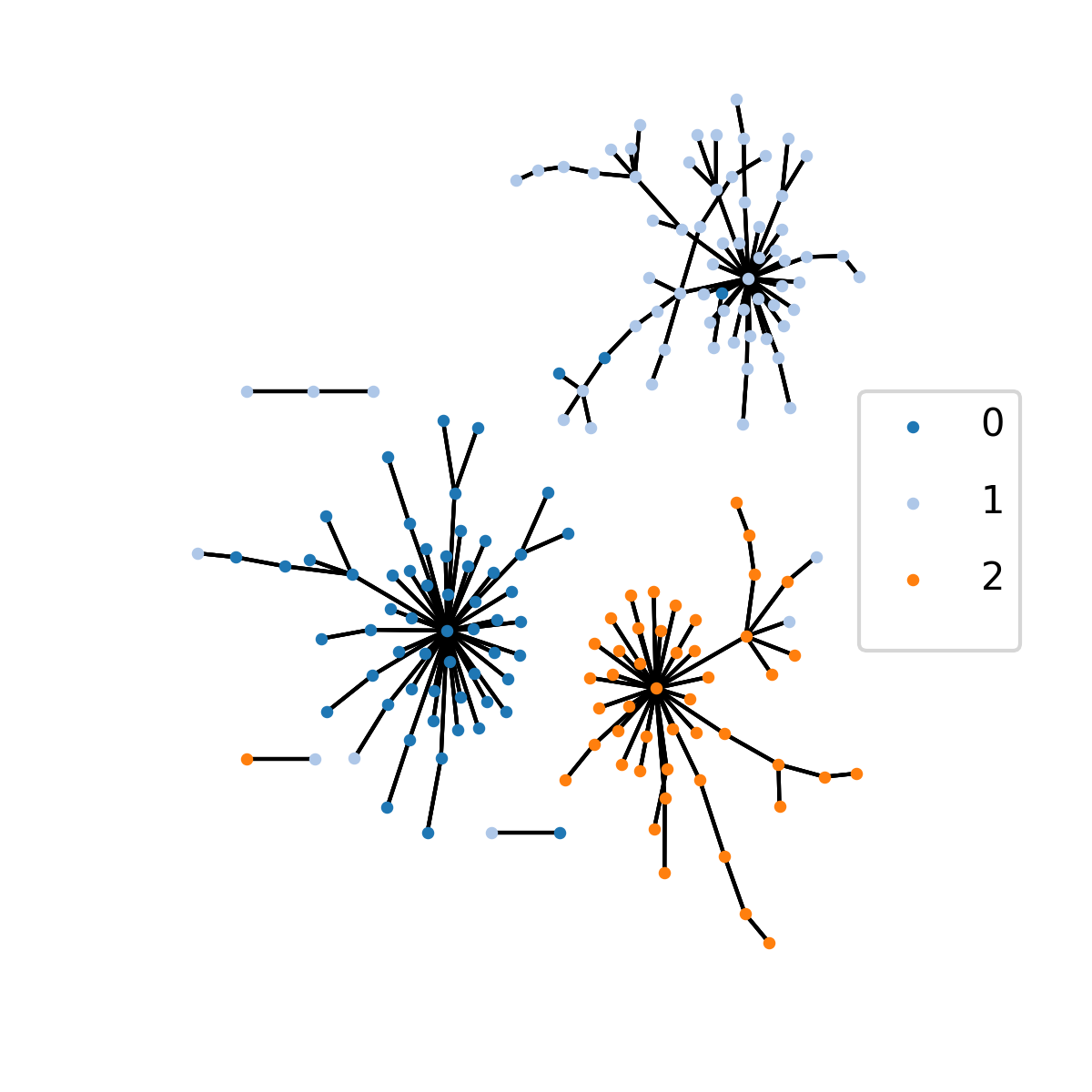}
		\caption{SPC}
		\label{fig:win_sa}
	\end{subfigure}
	
	\begin{subfigure}[b]{0.45\textwidth}
		\includegraphics[width=\textwidth]{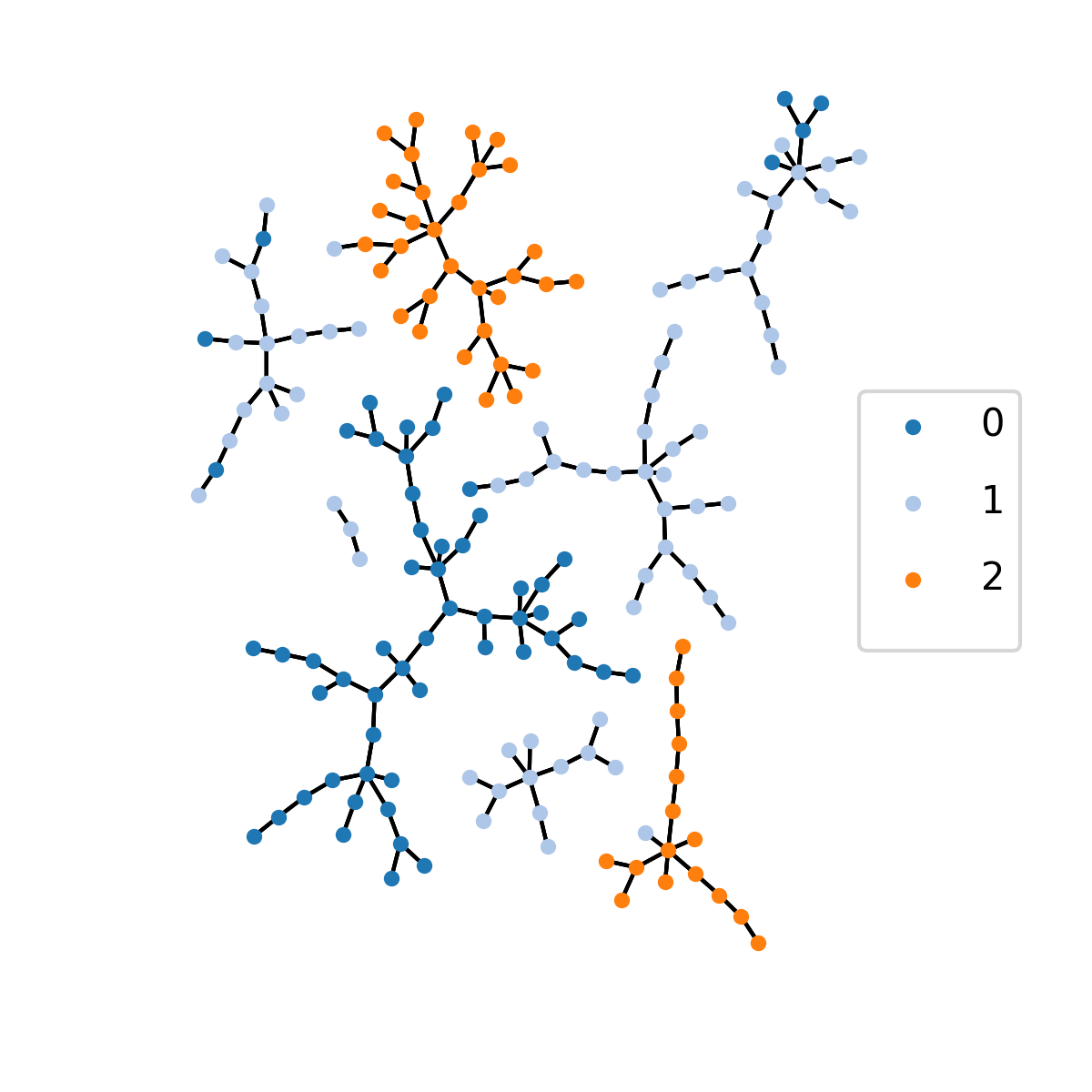}
		\caption{f-SPC}
		\label{fig:win_mle}
	\end{subfigure}
	\caption{ In a) 3 wines (Sec. \ref{ssec:wine}) MST of SPC's solution at $T=0.147$: The 3 largest clusters respectively contain most of the observations from the original wine groups except for few unclassified or misclassified samples. in b) f-SPC's solution: There are 7 clusters: 1 for Wine 0, 4 for Wine 1, and 2 for Wine 2. }\label{fig:win_sol}
\end{figure}

Between $T = 0.147$ and $T \approx 0.22$ we observe three clusters, and the best classification recovered in Fig. \eqref{fig:win_sa} provides the MST of the SPC's solution with an ARI of 0.82.

Figure Fig. \eqref{fig:win_mle} presents $L_c$'s solution with a likelihood of 83.94, an ARI of 0.51, and an expected likelihood of 66.97. As with the circle problem our solution's $L_c$ is higher than our expectation, and it has 7 clusters instead of 3. 1 cluster contains observations of cluster 0, while clusters 1 and 2 are split in smaller ones without much misclassification. In comparison K-Means and DBSCAN respectively achieve ARI of 0.85, and 0.42.

\subsection{{\it Sci-Kit learn}: Fishers Iris data } \label{ssec:iris}

Fisher' Iris Data using \texttt{Sci-kit learn} loader \footnote{D. Cournapeau, F. Pedregosa, O. Grisel 'load\_iris', 2007-2010. [Online]. Available: \url{http://scikit-learn.org/stable/modules/generated/sklearn.datasets.load_iris.html}. [Accessed: 12-Jun-2018]} which includes individuals from 3 species: Iris Setosa, Virginica, and Versicolor. $N=150$, $D=4$, and there are 50 nodes per cluster.

\begin{figure*}
	\centering
	\begin{subfigure}[b]{0.45\textwidth}
		\includegraphics[width=\textwidth]{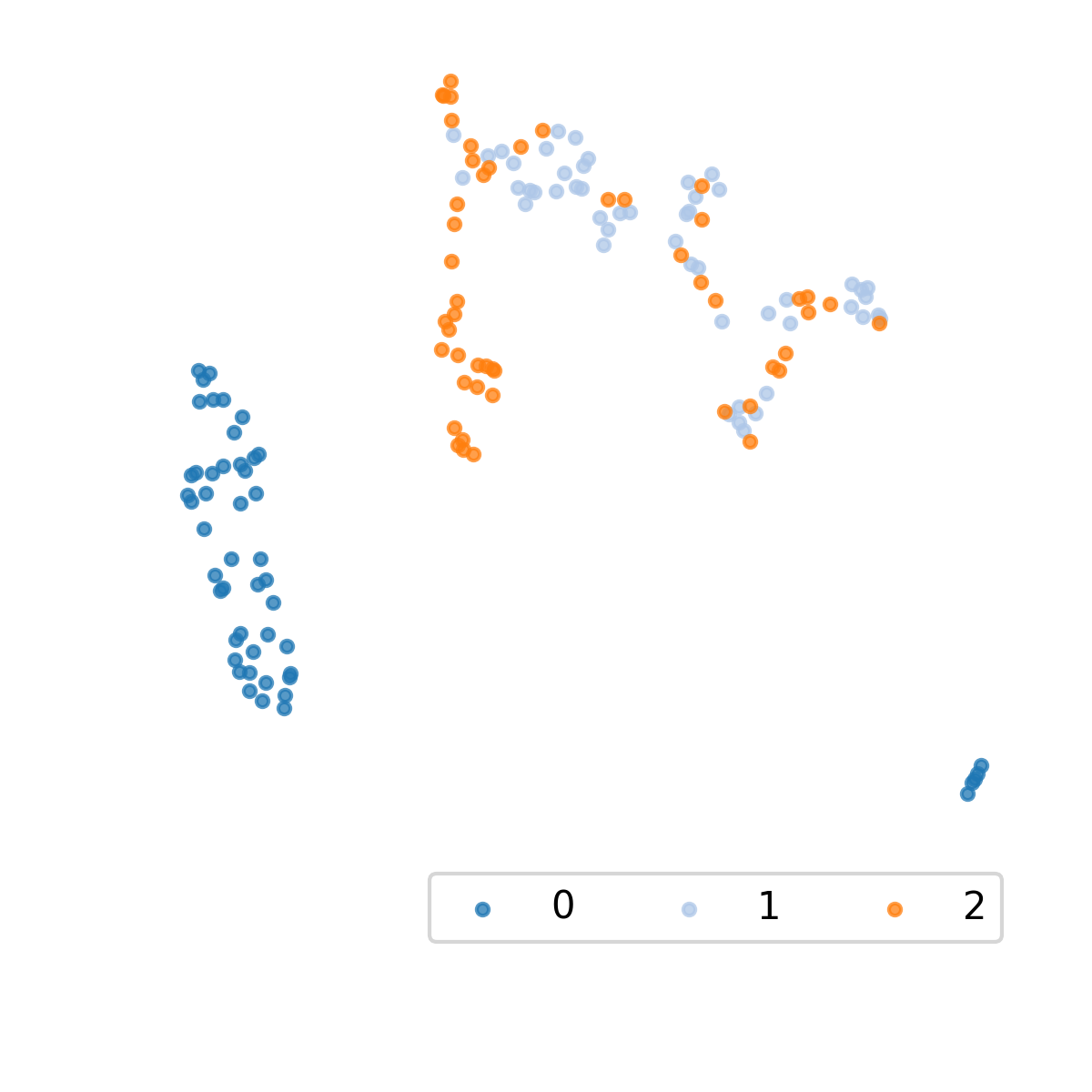}
		\caption{Scatter Plot}
		\label{fig:iris_clus}
	\end{subfigure}
	~
	\begin{subfigure}[b]{0.45\textwidth}
		\includegraphics[width=\textwidth]{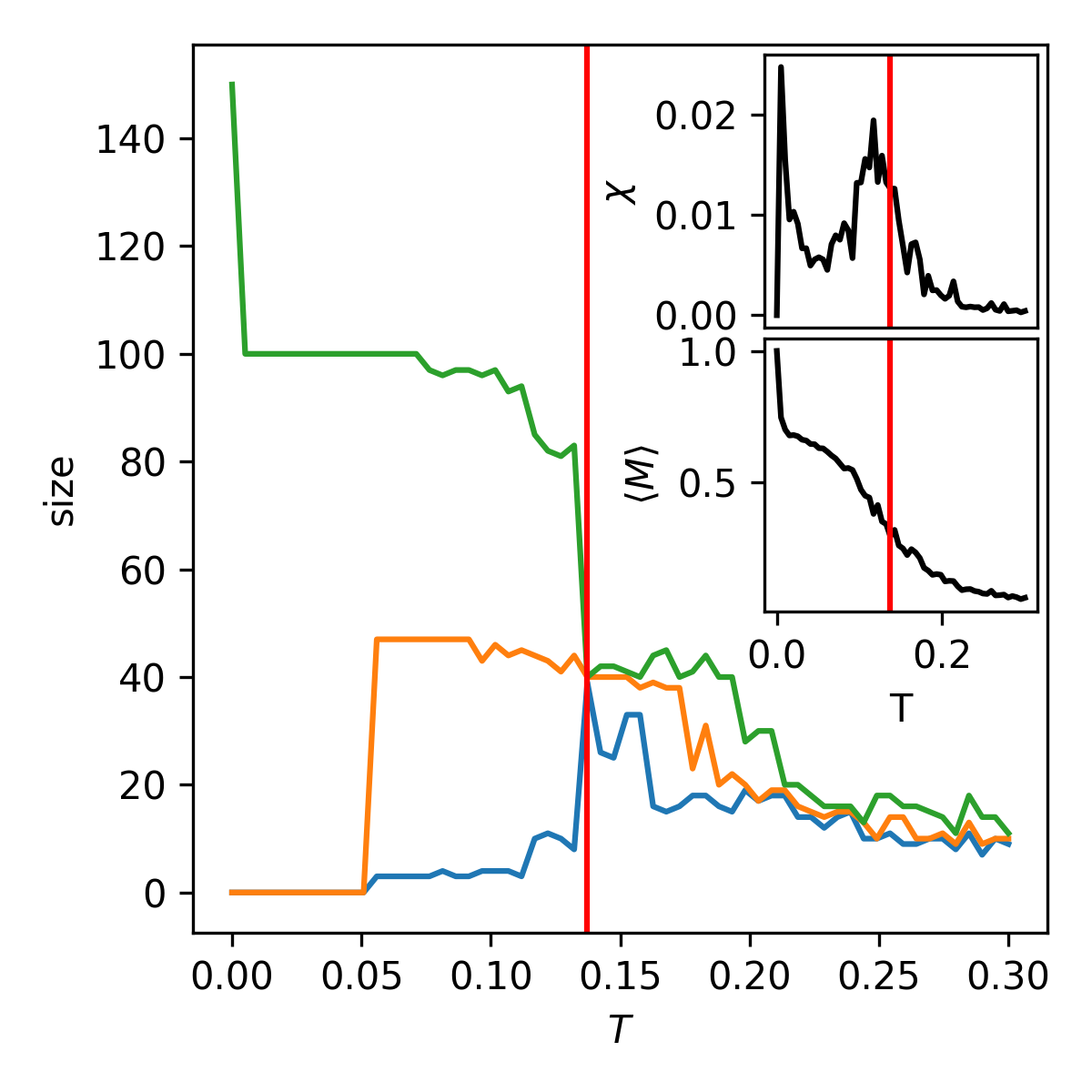}
		\caption{Cluster sizes vs Temperature $T$}
		\label{fig:iris_plot}
	\end{subfigure}
	\caption{In a) Iris 3 species (Cluster 0 for ``Setosa'', 1 for ``Versicolor'', and 2 for ``Virginica'') (Sec. \ref{ssec:iris}) : 2D plot of dimensionality reduction of 4 MinMax Scaled features. The Setosa, Versicolor, and Virginica clusters are respectively clusters 0, 1, and 2. Clusters 1 and 2 are not linearly separable whereas Cluster 0 is. in b) Cluster sizes vs Temperature using SPC Sec. \ref{sssec:maxent}: 1 Cluster starting at $T=0$, 2 cluster at $T = 0.05$, and 3 clusters at $T \approx 0.14$. Around $T \approx 0.16$, the system transitions into the Paramagnetic Phase, and clusters start dissolving. Insets: (in a) above right) Susceptibility $\chi(T)$, and (in a) below right) Average Magnetization $\langle M \rangle$ at $T \approx 0.14$. $\chi$ peaks first at $T = 0.007$, and a second time at $T \approx 0.12$: at each transition one or more clusters detach from the giant component.}\label{fig:iris}
\end{figure*}

\begin{figure*}
	\centering
	\begin{subfigure}[b]{0.45\textwidth}
		\includegraphics[width=\textwidth]{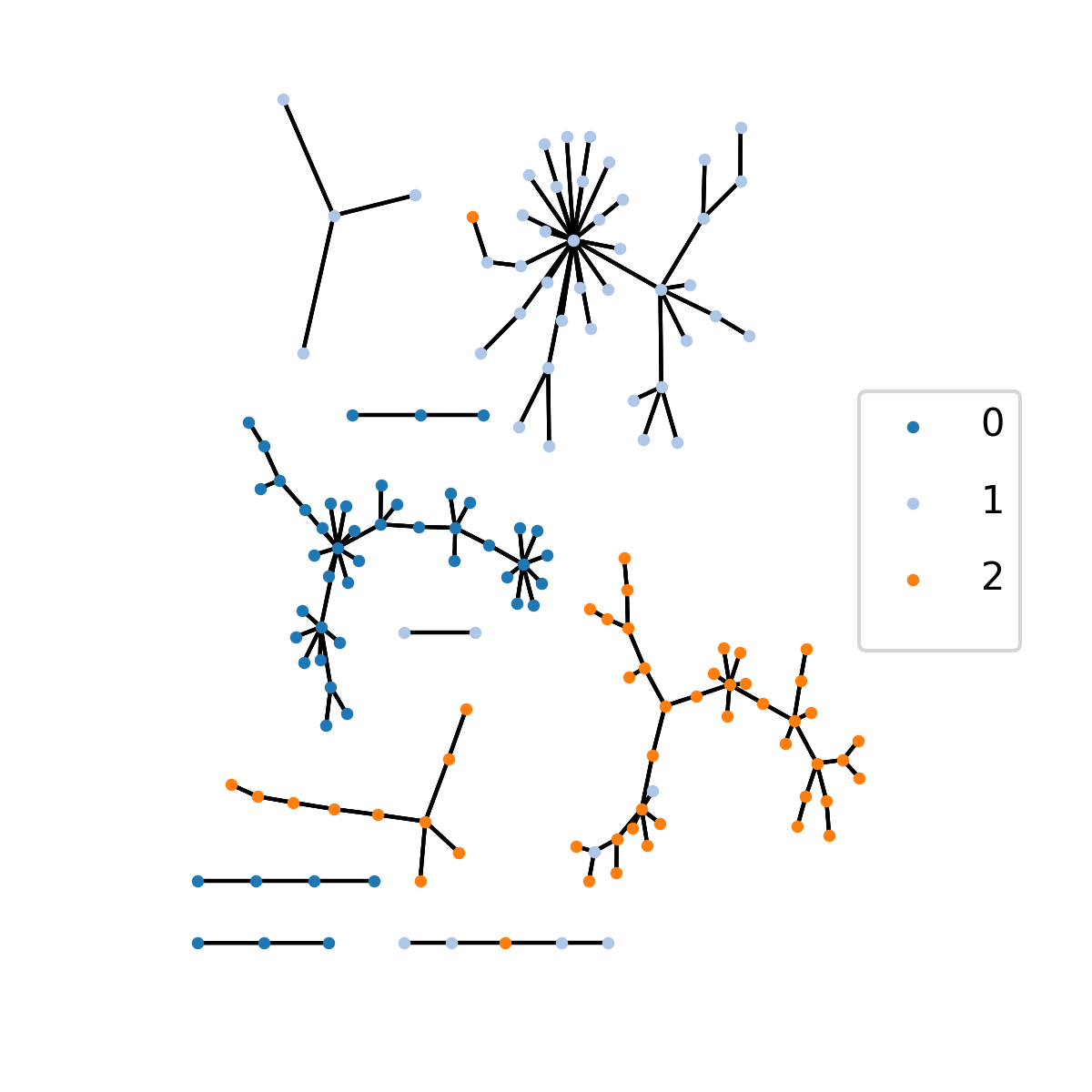}
		\caption{SPC}
		\label{fig:iris_sa}
	\end{subfigure}
	~
	\begin{subfigure}[b]{0.45\textwidth}
		\includegraphics[width=\textwidth]{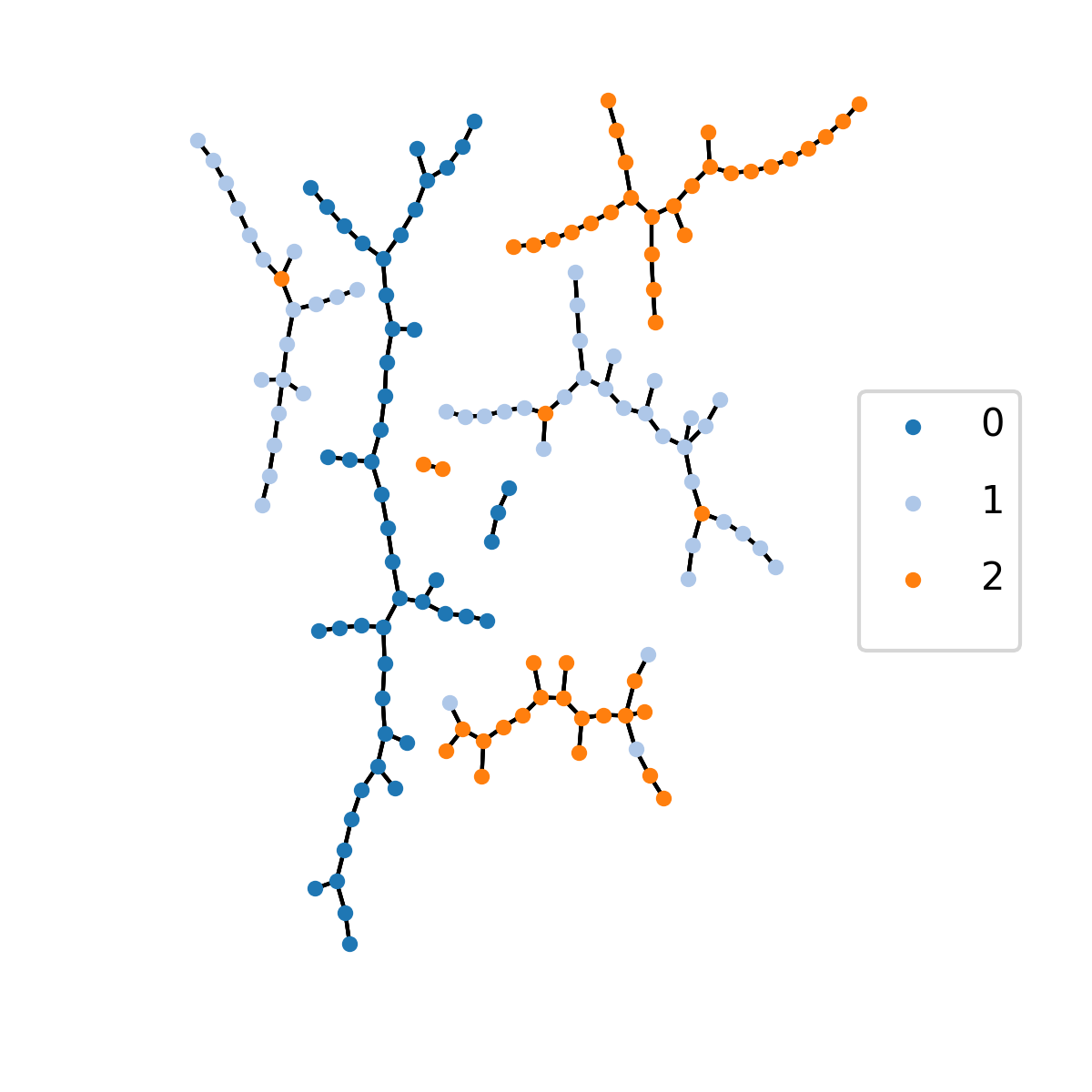}
		\caption{f-SPC}
		\label{fig:iris_mle}
	\end{subfigure}
	\caption{in a) Iris 3 species (Sec. \ref{ssec:iris}) : MST of SPC's solution at $T=0.137$. 3 large clusters representing the original Iris species (Cluster 0 for ``Setosa'', 1 for ``Versicolor'', and 2 for ``Virginica''), and 6 smaller ones. in b ) f-SPC's solution after 10000 generations. 5 large clusters: 1 for Setosa, 2 for Versicolor, and 2 for Virginica.}\label{fig:iris_sol}
\end{figure*}

As we can see in Fig. \eqref{fig:iris_plot}, It's one of the more challenging toy problems because two of the three clusters, Virginica, and Versicolor, are not linearly separable. We set $K=7$, and observe two phases in Fig. \eqref{fig:iris_clus}: for $0.05 < T < 0.137$ there are 2 clusters. The largest contains the Virginica, and Versicolor nodes while the smaller one includes most Seratosa nodes. The $2^{nd}$ phase transition occurs at right before $T=0.137$, and is followed by the separation of most Virginica nodes into their own cluster. This SPC solution Fig. \eqref{fig:iris_sa} has an ARI of 0.65.

The $L_c$ solution in Fig. \eqref{fig:iris_mle} has an ARI of 0.627, a likelihood of 132, and our expected $L_c$ is 104. In comparison K-Means and DBSCAN respectively achieve ARI of 0.73, and 0.52. Similarly to the precedent examples, we recover five large clusters: Cluster 1 contains Seratosa individuals, while the Virginica, and Versicolor clusters are split into 4 smaller ones with minimal misclassification.

\subsection{{\it Sci-Kit learn}: MNIST digits} \label{ssec:mnist}

The hand-written digits dataset, generated with \texttt{Sci-kit learn} loader \footnote{D. Cournapeau, F. Pedregosa, O. Grisel 'load\_digits', 2007-2010. [Online]. Available: \url{http://scikit-learn.org/stable/modules/generated/sklearn.datasets.load_digits.html}. [Accessed: 12-Jun-2018]}, is mainly used to test classification algorithms in supervised learning but we are interested in how well both SPC, and f-SPC deal with the nonlinear nature of hand-writing. The data contains 10 classes of digits ranging from 0 to 9. The full set has close to 2000 nodes from which we select 500, and 50 of each class. The images are 8 by 8, and $D = 64$.

\begin{figure*}
	\centering
	\begin{subfigure}[b]{0.31\textwidth}
		\includegraphics[width=\textwidth]{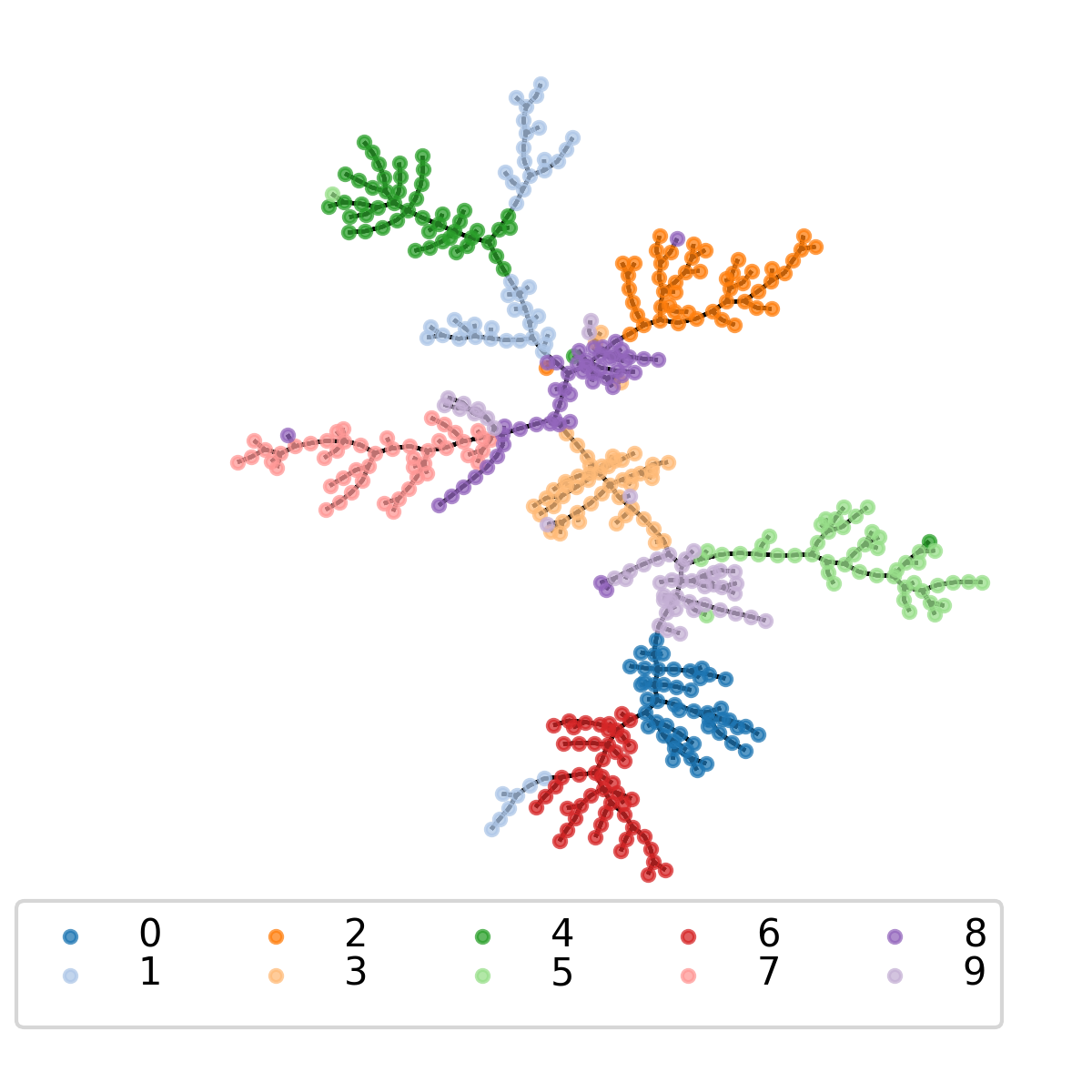}
		\caption{MST}
		\label{fig:dig_mst}
	\end{subfigure}
	~ 
	\begin{subfigure}[b]{0.31\textwidth}
		\includegraphics[width=\textwidth]{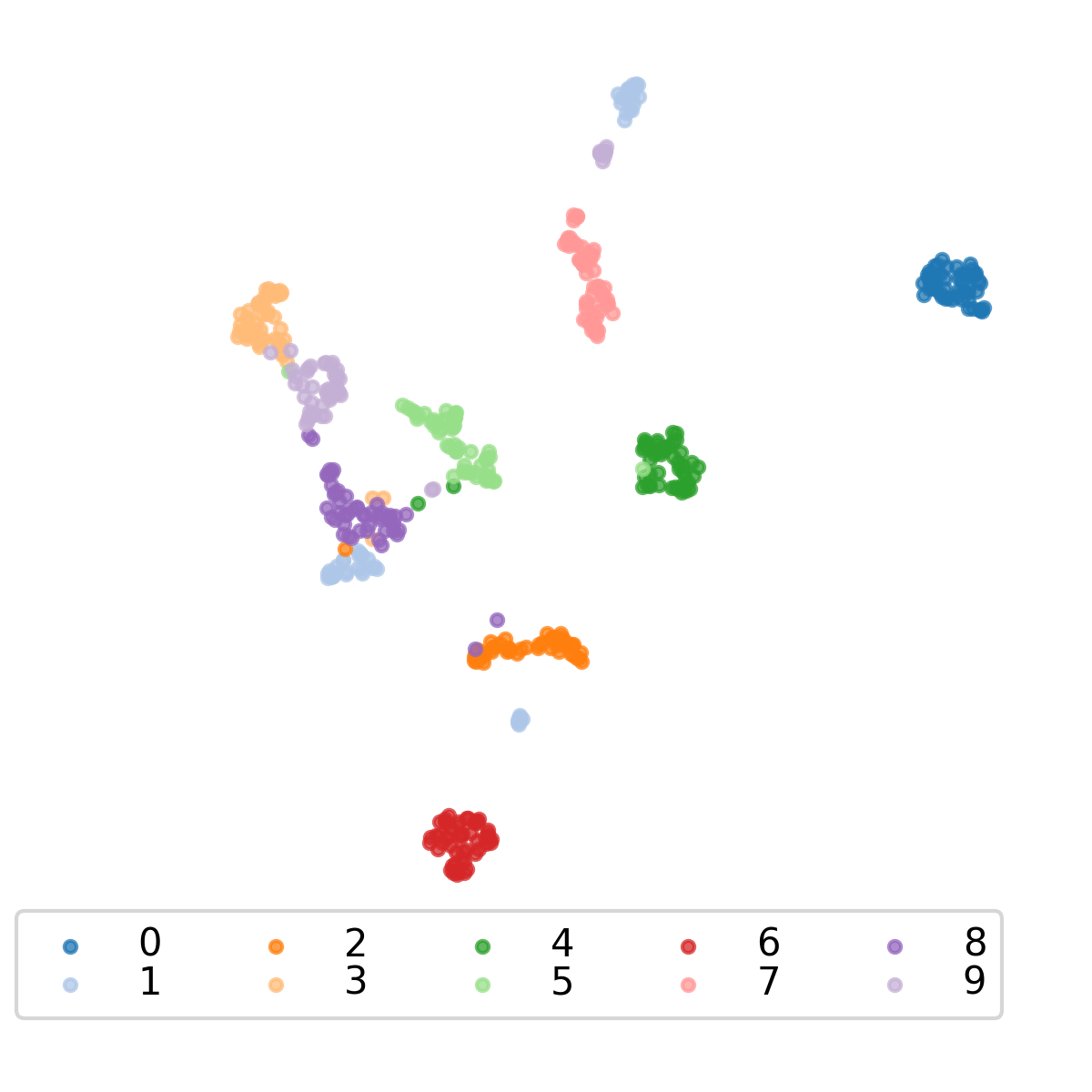}
		\caption{Scatter Plot}
		\label{fig:dig_plot}
	\end{subfigure}
	~
	\begin{subfigure}[b]{0.31\textwidth}
		\includegraphics[width=\textwidth]{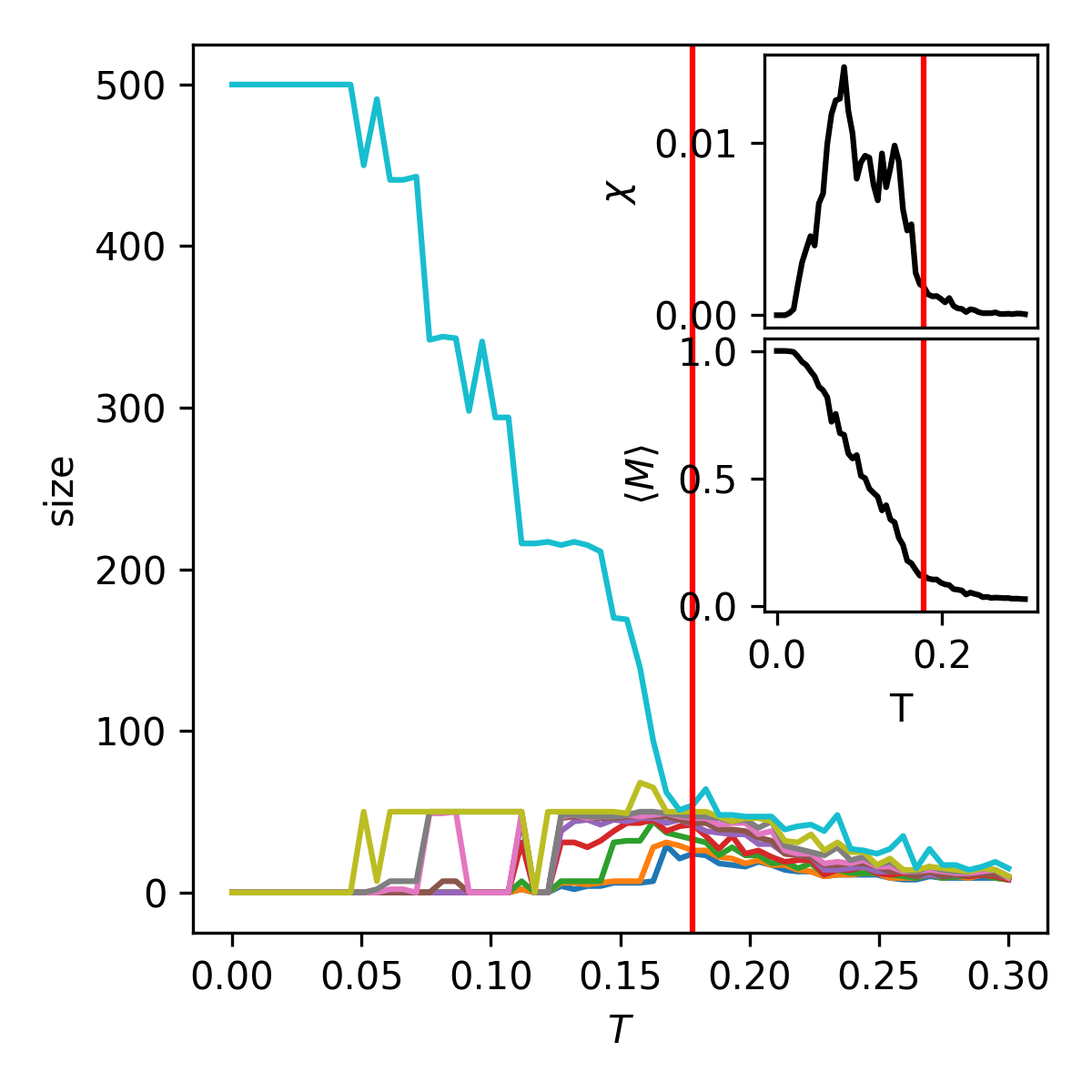}
		\caption{Cluster sizes vs Temperature $T$}
		\label{fig:dig_clus}
	\end{subfigure}
	\caption{in a) MNIST hand-written digits (Sec. \ref{ssec:mnist}) : 2D plot of dimensionality reduction of the 64 features using UMAP \cite{mcinnes2018umap}. $N=500$. 10 classes from 0 to 9. in b) the MST: Overall numbers of the same digit class are close. in c) Cluster sizes vs Temperature $T$ using SPC (Sec. \ref{sssec:maxent}) at $T \approx 0.18$. Insets: (in a) above right)Susceptibility $\chi(T)$, and (in a) below right) Average Magnetization $\langle M \rangle$ at $T \approx 0.18$.}\label{fig:dig}
\end{figure*}

The MST in Fig. \eqref{fig:dig_mst}, and the UMAP \cite{mcinnes2018umap} plot in Fig. \eqref{fig:dig_plot} show us all digit classes are linearly separable, the data contains outliers especially ``1'''s and ``9'''s which may be the result of different writing styles.

\begin{figure}
	\centering
	\begin{subfigure}[b]{0.45\textwidth}
		\includegraphics[width=\textwidth]{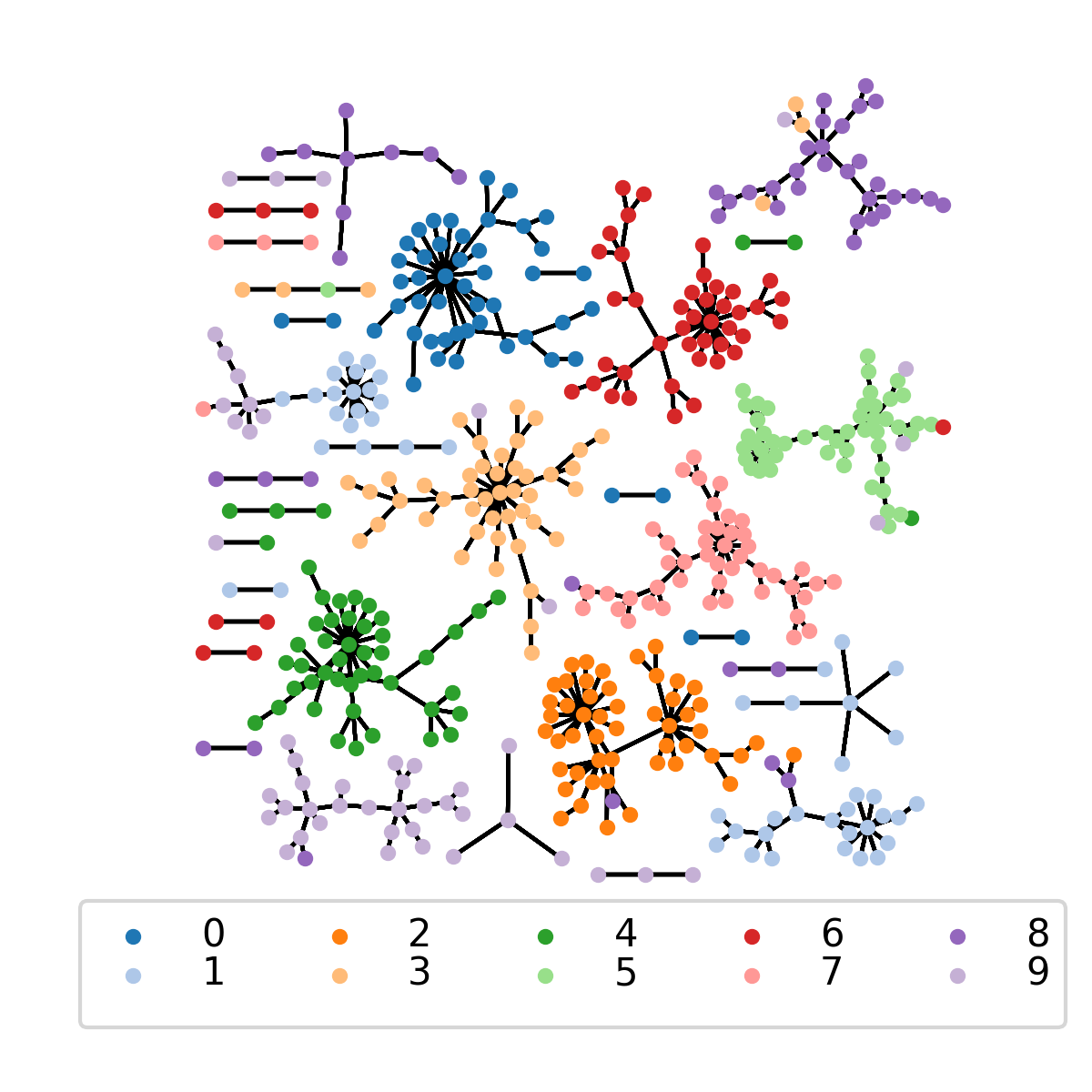}
		\caption{SPC}
		\label{fig:dig_sa}
	\end{subfigure}
	
	\begin{subfigure}[b]{0.45\textwidth}
		\includegraphics[width=\textwidth]{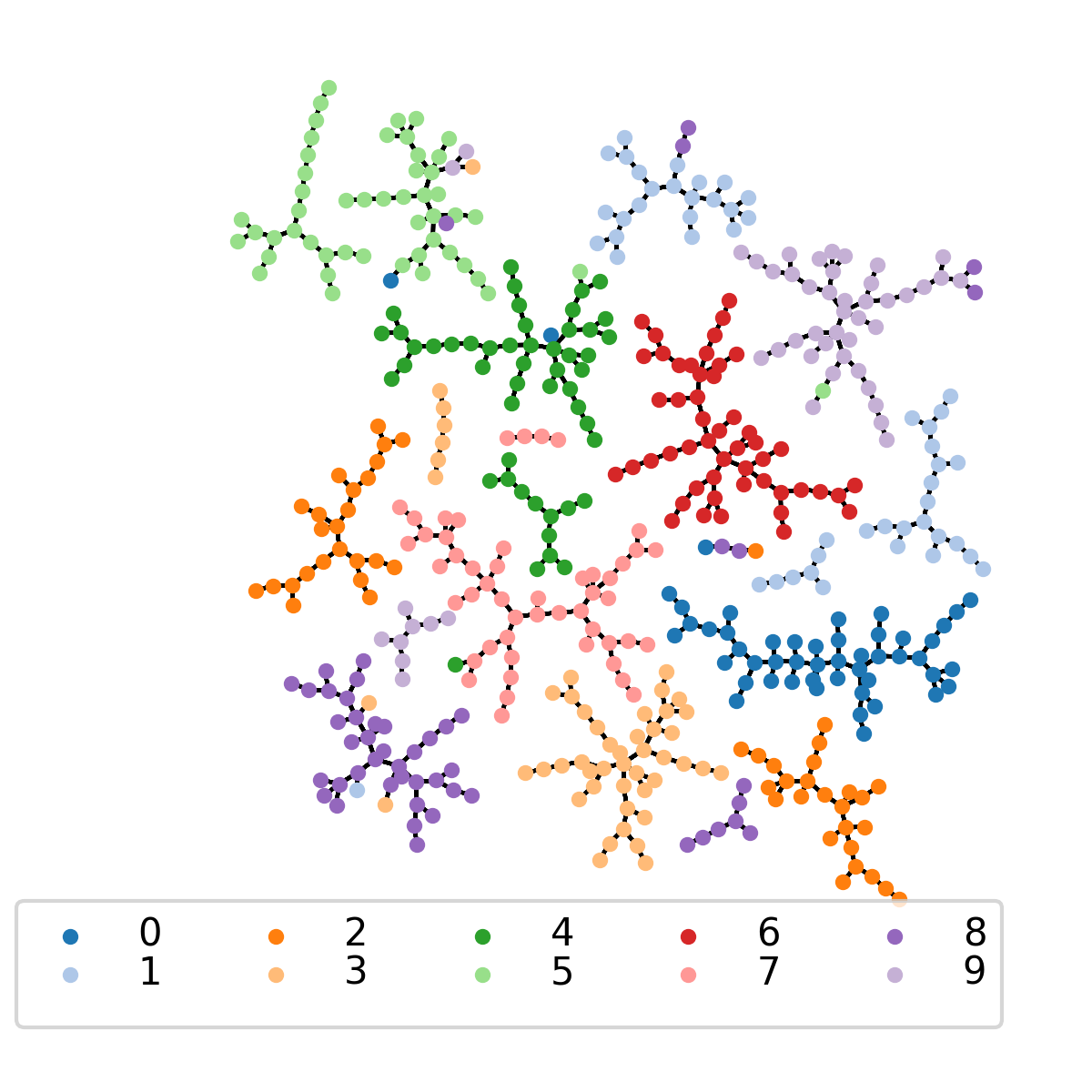}
		\caption{f-SPC}
		\label{fig:dig_mle}
	\end{subfigure}
	\caption{in a) MNIST hand-written digits (Sec. \ref{ssec:mnist}): MST of SPC's solution at $T \approx 0.18$. 10 classes recovered: Cluster 8 is split in two, and some observations from cluster 9 \& 1 are found in one mixed cluster. There are non-linearities in how digits are drawn which may explain the closeness of 9s and 1s. in b) f-SPC's solution: 10 classes recovered: 1 cluster for 0s and 6s, 3 for 1s, 2 for 2s, 3s, 4s, 5s, 7s, 8s and 9s.}\label{fig:dig_sol}
\end{figure}

We see that the big cluster breaks down in multiple steps Fig. \eqref{fig:dig_clus} due to the differing densities of clusters present in the data. Fig. \eqref{fig:dig_clus} shows that at $T = 0.178$ right after the final $\chi$ peak the configuration's clusters in Fig. \eqref{fig:dig_sa} has an ARI $= 0.75$, and show no significant misclassification.

$L_c$'s solution in Fig. \eqref{fig:dig_mle}, after 25k generations, has a likelihood of 149.47, an ARI of 0.747, and an expected $L_c$ of 135. Once again we encounter similar results as with the previous cases with the higher likelihood, and the number of clusters. The $L_c$ solution has close to 20 clusters, and while there is one main cluster per digit which is the case for digits 0 and 6, and mostly for 3, 7, 8, and 9, the digits 1, 2, 4, and 5 are all split in two clusters. We explain this by the inconsistent nature of hand-writing which produces different writing styles.
In comparison K-Means and DBSCAN respectively achieve ARI of 0.56, and 0. There are many reasons why DBSCAN fails this problem: DBSCAN classifies some observations as noise into one cluster, it also has issues tackling problems with clusters of different densities.

\subsection{ {\it Kaggle}: NYSE Data } \label{ssec:financial}

We obtained publicly available NYSE stock market data on Kaggle\footnote{C. Nugent 'S\&P 500 stock data - Historical stock data for all current S\&P 500 companies', 2017-2018. [Online]. Available: \url{https://www.kaggle.com/camnugent/sandp500}. [Accessed: 01-Dec-2017]}. The original data contains daily open, high, low, closing prices, and volume from 8/13/2012 to 8/11/2017. Because not all stocks traded for the whole duration we only select the stocks which did for the last 1250 days ($\approx 5$ years) which left us with 447 stocks, and furthermore we, in this case, were interested in a time horizon of 5 years in trading days from 8/23/2012 to 8/11/2017. We consider the daily trading closing prices which are use to compute the daily returns such that :

\begin{equation} \label{eq:15} r(t) = \ln\left(P_{t+1}\right) - \ln\left(P_{t}\right) \end{equation}

The final data set has a time-series Length $D = 1249$. Using Eqn. \eqref{eq:15} we consider three cases for the correlation matrix: i.) the full correlations, ii.) denoising using IMN (See Sec. \ref{sssec:succnorm}), and iii.) cleaning the matrix using a RMT method (See Sec. \ref{sssec:rmt}).

\begin{figure*}
	\centering
	\begin{subfigure}[b]{0.31\textwidth}
		\includegraphics[width=\textwidth]{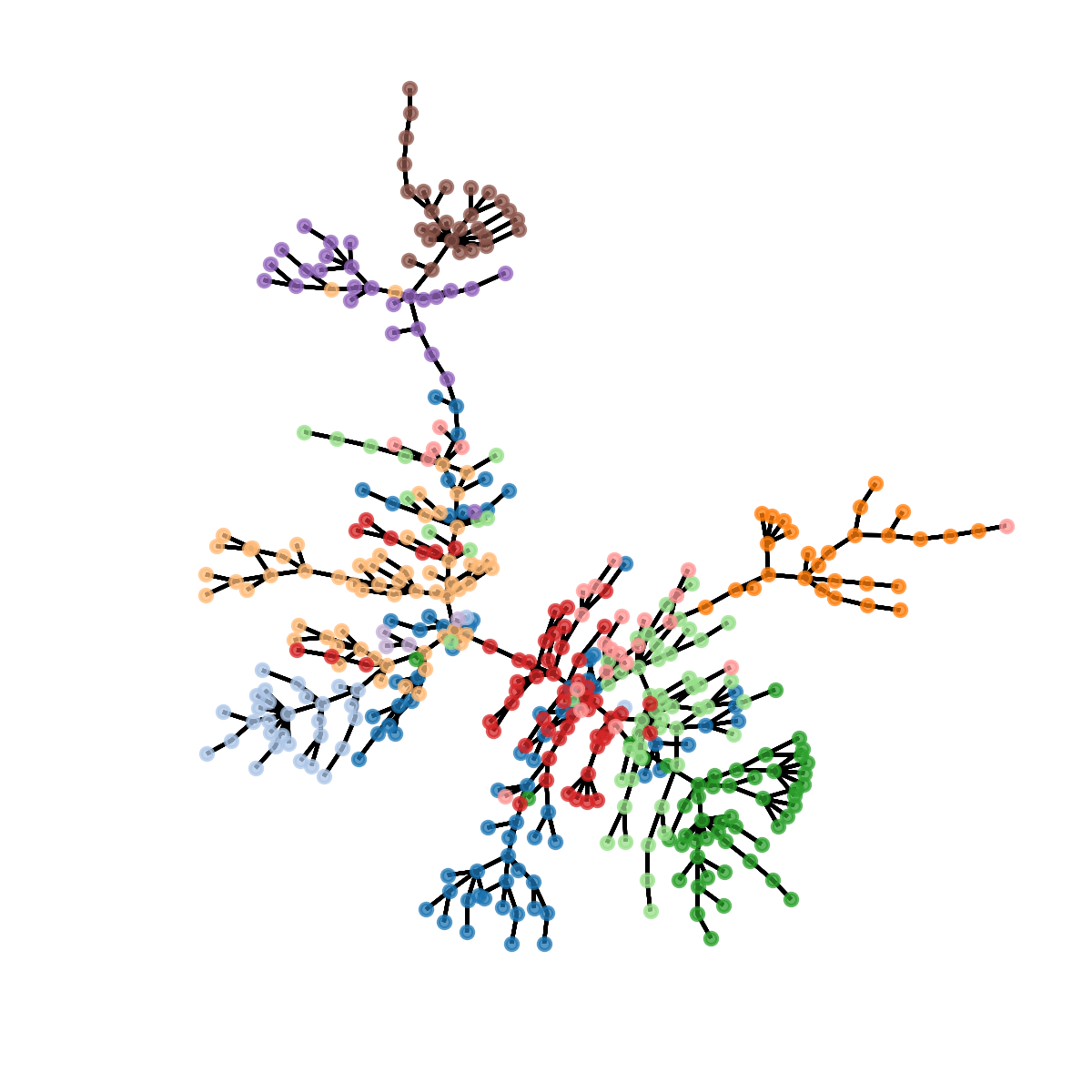}
		\caption{Market Mode}
		\label{fig:full_mst}
	\end{subfigure}
	~ 
	\begin{subfigure}[b]{0.31\textwidth}
		\includegraphics[width=\textwidth]{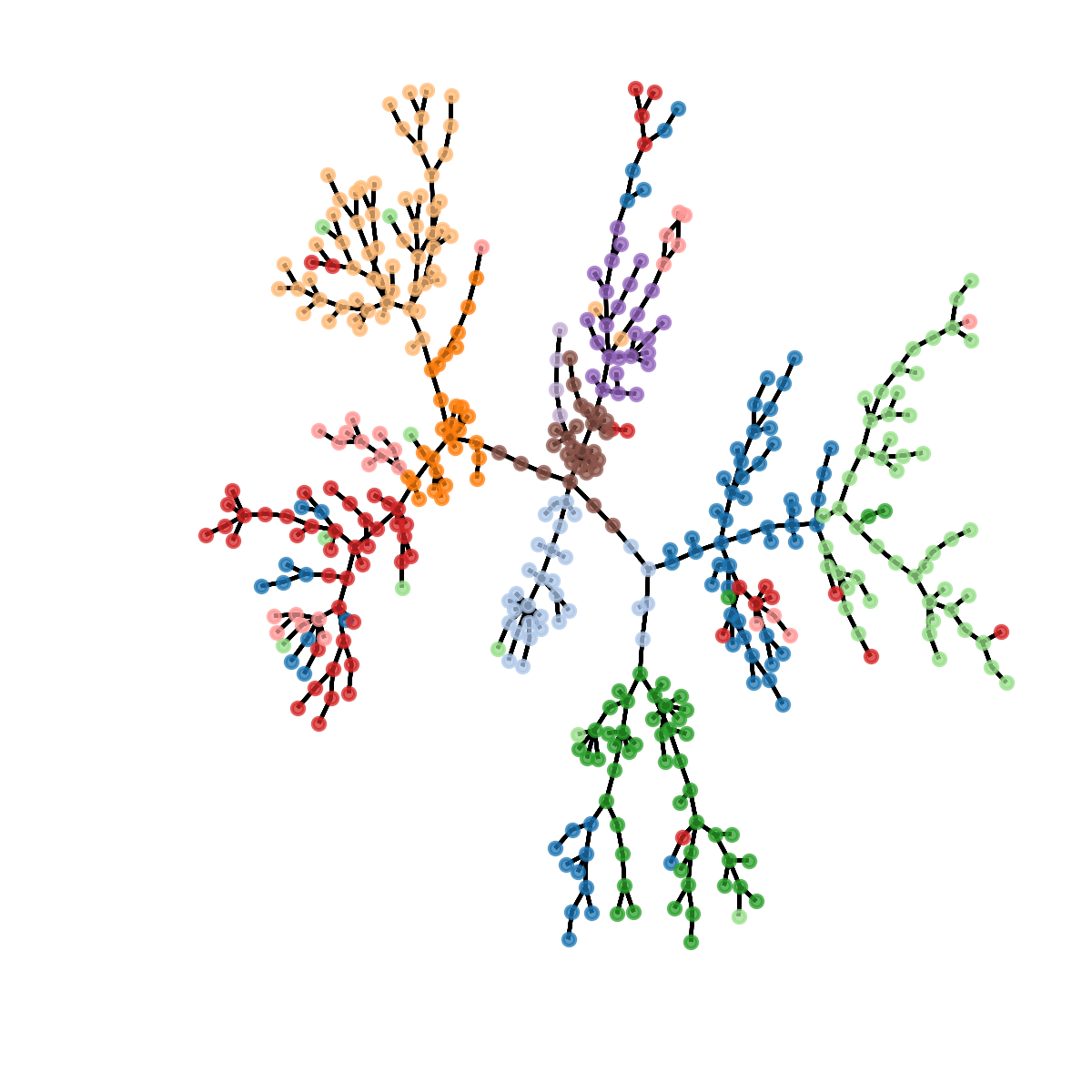}
		\caption{IMN}
		\label{fig:norm_mst}
	\end{subfigure}
	~
	\begin{subfigure}[b]{0.31\textwidth}
		\includegraphics[width=\textwidth]{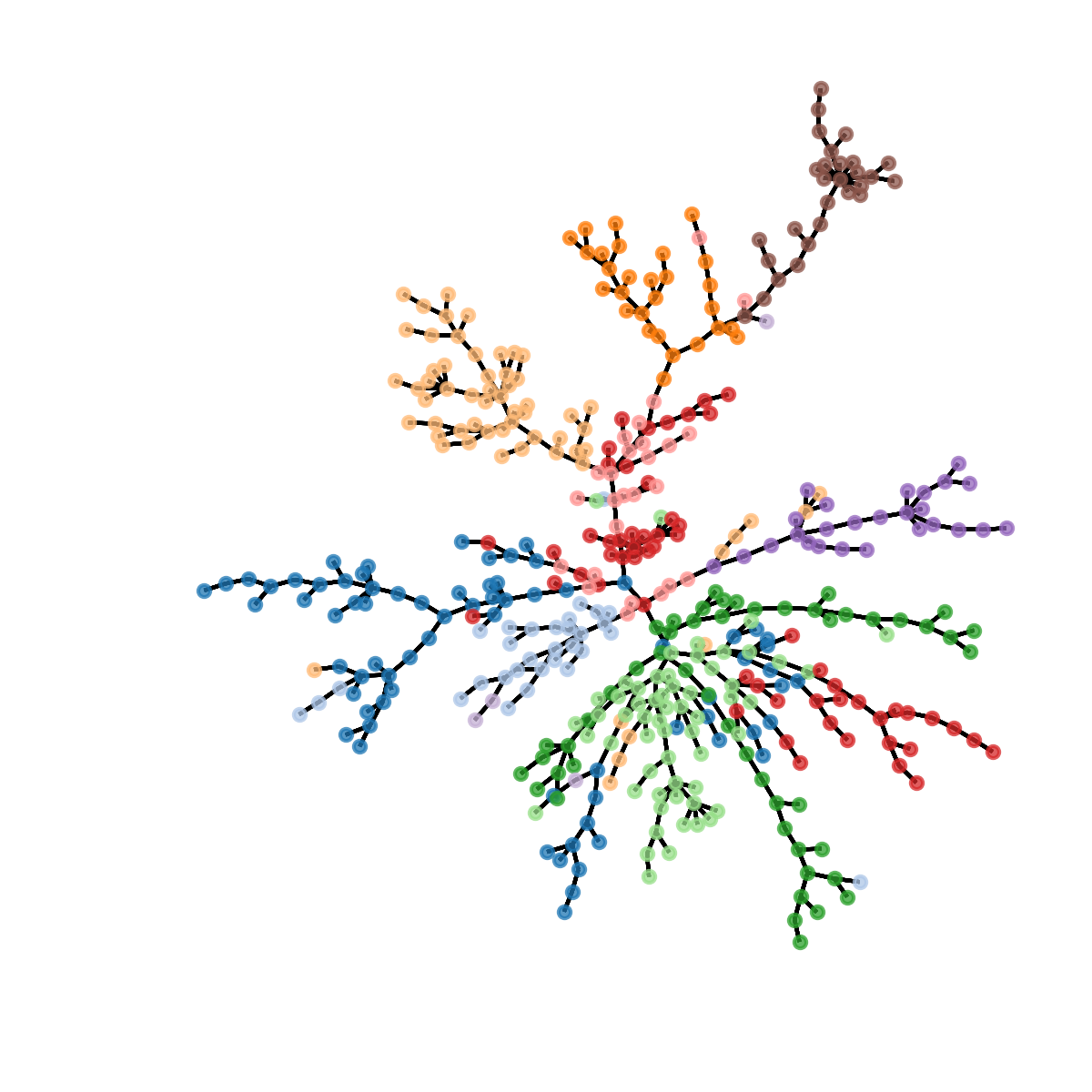}
		\caption{RMT}
		\label{fig:rmt_mst}
	\end{subfigure}
	\caption{in a) S\&P500: Market Mode Correlation-based MST of 447 stocks over 1249 trading days (Sec. \ref{ssec:financial}). in b) The Market Mode was removed using IMN (Sec. \ref{sssec:succnorm}), and in c) using RMT (Sec. \ref{sssec:rmt}). Colors refer to GICS sectors (See footnote \ref{fn:color} ) }\label{fig:sp500_mst}
\end{figure*}	

Financial markets are perpetually evolving living ecosystems, and this is illustrated in the lack of available true sectoral classification of publicly traded companies. In the process of clustering stock market data \footnote{A very nice review of clustering methods applied to financial datasets is available at \cite{marti2017review} }, we wanted to compare the results of our algorithms with industry standard classification but we faced the following difficulties:

We consider the following industry classifications \footnote{ There is no consensus industry classification used in the financial services industry.}: The New York Stock Exchange (NYSE) uses the Global Industry Classification Standard (GICS) \footnote{ The GICS counts 11 sectors, 24 industry groups, 68 industries, 157 sub-industries, and is updated annually. For more details on their hierarchical industry classification system \url{https://www.msci.com/documents/10199/4547797/GICS+Structure+effective+Sep+1\%2C+2016.xls}  } (which we use here). The National Association of Securities Dealers Automated Quotations (NASDAQ) and the London Stock Exchange (LSE) both use the Industry Classification Benchmark (ICB)\footnote{ The ICB counts 10 industries, 19 super-sectors, 41 sectors, 114 sub-sectors, and is For details on their hierarchical industry classification system \url{https://www.ftse.com/products/downloads/ICB_Rules.pdf}}.Industry classifications have sectoral, industrial and sub-industrial levels. Although commonalities exists one is left to determine the equivalences when information aggregation is required across different markets.

GICS, and ICB are static classifications which are updated at irregular intervals (i.e. GICS every year, ICB from weekly to yearly updates). The focus of these companies is to provide long term structural trends of financial markets. As such they lose their usefulness if one wants to consider the impact of rare events such as financial crashes which significantly alter the behavior of businesses. They also do not consider how the diversification of investments and activities affect their respective classifications. The case of Amazon can be argued to illustrate the idea behind the Adaptive Market Hypothesis (AMH) \cite{lo2004adaptive}: Amazon's GICS' sector is Consumer Discretionary. GICS uses this sector to classify companies whose activity they deem ``most sensitive to economic cycles'' \footnote{ A description of GICS sector is available at \url{https://www.msci.com/documents/10199/4547797/GICS+Sector+definitions-Sep+2016.pdf} }. It is unclear what is meant by ``sensitive'' in this instance as there are many possible interpretations, and this sector is very heterogeneous. Perhaps it highlights the adaptive nature of Amazon's business interests which started first as an order-to-delivery e-commerce bookstore but based on Fig. \eqref{fig:full_mst} is now closest to the Information Technology sector.

The life cycle of publicly traded companies can be short. Firms go public and private at relatively high frequency when compared to biological evolution on a human timescale as motivated by Farmer in \cite{farmer2009virtues}. The inclusion or exclusion of individuals in an ecosystem can and should have an impact on its structure based on how important the individuals are to the groups. When we looked for GICS data for our time-series, a number of companies had gone private since Aug 2017, and GICS classification had been updated without reflecting these new changes for these companies. Gathering data on these companies which translated into the newer nomenclatures was thus rendered more difficult. Yet again illustrating the need for expert-free unsupervised methods.

Finally, while as previously stated, GICS and ICB intend on providing data which capture long term trends. Financial markets are populated with participants (i.e. pension funds, high frequency trader, asset managers etc...) each holding a diverse set of objectives, who do not necessarily operate on the same time scales or have the same investment horizons. If one goal is to provide comprehensive analyses of the multiple existing dynamics in markets, tools which capture these trends, and methods which subsequently find relations between them should be prioritized.

This motivates us to argue that the highly dynamic nature of financial markets renders the use of static classifications problematic to a certain extent.

We use GICS's 11 sectors as the ``true'' economic sectors of the US financial market. These include Consumer Discretionary (74 stocks), Consumer Staples (31 stocks), Energy (28 stocks), Financials (62 stocks), Health Care (51 stocks), Information Technology (IT) (59 stocks), Industrials (58 stocks), Materials (26 stocks), Real Estate (26 stocks), Telecoms (4 stocks), and Utilities (28 stocks)\footnote{ \label{fn:color} Colors used for the 11 GICS economic sectors: Consumer Discretionary (royal blue), Consumer Staples (sky blue), Energy (orange) , Financials (beige), Health Care (dark green), Information Technology (light green), Industrials (red), Materials (pink), Real Estate (purple), Telecom (magenta), Utilities (brown).}. Although as previously mentioned we do not believe this classification to be valid, here we make use of it as benchmark. 

Looking at MSTs in figures \eqref{fig:full_mst}, \eqref{fig:norm_mst}, and \eqref{fig:rmt_mst}, and aided by the GICS classification as legend, we notice nodes belonging to the same economic sectors are mostly located in proximity of each other as one would expect in a static world or over time-scales where the static model is a reasonable approximation.

\begin{figure*}
	\centering
	\begin{subfigure}[b]{0.35\textwidth}
		\includegraphics[width=\textwidth]{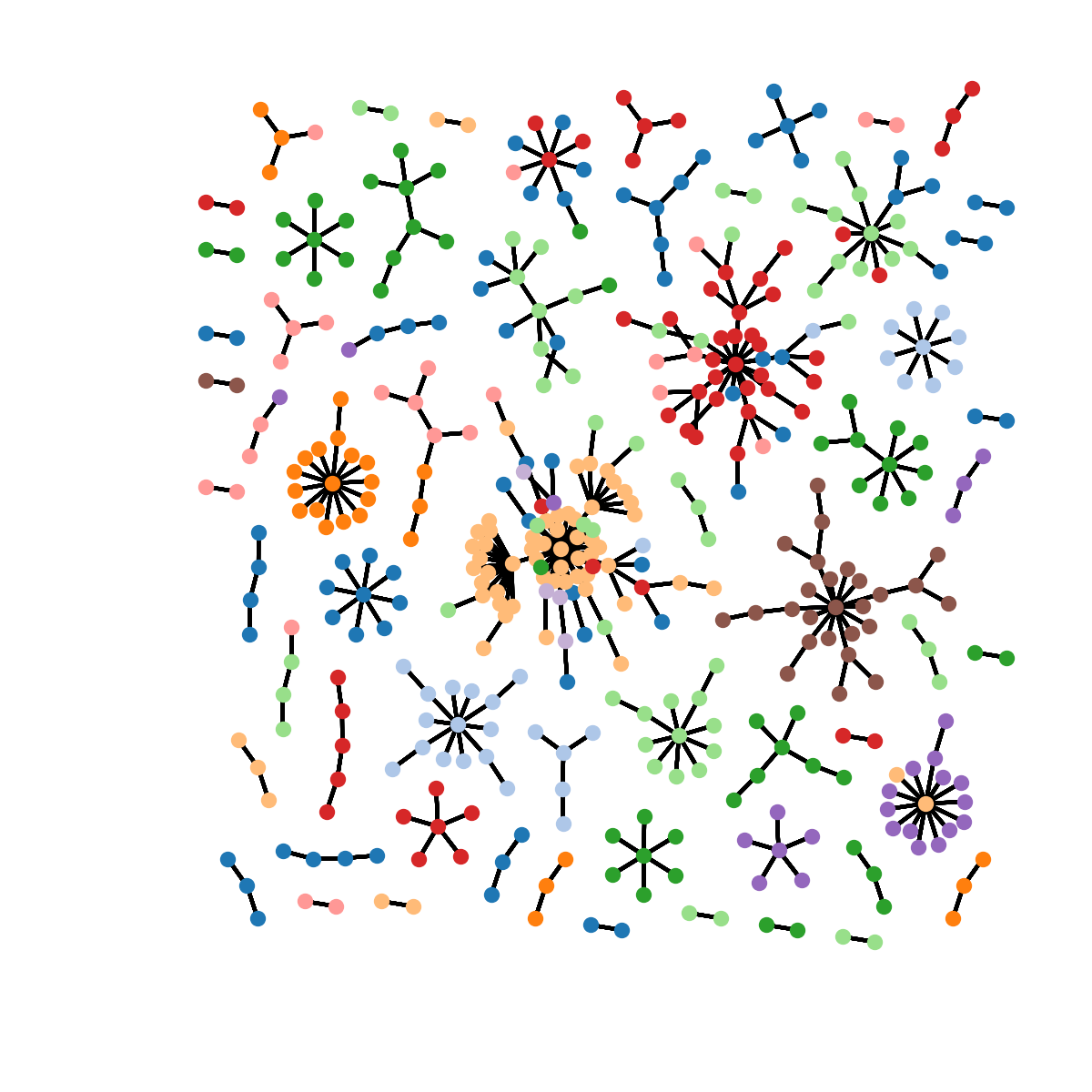}
		\caption{Market Mode - SPC}
		\label{fig:full_sa}
	\end{subfigure}
	~ 
	\begin{subfigure}[b]{0.35\textwidth}
		\includegraphics[width=\textwidth]{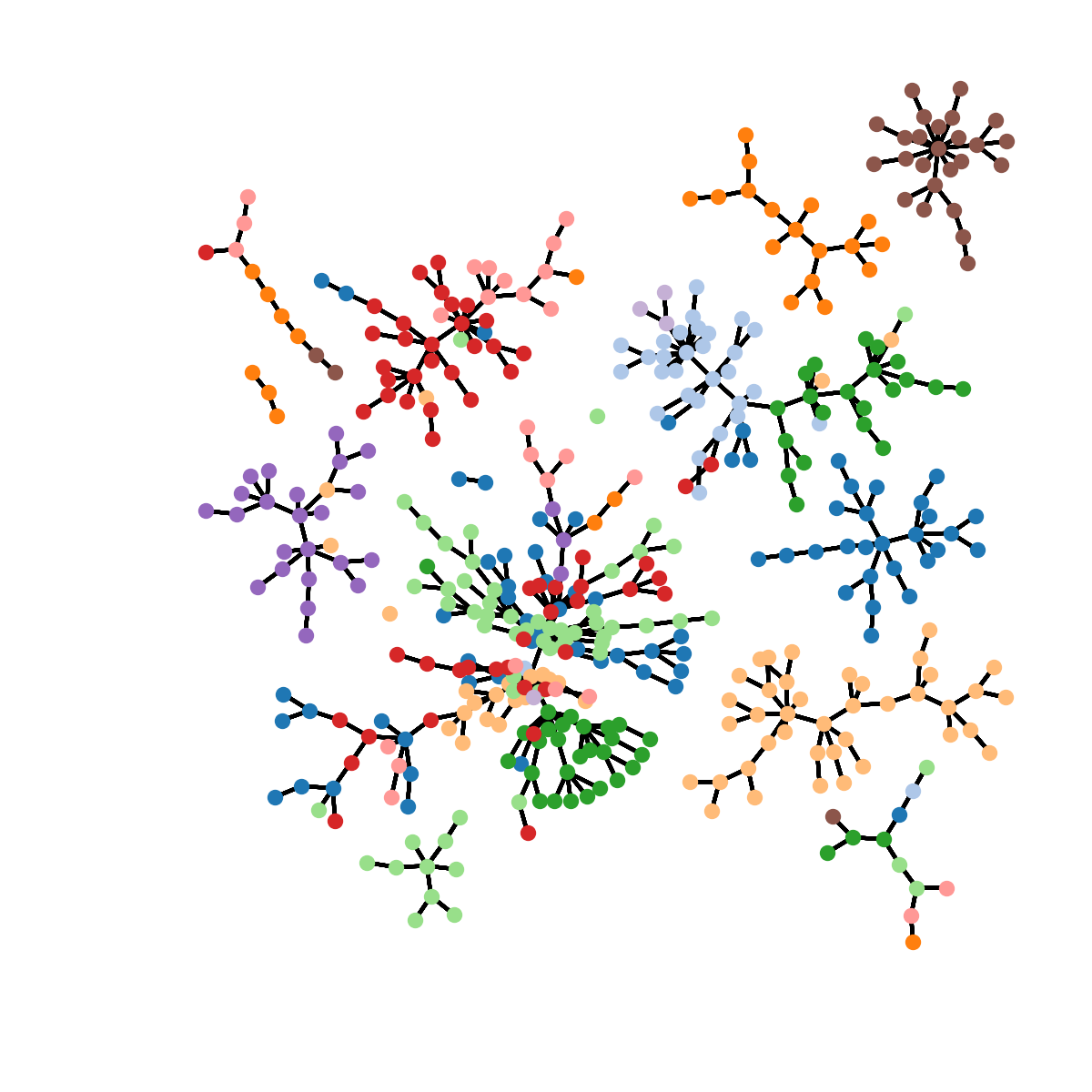}
		\caption{Market Mode - f-SPC}
		\label{fig:full_mle}
	\end{subfigure}
	
	\begin{subfigure}[b]{0.35\textwidth}
		\includegraphics[width=\textwidth]{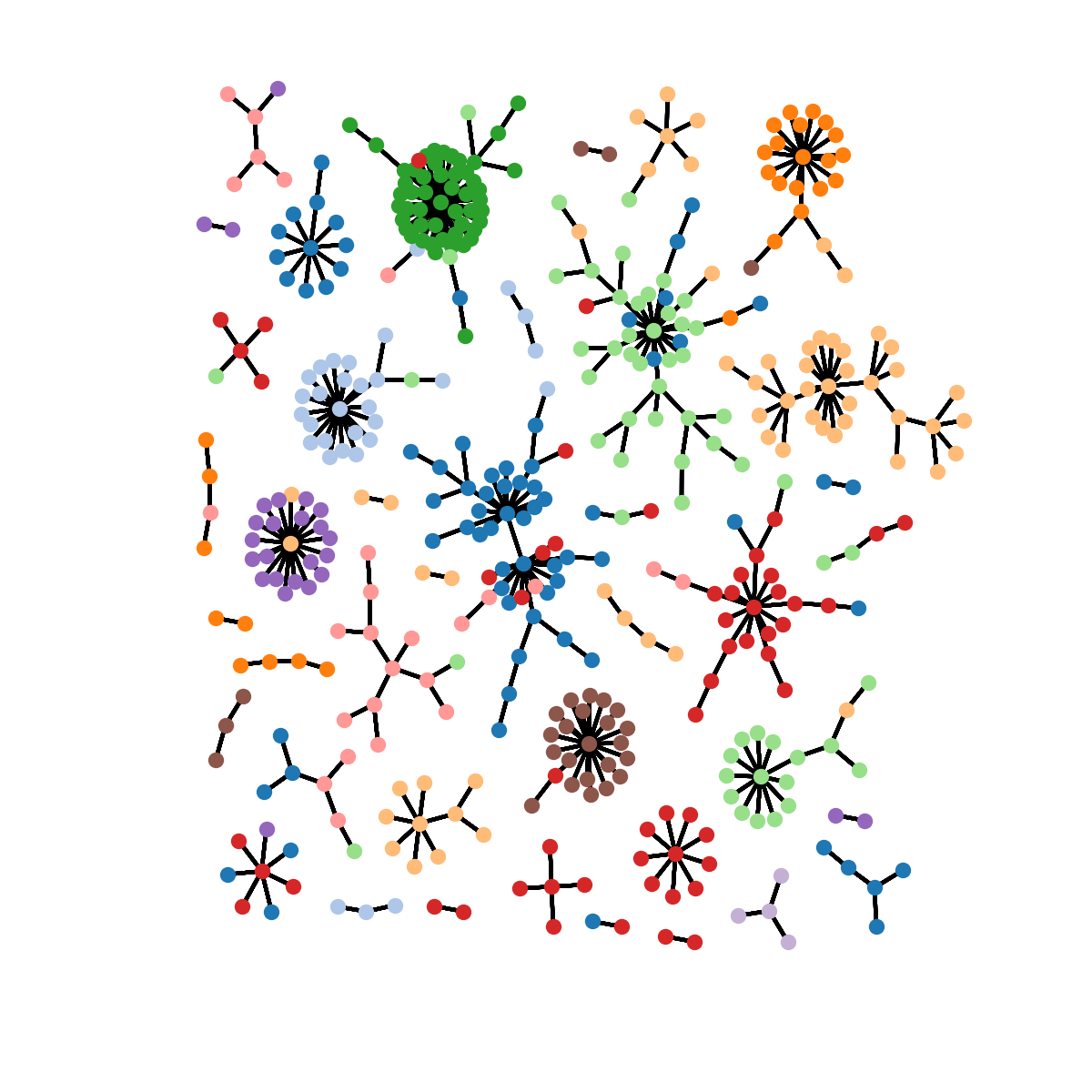}
		\caption{IMN - SPC}
		\label{fig:norm_sa}
	\end{subfigure}
	~
	\begin{subfigure}[b]{0.35\textwidth}
		\includegraphics[width=\textwidth]{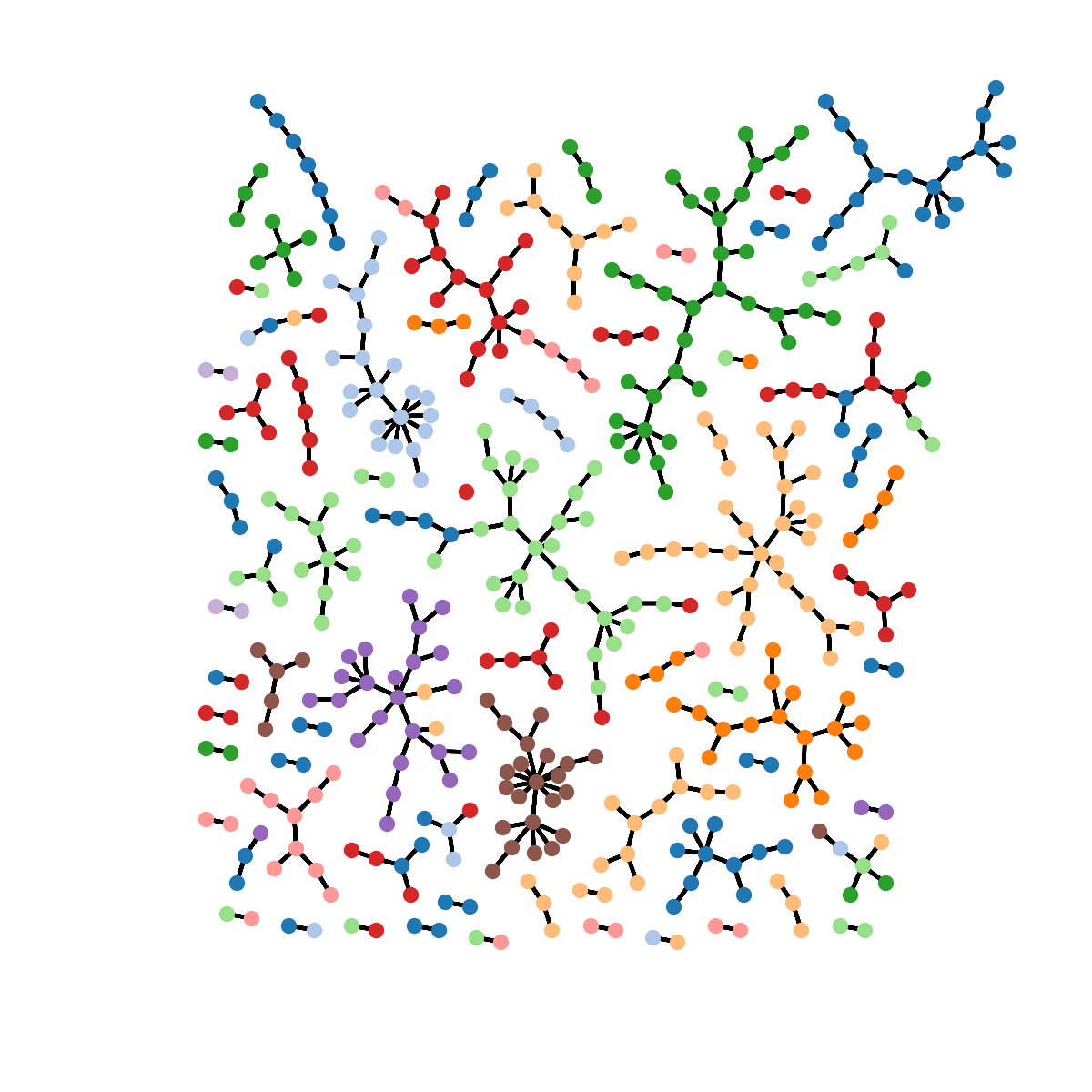}
		\caption{IMN - f-SPC}
		\label{fig:norm_mle}
	\end{subfigure}
	
	\begin{subfigure}[b]{0.35\textwidth}
		\includegraphics[width=\textwidth]{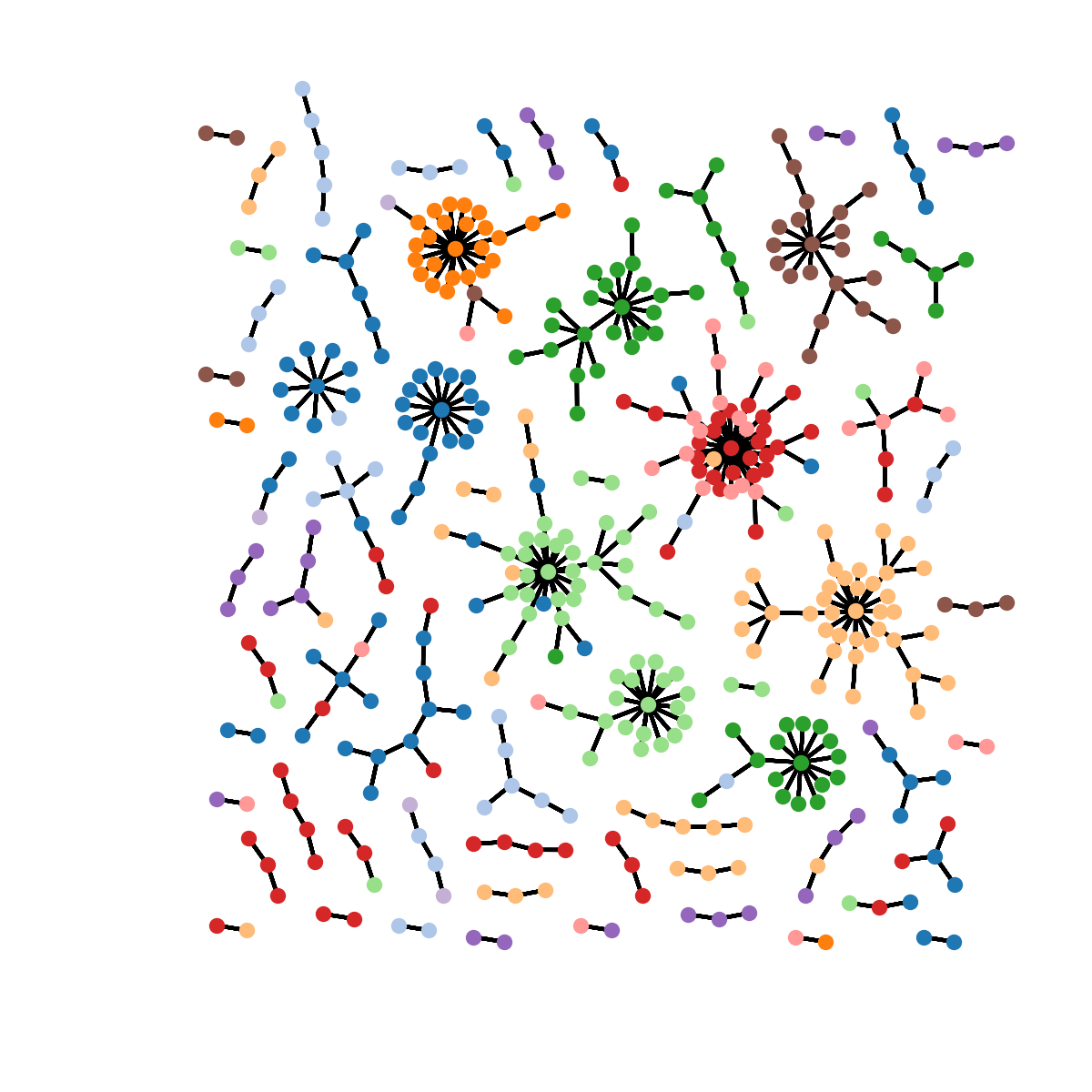}
		\caption{RMT - SPC}
		\label{fig:rmt_sa}
	\end{subfigure}
	~ 
	\begin{subfigure}[b]{0.35\textwidth}
		\includegraphics[width=\textwidth]{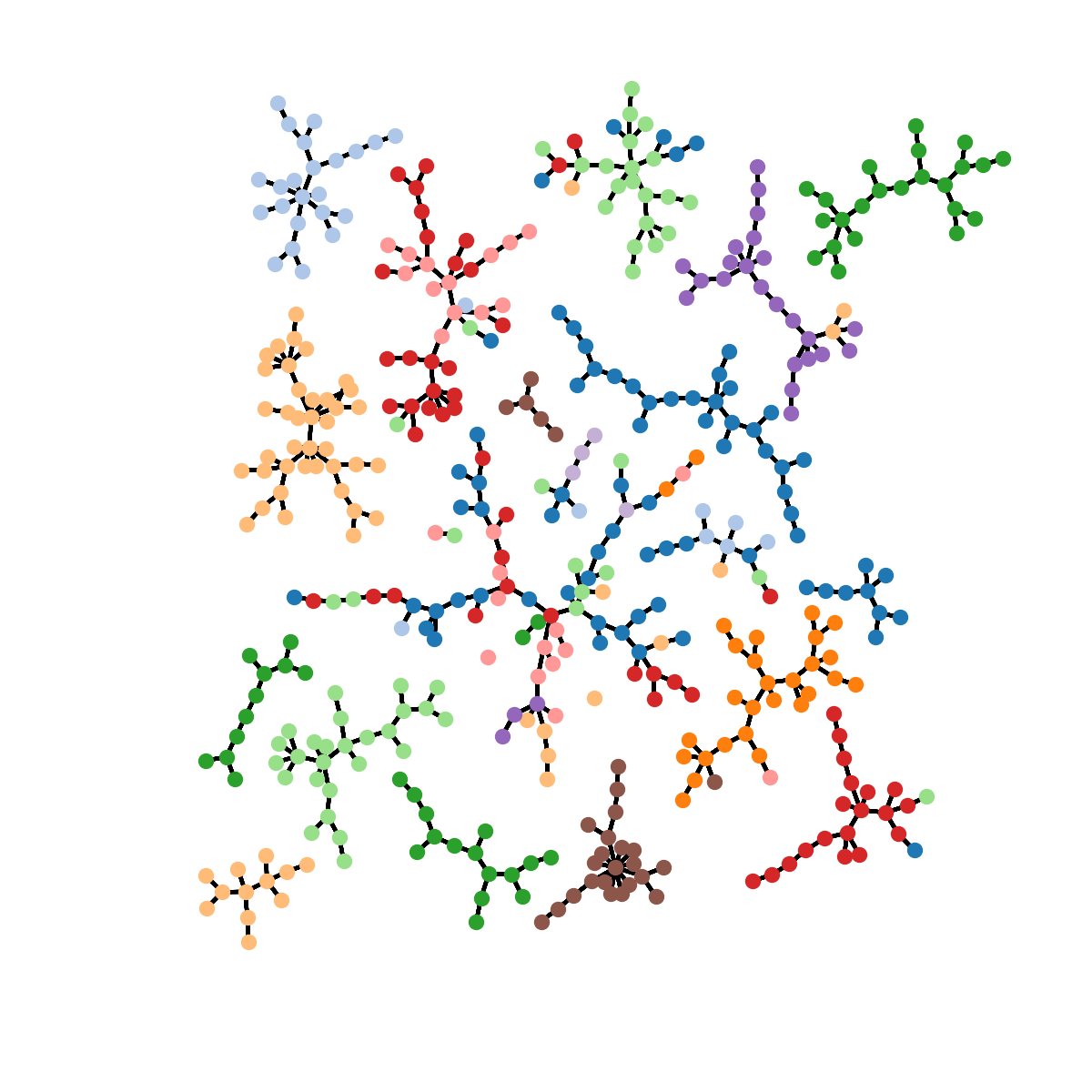}
		\caption{RMT - f-SPC}
		\label{fig:rmt_mle}
	\end{subfigure}
	
	\caption{S\&P500: $N=447$ stocks traded over 1249 days (Sec. \ref{ssec:financial}). in a), c), and e) SPC's solution at $T=0.081$, $T=0.071$, and $T=0.119$ respectively for the Full ``Market Mode'' sample Correlation Matrix, the iteratively normalized (Sec. \ref{sssec:succnorm}) , and Noise cleaned RMT (Sec. \ref{sssec:rmt}) cases. And in b), d), and f) the f-SPC's solutions after 25k generations. Colors refer to GICS sectors (See footnote \ref{fn:color} ).}\label{fig:sp500_sol}
\end{figure*}

We report SPC results in figures \eqref{fig:full_sa}, \eqref{fig:norm_sa}, and \eqref{fig:rmt_sa} respectively at $T = 0.081$,$T = 0.071$, and $T = 0.129$ for the full (K=5), normalized, and RMT cases.

We briefly mention again one of SPC's features which consists in linking a node to its closest neighbor based on the spin-spin correlations. Using the condition $\theta > 0.5$ we construct a graph but in the case where a node has no correlations meeting our condition, it is linked to its neighbor of highest spin-spin correlation. This feature forces SPC to produce graphs without isolated nodes. At the same time, and because of this fact, we consider small size clusters are equivalent to noisy, insufficiently correlated, or unclassified observations.

SPC solutions recover GICS information as seen by their respective ARI: 0.317, 0.479, and 0.33. The solution with highest number of noisy or unclustered stocks is Fig. \eqref{fig:full_sa}, The financial sector is merged with many other stocks from other sectors, whereas most industries are found in one or two clusters. The complexity goes down when we move to Fig. \eqref{fig:norm_sa} where every sector have mostly separated into their own unique cluster, and Fig. \eqref{fig:rmt_sa} which gives a similar picture although with more smaller unclassified clusters present.

$L_c$ results were simulated for 25k generations, and we obtained $L_c$ values of 113.92, 54.13, and 367.93 respectively for the Full Fig. \eqref{fig:full_mle}, the Normalized Fig. \eqref{fig:norm_mle}, and the RMT Correlations Fig. \eqref{fig:rmt_mle}. Their economic GICS information recovered via the ARI were, following the same order, 0.25, 0.35 and 0.41.
While Fig. \eqref{fig:full_mle} has the smallest number of clusters 15, we find clusters which mostly contain firms from single industries such as the financial, utilities, Real Estate, and Energy sectors. The other clusters more or less mixed including a very large one which we could refer to as the ``market''. Recall that Fig. \eqref{fig:rho_full} shows the correlations of the ``Market Mode'' are mostly positive, and one can easily infer this kind of result.  Fig. \eqref{fig:norm_mle} and Fig. \eqref{fig:rmt_mle} provide a cleaner pictures of the market: in Fig. \eqref{fig:norm_mle} there is no large ``market'' cluster and every industry is mostly represented in their own respective clusters. Firms, previously found in the ``market'' are now for most of them located in clusters representative of their respective industries. Similar situation in Fig. \eqref{fig:rmt_mle} except the industry sectors have a better definition while a large mixed cluster remains present similarly to Fig. \eqref{fig:full_mle}.

Th neighborhood search SPC performs constrains the scope of the optimization. In SPC's case, SA can only minimize the Hamiltonian $H_S$ over the neighborhoods necessitates an additional decision in picking the neighborhood size $K$ which acts as a hyper-parameter. The likelihood $L_c$ is optimized over the whole range of observations effectively removing such need, and making the optimization fully unsupervised. This also means that there exists a possibility that nodes which wouldn't cluster together, because of neighborhood limitation, would in this particular case. It is unclear which is the best way to proceed.

Our second goal is to explore clustering differences which arise in our 3 cases. In Fig. \eqref{fig:full_sa}, and Fig. \eqref{fig:full_mle} The biggest clusters have significant overlap with economic sectors except for a few large ones such as the financials cluster which also houses stocks from other sectors. The picture gets cleaner once we look at Fig. \eqref{fig:norm_sa}, and Fig. \eqref{fig:norm_mle} where we now have less mixing in most clusters, and finally in Fig. \eqref{fig:rmt_sa}, and Fig. \eqref{fig:rmt_mle} some of the clusters such as the Real Estate, Utilities, Health Care, and Consumer Staples found in the Normalized case are split. We noticed a similar result in Sec. \ref{ssec:iris} where the $L_c$ solution had a higher number of clusters, but they were essentially subgroup within the ones found by SPC.

\subsection{BRICS data} \label{ssec:brics}

We obtained publicly available BRICS (Brazil, Russia, India, China and South Africa) stock market data \footnote{C. Nugent 'S\&P 500 stock data - Historical stock data for all current S\&P 500 companies', 2017-2018. [Online]. Available: \url{https://www.kaggle.com/camnugent/sandp500}. [Accessed: 01-Dec-2017]}. The original data contains daily closing prices of 226 stocks:Brazil (60), China (50), India (30), Russia (43), and South Africa (43). We will refer to BRICS as a way of listing the mentioned countries in the previously given specific order. The window spans 2005 to 2015 from which we retained the last 5 years of daily trading. The data set suffers from a missing data problem which we would make it impossible to compute correlation matrices. We deal with the problem by using the time-series missing-data  which consists on computing correlations only on overlapping sections of time-series. The resulting correlation matrix is then made positive definite, and cleaned using IMN (See Sec \ref{sssec:succnorm} ).

\begin{figure*}
	\begin{subfigure}[b]{0.45\textwidth}
		\includegraphics[width=\textwidth]{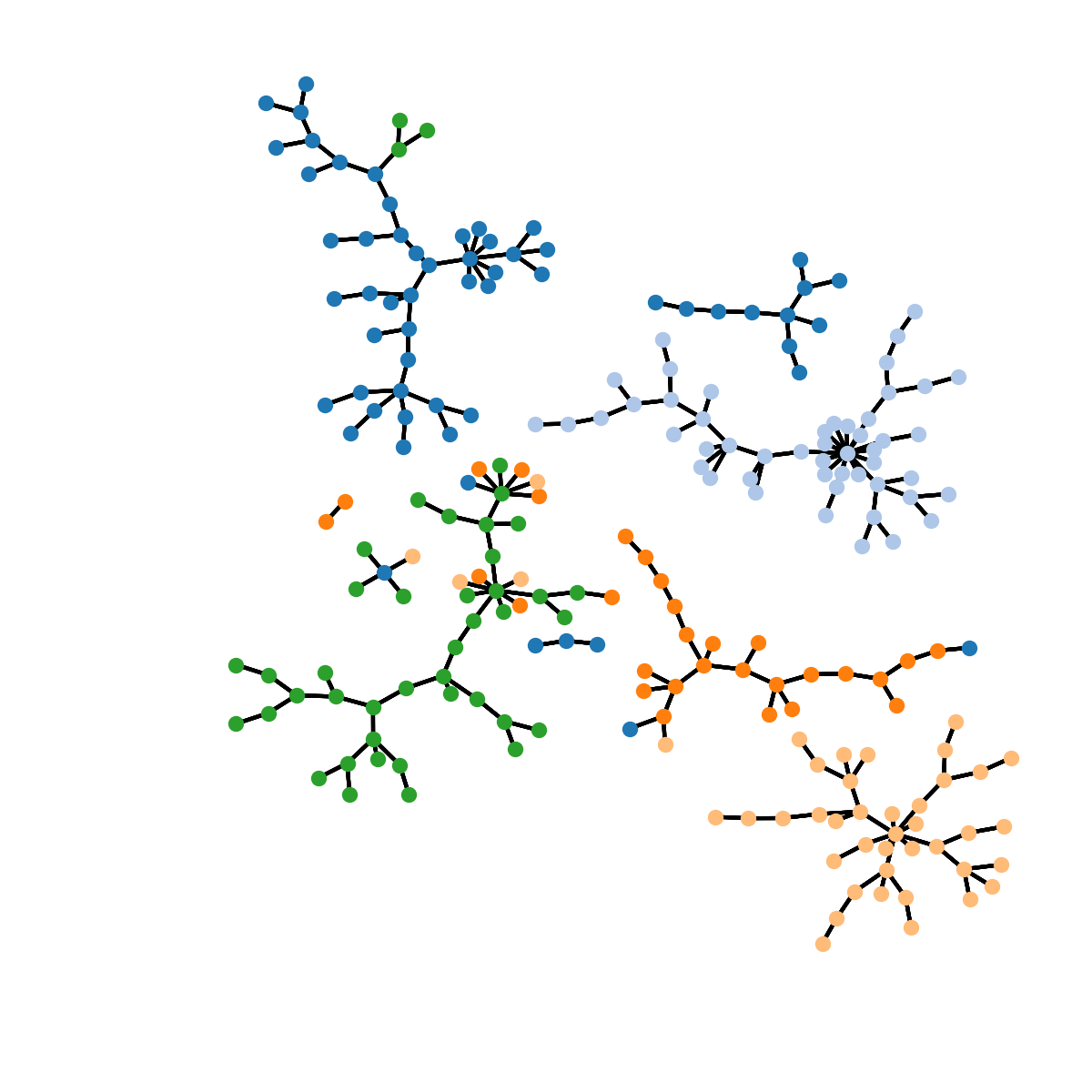}

		\caption{SPC BRICS}\label{fig:brics_spc}
	\end{subfigure}
	~
	\begin{subfigure}[b]{0.45\textwidth}
		
		\includegraphics[width=\textwidth]{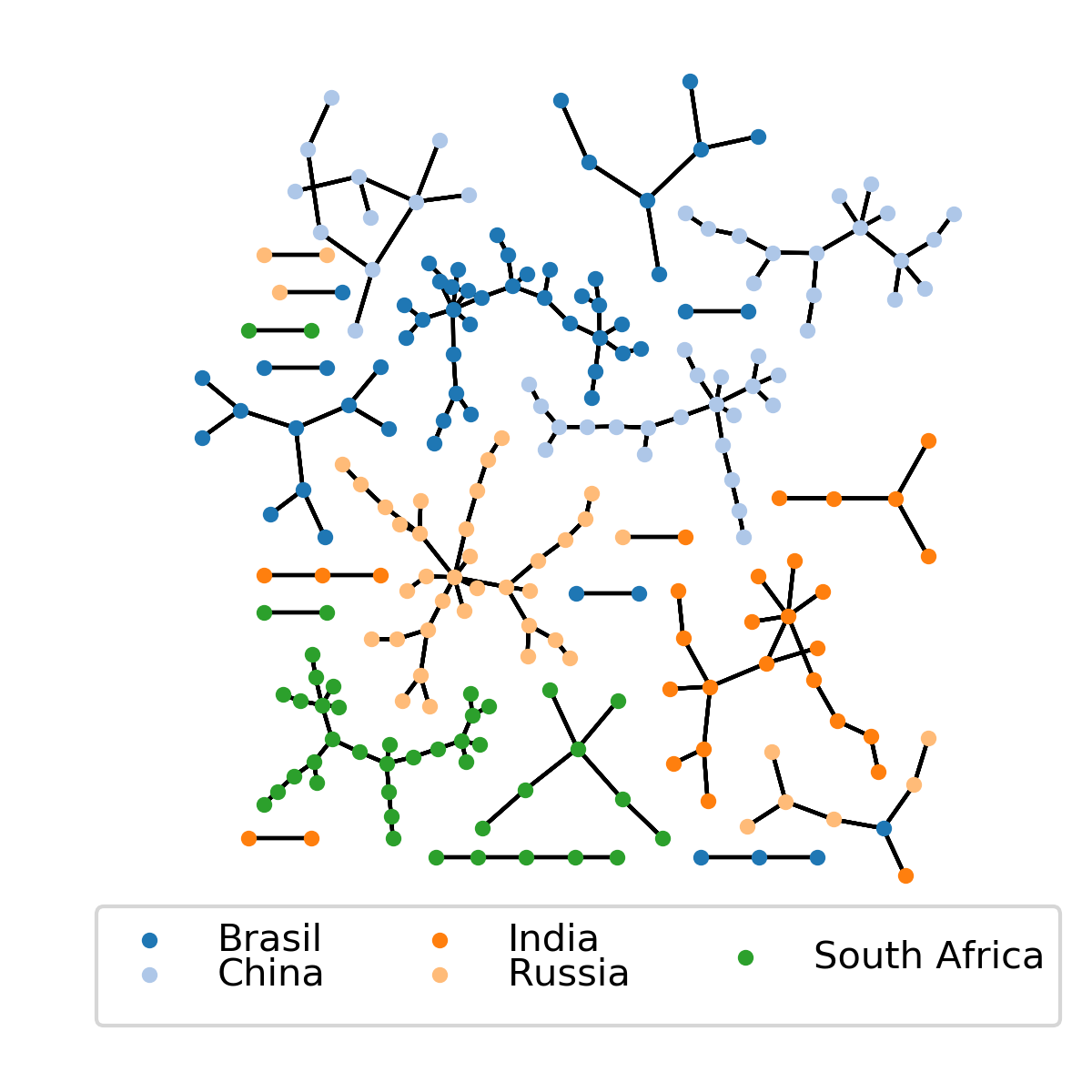}

		\caption{FSPC BRICS}\label{fig:brics_louvain}
	\end{subfigure}
	\caption{226 BRICS stocks. Data cleaned using IMN (See Sec. \eqref{sssec:succnorm}). in a) SPC solution at $T = 0.12 $, and in b) f-SPC's solution with $L_c = 7.56$}
\end{figure*}

Using MR the Likelihood's local maxima reached was 7.56, with 24 clusters: very little mixing, and the 5 countries are concentrated inside 1 to 3 clusters per country (see Fig. \eqref{fig:brics_louvain} ). This confirms our expectations which were that same country stocks should mostly belong to the same clusters, and the existence of multiple clusters per country as evidence of meso-scale industry/sectoral level classification.

We continue with the SPC result given for $T=0.12$ in Figures \eqref{fig:brics_spc}, and \eqref{fig:brics_clusters_verticals} with 9 clusters.
The ARI between the SPC and f-SPC solutions is 0.5. Globally speaking the same clusters are recovered except for their size being slightly smaller for the MR solution, and clusters ``purity'' is higher in the MR candidate potentially (stocks which belong to different countries do not mix). In both candidates Brazil financial market is divided in 2 clusters which upon a more detailed cluster analysis could reveal industry (or sector) economic subdivisions.

\section{Discussion} \label{sec:disc}


In this paper, we were able to successfully implement SPC, and f-SPC, and we tested those methods on synthetic, and real data. If there exists significantly different structures within the data, SPC will exhibit multiple transitions within the SP-phase. As the temperature is varied, the couple Susceptibility $\chi$ and Average Magnetization $\langle m \rangle$ signal the occurrence of phase transitions. The spin-spin correlation $G$ is indirectly linked to the interaction strength $J$ and the densities found in the data. The method has the advantage of being unsupervised for the most part, it does not necessitate {\it a-priori} knowledge of the number of clusters, and makes no assumption about the distributions of the data. While SPC performs a neighborhood search, which requires picking a value for $K$, it does not affect the simulation significantly for large data sets; as was previously seen in \cite{blatt1997data}. The parameter $\theta$ is set to 0.5 and helps decide clusters membership. We clustered at every temperature within a pre-determined range such that we do not need to identify ``clustering temperatures $T_{clus}$'' like in \cite{blatt1996superparamagnetic} but we do so at the expense of additional computational cost. We note that in the literature there are modifications of the Potts model clustering which automate parameter selections for the clustering temperature $T_{clus}$, the local length scale $a$, and the cluster membership threshold $theta$ through validation based calibration. \cite{murua2014conditional}.

Once we have a hierarchy of configurations, such as in Fig. \eqref{fig:dig_clus}, one needs to select appropriate clusters representative of the different regimens of the SP-phase. This is an easy task as can be seen in Fig. \eqref{fig:win_clus} when the number of cluster is low, and the clusters have similar densities which establishes a stable phase for a relatively wide temperature range. On the other hand if the number of clusters is high, and the data is composed of clusters of different densities, the susceptibility, which tracks the variance of biggest cluster has its limitations \cite{kullmann2000identification}. As the size of the data-set increases the susceptibility is useful as a tool to locate the final transition and lowest clustering level.

f-SPC only requires the correlation matrix and is completely unsupervised. The randomly generated population is diversified at every iteration by applying as many as 7 different mutations. It's a fast and deterministic algorithm while SPC is MCMC based and requires statistical averages. The computation time is affected by the order of the observations in the data \cite{blondel2008fast}. We noticed that ordering our data based on the order of one of the observations' closest ``neighbors'' produced better results which should motivate further exploration of potential heuristics dealing with this issue.

$L_c$ measures the quality of cluster configurations: its value is computed from the clusters sizes $n_s$ and the intra-cluster correlations $c_s$. The optimization is global which, as opposed to SPC, avoids the need to determine a neighborhood size $K$. There exist problems where choosing a sufficiently big $K$ has an non-trivial impact on SPC's solutions. One such example would be the existence of a relatively low density and sparse cluster in a data set mostly composed of high density clusters. Low values of $K$ would fail to recover the low density clusters which would remain unclassified whereas this isn't an issue for f-SPC which would perform much better.

f-SPC results are consistent for high dimensionality data-sets. if we consider the metric used to evaluate the noise in correlation matrices $q = \frac{N}{D}$ as the ratio of the dimension over the number of observations in the limit of $N \rightarrow \infty$. We recall that in \cite{wilcox2007analysis}, $q$ encodes the noise level of the eigenvalues of the sample covariance matrix. $q$ values for our problems are 250 for the two circles, 13.69 for the wines, 166.66 for the 3D blobs, 1 for the 500D blobs, 37.5 for Fisher's Iris, 7.81 for the MNIST digits, and 0.35 for the NYSE Kaggle data. The $L_c$ results consistent with SPC were the 500D blobs, MNIST, and the NYSE stock data which all confirm that a low $q$ is necessary to compute appropriate correlation matrices. We want $q$ to be as small as possible, and if possible close to 0. This is not always the case, and we have tested ways to de-noise the correlation matrix (Sec. \ref{sssec:succnorm}, and \ref{sssec:rmt}) in the case of financial time-series but it is unclear at this time what would the solutions be in other cases. In \cite{giada2001data} Marsili and Giada derive $L_c$, and along the way they assume that $D\rightarrow \infty$ which in turn means one has to consider the finite size effects of the method. Fig. \eqref{fig:b_lc3d} and Fig. \eqref{fig:b_lc3d} tell us that if we were to visualize the $L_c$'s objective surface, depending on dimensionality of the problem we could face a ``rough'' space. One could consider $L_c$ as a sort of modularity function just like in the Network Science literature. One major Network Science problem is the efficient detection of communities inside networks. Similar to our work, cluster configurations are the input of the modularity function $Q$ which, through diverse heuristics, is maximized. However it is well known \cite{good2010performance} to Network Scientists that modularity objective surfaces are degenerate: many significantly different clustering results have similar modularity, or in our case, higher likelihood than the true clustering.

We compare this to the modus operandi of SPC: The generative model of the SPC is the Gibbs-Boltzmann distribution which not only validates clusters locally using Eqn. \eqref{eq:bolt_link_dis}. It is a bottom up approach as opposed to global optimization methods which are top down. One assumes that there exists multiple realizations (micro-states) of the generative model, the so called ``equivalence classes'' which are valid representation of the data. In order to link micro and macro-state one could pick any one micro-state translating into the desired macro-state: Maximum Likelihood methods essentially achieve this feature by searching the space of solutions for any candidates meeting the global objective. We argue that in complex systems, the existence of equivalence classes as illustrated by the degeneracy of clustering objective surfaces leads to the Maximum Entropy principle \cite{jaynes2003probability} as an alternative optimization device. The generative model generates equivalence classes (among which are included maximum likelihood candidates) each with differing probabilities, and one then needs to probabilistically combine them to achieve some sort of representative weighted average.


We suspect a way to deal with cases where $D$ is small compared to $N$, and indirectly $q \gg 1$, would be a modification of $L_c$ by adding an additional term acting as a regularizer which could account for the number of clusters. Our rationale follows that $L_c$ as an objective function is degenerate with multiple spin configurations whose likelihood are equal or very close. This degeneracy comes from, if we assume the minimum number size of clusters to be 2 (no singletons), the number of possible configurations $(\frac{N}{2})^N$ which for a case $N = 100$ would be on the other of $10^{169}$. 

Finally one is left to decide which de-noising method is deemed optimal and as a consequence which clustering one prefers. The assumptions in both methods have their validity and should be carefully considered. Whereas IMN (Sec. \ref{sssec:succnorm}) consider the covariance matrix as IID normal random variables, RMT (Sec. \ref{sssec:rmt}) predicts a spectrum of random matrices eigenvalues exists which is pure noisy signal. The noise is removed by reconstructing the data without the noisy eigenvalues whose number increases with dimensionality. We suspect a proper way of deciding which method is optimal is the implementation of such methods as bases of trading strategies.

\section{Conclusion} \label{sec:conclusion}

In this paper we have presented two unsupervised data clustering algorithms inspired by the Potts Model \cite{wu1982potts}. Using SA (SPC Sec. \ref{sssec:maxent}) optimizes the Hamiltonian of a thermodynamics systems, and the ground state energy solution recovered provides the best clustering structure present in our data. We show that the parameter-free Marsili-Giada (Sec. \ref{sssec:mle}) maximum likelihood method implemented with a modified version of Hendricks {\it et al.}\cite{hendricks2016detecting,hendricks2016high} Parallelized Genetic Algorithm recover solutions similar to those found in the super-paramagentic phase Fig. \eqref{fig:sp500_persistence}. 

This was done by comparing the SPC solutions to the f-SPC one using the ARI \cite{lloyd1982least}. In addition to this we compare the Likelihood of SPC solutions to f-SPC ones to show that f-SPC have higher likelihood; this prompts an additional discussion on implication for statistical inference in complex systems. The methods were tested both on several standard toy test cases as well as real stock market time-series data; this illustrates their universality provided an appropriate similarity metric is selected, such as the Pearson correlation coefficients or the correlation distance. We are able to show that the results are similar to the 11 standard GICS economic sectors, however the differences in the number of clusters, and their composition should be cause for concerns with respect to the use of GICS classification. Hence, we question the effectiveness of naively using GICS classifications for risk-management and for investment decision making on both the medium and short-term. 

Building on the work presented in this paper, and original aim, as in \cite{hendricks2016detecting}, we would have liked to perform cluster analysis and compared configuration of stock market intra-day time-series. The last thirty years have seen a magnificent increase in technological power which enabled simultaneous trading on multiple time scales. The so called High Frequency Trading (HFT) paradigm increased tenfold the amount of stock market data, and has significantly impacted the market participants behaviors at shorter time scales. It is therefore natural to consider market participants all having different operating time horizons and for this to somehow be reflected in the non-trivial composition of clusters. We assume traders use all information available to make decisions however the rate of economic information released about publicly traded firms can range from once a month to once a year. This rate is significantly lower than that of HFT which leads us to think trading is not solely based on economic information, and because of the different time-scales one can suspect different objectives may be at play \cite{WilcoxGebbie2014}. The key motivation here was to build and test high-speed classification methods towards the goal of unsupervised classification of traders from live market data. It is notoriously difficult to obtain trading data linked to individual market participants accounts. This kind of data would be extremely useful to directly, not only study traders' behavior, but begin to understand the kind of ecosystem a financial market is. Unfortunately one is only left with the possibility proxy studies through the dynamics of the traded securities. 

One possible approach is the unsupervised clustering of simulated technical trading agents implementing investment strategies in an artificial financial market. This has been a key motivation in the present work. One logical next step is what we call Dynamical Cluster Analysis (DCA): Events such as financial crises like the one which preceded the 2008 Great Recession can be investigated at the intra-day scale. Here again we conjecture shocks to the system irremediably affect strategies, and clustering structures are less persistent with time as in \cite{musmeci2014risk}. Ultimately some sort of quantification of clustering on different temporal scales could be useful towards probing potential hierarchical causal affects given that different effective theories may dominate at different scales \cite{WilcoxGebbie2014}.

We have so far worked with changes in price returns as our single factor model. The derivation and formulation of the Giada-Marsili $L_c$ allows for multivariate clustering: We can use F by N by N Correlation matrices where $F$ is the number of factors we want to include. Another challenge however would be that at this time we are not aware of an implementation of multi-factor correlation based SPC. In closing, we promote the idea that a promising future research direction would be quantized spin-models and ultimately building unsupervised learning algorithms that more effectively accommodate state-interference and phase information within some likelihood method, as well as additional optimization refinements to the algorithm, perhaps using community detection algorithms to enhance performance \cite{blondel2008fast}. 
 
\section{Acknowledgements} \label{sec:ackno}
 The authors thank Etienne Pienaar, Micheal Gant, Jeff Murugan, Nic Murphy and Diane Wilcox for discussions and comments. LY would like to thank Fangqiang Zhu for first introducing him to the elegance of Ising models and their simulation. TG acknowledges the University of Cape Town for FRC funding (fund 459282).

\appendix
	
\section{The Algorithms} \label{sec:algos}

\setcounter{table}{0}
\renewcommand{\thetable}{\arabic{table}}

The algorithms implemented in this paper have been coded in python and are available on a github repository at \cite{git:potts}.
\subsection{SPC Algorithms} \label{ssec:alg_spc}

We provide a pseudo-code for the SPC \cite{blatt1996superparamagnetic} algorithm introduced in Sec. \ref{sssec:maxent}. Given a distance matrix, and a neighborhood of size K, the algorithm uses the Swendsen-Wang \cite{swendsen1987nonuniversal} MCMC method to optimize the thermodynamic system.



	\begin{algorithm}[!htb]
	\caption{\label{alg:spc} SPC (Sec. \ref{ssec:spc}, and \cite{blatt1996superparamagnetic}), and function ``\texttt{runz}'' in {\tt``Super-Paramagnetic-clustering.py''} in \cite{git:potts}}
		\begin{algorithmic}[1]
						\STATE{\textbf{INPUT}: Strength Matrix $J$; Neighbor list \texttt{nodenext}; Temperature $T$; }
			\FOR { $k=0$ to $k=M$}
			
			\STATE{ Create edge configuration matrix ``\texttt{link}''}
			
			\STATE { Create Swendsen-Wang clusters }
			
			\STATE { Compute, and store thermodynamic quantities i.e. $m, c_{ij}$}
			\ENDFOR
						\STATE{\textbf{OUTPUT}: $\chi$;$\langle m \rangle $;$\langle H \rangle $;Spin-spin correlation matrix $ G_{ij} $}
		\end{algorithmic}
	\end{algorithm}


Hoshen-Kopelman \cite{al2003extension} is the mechanism behind clusters discovery in SPC. It allows for the graph to be traveled via its nodes and neighborhoods while clustering the system locally.
%
	
	\begin{algorithm}[!htb]
		\caption{\label{alg:hk} Extended Hoshen-Kopelman (Table \ref{tab:extendedHK}, and \cite{al2003extension}), and \cite{swendsen1987nonuniversal}), and function ``\texttt{eHK}'' in ``Super-Paramagnetic-clustering.py'' in \cite{git:potts}}

		\begin{algorithmic}[1]
			\STATE{\textbf{INPUT}: Matrix of links}
			
			\STATE {set label counter to 0, Initialize \texttt{nodel} Nx1 array}
			
			\STATE {Create \texttt{nodelp}, an empty array}
			
			\FOR {$i = 0$ to $i=N$}
			\IF {node i isn't linked at all}
			\STATE {Set label counter to i's label}
			\STATE { Store i's label to \texttt{nodelp}}
			\STATE {Increase label counter by 1}
			\ELSE 
			\STATE {Find i's linked neighbors, and store their \texttt{nodelp} labels}
			\IF {None are labeled}
			\STATE {Set label counter to i's label}
			\STATE { Store i's label to \texttt{nodelp}}
			\STATE {Increase label counter by 1}
			\ELSE
			\STATE {store the labels of the linked neighbors}
			\STATE {Store root of the labels of the linked neighbors}
			
			\STATE {Set min the smallest root label}
			
			\STATE {Set the \texttt{nodel} of i to min}
			\STATE {In \texttt{nodelp} change linked neighbors root labels to min}
			\ENDIF
			\ENDIF
			
			\COMMENT {Make \texttt{nodelp} sequential}
			\FOR { $y = 0$ to $y = \texttt{len(nodelp)}$}
			\STATE {$ n = y $}
			\WHILE { The root of n is less than n }
			\STATE {Set n to the root of n}
			\ENDWHILE
			\STATE { Set the root of y to n }
			\ENDFOR
			
			\COMMENT { Relabel the labels with their roots }
			\FOR { $i = 0$ to $i = \texttt{len(nodelp)}$ }
			\STATE { Find labels in \texttt{nodel} == i}
			\STATE { Update them with their root in \texttt{nodelp}}
			\ENDFOR
			\ENDFOR
			\STATE{\textbf{OUTPUT}: Spin configuration}
			
		\end{algorithmic}
		
	\end{algorithm}
%
%
\pagebreak
Swendsen-Wang flips clusters at each iteration, allowing the quick convergence toward a local maxima.


\begin{algorithm}[!htb]
	\caption{\label{alg:sw}Swendsen-Wang (Table \ref{tab:sw}, and \cite{swendsen1987nonuniversal}), and function ``\texttt{eHK}'' in ``Super-Paramagnetic-clustering.py'' in \cite{git:potts} }
	
	\begin{algorithmic}[1]
		\STATE{\textbf{INPUT}: Strength Matrix $J$}

		\STATE {Create new spin configuration using Hoshen-Kopelman}
		
		\STATE {Flip all clusters}
		\STATE{\textbf{OUTPUT}: Flipped Spin Configuration}
		
	\end{algorithmic} 
	
\end{algorithm}
%
\subsection{f-SPC Algorithms} \label{ssec:alg_fspc}

We discussed f-SPC in Sec. \ref{sssec:mle}, and here we provide a pseudo-code for the implementation of the parallelized genetic algorithm which generates clustering candidates, evaluates their likelihood $L_c$ \cite{giada2001data} using Eqn. \eqref{eq:lc}, selects the best candidates and discards the others.


	\begin{algorithm}[!htb]
	\caption{\label{alg:ga} f-SPC PGA (Sec. \ref{sssec:mle}, and \cite{hendricks2016detecting}), and {\tt``fast-SP-clustering.py''} in \cite{git:potts}}

		\begin{algorithmic}[1]
			
			\STATE{\textbf{INPUT}: N individuals; G generations; $L_c$ likelihood function; Mutation function; Recombination function; Correlation Matrix $\rho$ }

			\STATE{Create Initial Population of N individuals}
			
			\FOR { (In parallel ) All individuals}
			\STATE{Evaluate Fitness}
			\ENDFOR
			
			\FOR {Number of Generations G}
			
			\STATE{Create Offspring (from the entire population)}
			
			\STATE{Mutate offspring}

			\FOR { (In parallel ) All offspring}
			
			\STATE{ Evaluate Fitness}
			
			\ENDFOR
			
			\STATE{ Recombine Parents and Offspring }
			
			\STATE{ Select Next Population (N individuals) }
			
			\ENDFOR
			\STATE{\textbf{OUTPUT}: Generation with highest likelihoods}
			
		\end{algorithmic}
	\end{algorithm}

\newpage
\bibliographystyle{cas-model2-names}
\bibliography{lytg_fspc} 

\end{document}